%% file: MS_paper.tex
\documentclass[12pt]{amsart}
\usepackage{geometry}                
\usepackage[leqno]{amsmath}
\usepackage{amsthm}
\usepackage{graphicx}
\usepackage{xcolor}
\usepackage{setspace}
\usepackage[round]{natbib}
\usepackage{amssymb}
\usepackage{multirow}
\usepackage{array}
\usepackage{amsfonts}
\usepackage{mathtools}
\usepackage[foot]{amsaddr}
\usepackage{xr}
\usepackage{comment}

\usepackage{bm}
\usepackage{framed}
\usepackage{pbox}
\usepackage{pdfpages}

\usepackage{algorithm}
\usepackage{algorithmic}

\usepackage{tikz}
\usepackage[hidelinks]{hyperref}
\hypersetup{
  pdftitle={Binary Classification with the Maximum Score Model and Linear Programming},
  pdfauthor={Joel L. Horowitz and Sokbae Lee},
  pdfsubject={Maximum score estimation, partial identification, and binary classification},
  pdfkeywords={binary classification, maximum score, partial identification,
               linear programming, minimax regret}
}

\newtheorem{theorem}{Theorem}
\newtheorem{assumption}{Assumption}
\newtheorem{corollary}{Corollary}[section]
\newtheorem{lemma}{Lemma}

\newtheorem{proposition}[theorem]{Proposition}
\theoremstyle{remark}
\newtheorem{remark}{Remark}

\newtheorem*{example*}{Example}

\numberwithin{theorem}{section}
\numberwithin{lemma}{section}
\numberwithin{equation}{section}
\newcommand{\eps}{\varepsilon}

\newcommand{\G}{\mathcal G}
\newcommand{\Salpha}{\mathcal S_\alpha}
\newcommand{\Jhat}{\widehat{\mathcal J}_\alpha}
\newcommand{\Ehat}{\widehat{\mathcal E}_\alpha}
\newcommand{\Ealpha}{\mathsf E_\alpha}
\newcommand{\ReFP}{\mathrm{Reg}_{\mathrm{FP}}}
\newcommand{\ReFN}{\mathrm{Reg}_{\mathrm{FN}}}
\newcommand{\ReEmp}{\mathrm{Reg}_{\emptyset}}
\newcommand{\ReEmpp}{\mathrm{Reg}_{\emptyset}^{+}}
\newcommand{\ReEmpm}{\mathrm{Reg}_{\emptyset}^{-}}

\defcitealias{CL34}{Clopper--Pearson (1934)}

\begin{document}

\title[Binary Classification]{Binary Classification with the Maximum Score Model and Linear Programming}
\author[Horowitz]{Joel L. Horowitz$^1$}
\address{$^1$Department of Economics, Northwestern University, Evanston, IL 60208, USA.}
\email{joel-horowitz@northwestern.edu}

\author[Lee]{Sokbae Lee$^2$ $^3$}
\address{$^2$Centre for Microdata Methods and Practice,
Institute for Fiscal Studies, 7 Ridgmount Street, London, WC1E 7AE, UK.}
\address{$^3$Department of Economics, Columbia University, New York, NY 10027, USA.}
\email{sl3841@columbia.edu}

\thanks{We thank Kei Hirano, Chuck Manski, Adam Rosen, J\"{o}rg Stoye,
Elie Tamer, and three anonymous referees for helpful comments.
Part of this research was conducted while we were visiting cemmap and the
Department of Economics at University College London.}

\date{August 28, 2026}%

\begin{abstract}
This paper presents a computationally efficient method for binary classification using Manski's (1975, 1985) maximum score model when covariates are discretely distributed and parameters are partially but not point identified. We establish minimax-regret-optimal classification rules that take account of partial identification of the model's parameters. We bound misclassification probabilities and expected excess regret induced by sampling uncertainty. We also describe an extension of our method to continuous covariates. Our approach avoids the computational difficulty of maximum score estimation by reformulating the problem as two linear programs. Compared to parametric and nonparametric methods, our method balances extrapolation ability with minimal distributional assumptions. Monte Carlo simulations and empirical applications demonstrate its effectiveness and practical relevance. \\


\end{abstract}

\maketitle

\onehalfspacing


\section{Introduction}

This paper introduces a computationally efficient method for applying
Manski's (\citeyear{Manski:1975}, \citeyear{Manski:1985})
maximum score model of binary response to binary classification when covariates may be discretely distributed and parameters are partially but not point identified.
We establish minimax-regret-optimal classification rules that take account of partial identification of the model's parameters.
We give conditions under which these are equivalent to minimizing the worst-case cost of misclassification.
We also bound misclassification probabilities and expected excess regret induced by sampling uncertainty.

In Manski’s model, the binary dependent variable $Y$ is related to a finite-dimensional vector of explanatory variables $X$ as follows:
\begin{align}\label{model-intro}
Y=I\left(\beta^{\prime} X-U > 0\right), \quad \mathbb{P}(U < 0 \mid X)=\tau,
\end{align}
where $I(\cdot)$ is the indicator function, $\beta$ is an unknown vector of constants,
the conditional distribution of \(U\) given \(X\) is absolutely continuous
with respect to Lebesgue measure,
and $0<\tau<1$, implying that zero is the $\tau$-quantile of $U$ conditional on $X$. The distribution of $U$ is otherwise unrestricted and may depend on $X$ in an unknown way (i.e., heteroskedasticity of unknown structure).

Given an independent random sample
\(\{(Y_i,X_i):i=1,\ldots,n\}\), the corresponding maximum score
estimator of \(\beta\) is defined as
\begin{align}\label{def:mse}
\hat{\beta}\in
\arg\max_{b\in\mathbf B}
\sum_{i=1}^n (Y_i-\tau) I(b'X_i > 0),
\end{align}
where $\mathbf{B}$ is the parameter space.

If $\beta$ were known, an individual with covariate $X_*$ would be classified as $Y=1$ if $\beta' X_* > 0$ and $Y=0$ otherwise. As a tie-breaking rule, we assign $Y = 0$ if $\beta' X_* = 0$.
However, when $X$ follows a discrete distribution, $\beta$ and $\beta' X_*$ are not point identified. Under mild conditions, $\beta' X_*$ is contained in an identified interval, denoted by $[c_L, c_U]$ with $c_L \leq c_U$.
In this paper, we propose two simple classification rules. In one rule, an individual with covariate $X_*$ is classified as $Y=0$ if $c_U \leq 0$, as $Y=1$ if $c_L \geq 0$ and \(c_U>0\), and remains unclassified, for example while additional information is sought, if $c_L < 0 < c_U$. Here, the exact boundary case \(c_L=c_U=0\) is assigned to \(Y=0\), as this corresponds to $\beta' X_* = 0$.
This minimax-regret optimal classification rule relies on two elements:
non-classification (abstention) must be an available action, and, when the sign of \(\beta'X_*\) is ambiguous, abstention must have no larger worst-case regret than either possible binary classification rule.
In the second classification rule, abstention is not possible, and decisions are made via randomization.
In this case, we classify as $Y=0$ if $c_U \leq 0$, as $Y=1$ if $c_L \geq 0$ and \(c_U>0\), and randomize if $c_L < 0 < c_U$. We again establish conditions under which this randomized classification rule is minimax-regret optimal.

Our approach is useful when the researcher wants out-of-sample classifications, an interpretable linear index is plausible, and abstention or randomization has a natural decision-making interpretation. The value of \(\tau\) is part of the classification exercise. When it is not governed by the underlying decision problem, the sample
proportion of observations with \(Y=1\) provides a simple empirical
benchmark for choosing a reasonable value of \(\tau\) (see Section~\ref{sec:examples} for examples).

We further present estimators for $[c_L, c_U]$ and construct both finite-sample and asymptotic confidence intervals, denoted by $[\hat{c}_L, \hat{c}_U]$. The estimated classification rule follows the same logic:
we classify an individual as $Y=0$ if $\hat{c}_U \leq 0$ and as $Y=1$ if $\hat{c}_L \geq 0$ and \(\hat{c}_U>0\), with non-classification or random classification occurring when $\hat{c}_L < 0 < \hat{c}_U$.
We say that a misclassification occurs if the feasible
classification based on \([\hat c_L,\hat c_U]\) differs from the population
classification based on \([c_L,c_U]\). Thus misclassification here is a
sampling error relative to the population maximum-score classification rule.
Theoretically, we bound these misclassification probabilities and the
resulting expected excess regret.

Computing $\hat{\beta}$ (solving the optimization problem \eqref{def:mse}) is a notoriously difficult, NP-hard combinatorial optimization problem. In addition, $\hat{\beta}$ converges in probability to $\beta$ at the slow rate of $n^{-1 / 3}$ when $\beta$ is point identified. The objective of this paper, however, is estimation of the interval $\left[c_L, c_U\right]$ that contains $\beta^{\prime} X_*$, not point identification or estimation of $\beta$. We give conditions under which $\left[c_L, c_U\right]$ can be estimated by solving two linear programming problems, one for $c_L$ and one for $c_U$. Computationally efficient algorithms and software for solving linear programming problems are widely available, so we avoid the computational difficulty of solving
\eqref{def:mse}.

While the underlying model involves the parameter vector \(\beta\), our primary interest lies in conducting binary classification rather than in precisely estimating \(\beta\) itself. Although the bounds for \(\beta\) may be wide, informative classification remains possible. By focusing on the signs of linear combinations of covariates rather than on point estimates of coefficients, we leverage the structure of the identified set to assign classifications reliably. Thus, even when parameter uncertainty is substantial, the classification rule can yield meaningful and robust decisions.

Various methods exist for binary classification. Fully nonparametric methods, such as estimating $\mathbb{P}(Y=1 \mid X=X_*)$ nonparametrically, where $X_*$ is an out-of-sample value of $X$, and classifying based on whether this probability exceeds $\tau$, make minimal assumptions but suffer from poor extrapolation ability. If the covariates follow a discrete distribution, these methods fail to classify individuals whose $X_*$ is not observed in the sample and are imprecise if there are few observations of $X_*$.
Parametric methods, including logit, probit, and discriminant analysis, allow extrapolation but rely on restrictive distributional assumptions. The method we propose balances these trade-offs: it permits extrapolation while being less restrictive than fully parametric approaches and some semiparametric models. For example, semiparametric single-index models satisfying the quantile restriction are special cases of the maximum score model. We acknowledge, however, that despite its advantages relative to many other models, the maximum score model can be misspecified, which may cause misclassification.

The remainder of this paper is structured as follows.
Section~\ref{sec:lit-review} reviews the literature on maximum score estimation.
Section~\ref{sec:model} formally describes the model.
Section~\ref{sec:decision:theory} provides a framework for optimal classification.
Section~\ref{sec:estimation:inference} outlines our estimation procedure and discusses finite-sample and asymptotic inference with discretely distributed covariates.
Finite-sample inference typically yields conservative bounds in moderate-sized samples common in economics applications, as it imposes only weak constraints on the distribution of the underlying random variables and ensures validity for the worst-case scenario under those constraints.
In contrast, asymptotic inference yields tighter bounds but does not guarantee validity in finite samples.
Section~\ref{sec:computation} details our computational algorithms.
Section~\ref{sec:regret} studies misclassification induced by random sampling
error in the estimated confidence region.
Section~\ref{sec:hypercube-xv} extends the framework to accommodate continuous covariates, specifically incorporating \citet{manski2002inference}-type bounds.
Section~\ref{sec:examples} provides empirical illustrations:
Section~\ref{sec:KLS} presents an empirical example using data from \citet{KLS};
Section~\ref{example:gps:grouped} illustrates the extension discussed in Section~\ref{sec:hypercube-xv}, again using the same data;
Section~\ref{sec:CHV} presents another empirical application using data from \citet{Corno:ECMA:2020}.
Section~\ref{sec:mc:class-errors} reports a Monte Carlo experiment demonstrating the effectiveness of our classification rules.
Section~\ref{sec:conclusions} concludes with a discussion of key findings and potential directions for future research.
Appendix~\ref{sec:proofs} contains the proofs of the theoretical results presented in the main text.
Supplemental Appendix~\ref{sec:sensitivity-eps} reports sensitivity analyses in the estimation and computation steps.
Supplemental Appendix~\ref{subsec:refined-misclassification} gives
refinements of the misclassification bounds and extends the
analysis to expected excess regret.
Supplemental Appendix~\ref{sec:cluster} extends our framework
to clustered dependence and survey sampling weights. 
Supplemental Appendix~\ref{sec:MC} presents Monte Carlo simulations comparing finite-sample and asymptotic inference methods.

\section{Literature Review}\label{sec:lit-review}

The literature on the maximum score model is concerned mainly with identification and estimation of $\beta$,
methods for computing \(\hat{\beta}\) in \eqref{def:mse}, and inference.
The literature is large, and we cite only a limited set of papers that treat these topics.

\subsection{Identification, Asymptotic Properties, and Related Methods}

\citet{Manski:1988} provided classical results for point identification. Subsequently, advances have been made in terms of partial identification: see, e.g.,
\citet{manski2002inference}; \citet{komarova2013}; \citet{blevins2015}; \citet{chen2015}; and \citet{KKN:2024} among others.

When $\beta$ is point identified, $\hat{\beta}$ converges in probability to $\beta$ at the rate $n^{-1 / 3}$, and $n^{1 / 3}(\hat{\beta}-\beta)$ has an intractable asymptotic distribution \citep[see, e.g.,][]{kim1990,Seo2016}.
For inference, subsampling and bootstrap methods have been considered by,
among others, \citet{delgado2001}, \citet{abrevaya2005},
\citet{Lee:Pun:06}, \citet{Patra2015}, and
\citet{Cattaneo2020bootstrap}. These methods are computationally demanding
because they require solving \eqref{def:mse} repeatedly.

\citet{Horowitz:92,horowitz2002} developed smoothed maximum score estimation to mitigate the difficulties coming from the maximum score objective function.
The smoothed maximum score estimator is asymptotically normal with a faster rate of convergence and enables the bootstrap to provide asymptotic refinements.
Other approaches include:
finite-sample inference \citep{Rosen:Ura:2024};
Bayesian methods \citep{benoit2012,Walker:2024};
non-Bayesian Laplace-type approaches \citep{JPW15, JPW14};
and
two-stage methods \citep{GXX:2022}
 among many others.

\subsection{Computational Challenges}

\leavevmode
One major challenge in applying maximum score estimation is computation. The
maximum score objective is piecewise constant, has a finite range, and may have
multiple maximizers. Computing an exact maximum score estimator is
computationally challenging and, in general, NP-hard
\citep[see, e.g.,][]{Johnson:78}.
Early algorithms for addressing maximum score estimation were developed by \citet{Manski:Thompson:1986} and \citet{PINKSE1993}. Second-generation approaches leverage advances in mixed integer programming (MIP) solvers, along with significant increases in computing power since the development of the first-generation methods. \citet{Florios:Skouras:08} presented strong numerical evidence that an MIP-based method outperforms earlier techniques. Similarly, \citet{Kitagawa:Tetenov:2015} employed an alternative MIP formulation to solve a maximum score-type problem for treatment choice rules, where they maximized an empirical welfare criterion resembling the maximum score function. \citet{Chen:Lee:18,Chen:Lee:20} extended this approach to $\ell_0$-constrained and $\ell_0$-penalized empirical risk minimization for high-dimensional binary classification. However, the MIP approach faces intrinsic scalability limitations. To the best of our knowledge, numerical studies applying MIP to maximum score estimation typically remain confined to datasets with a modest sample size (e.g., around 1,000) and a limited number of parameters. In contrast, our approach is based on linear programming, which is considerably more scalable and capable of handling larger datasets with a greater number of parameters. 

\subsection{Support Vector Machines}

Our paper is also related to the large literature on binary classification
methods in machine learning and statistics. Support vector
machines (SVMs) are particularly relevant because they also seek a linear
decision boundary for binary classification; see \citet{steinwart2008support}
for a textbook treatment.

The maximum score method is built from the binary response
model and its quantile restriction, and can be interpreted as minimizing
a weighted classification loss: false negatives receive weight
\((1-\tau)\), and false positives receive weight \(\tau\). This corresponds to
the weighted binary classification loss in \citet[pp.~23--24]{steinwart2008support}
and to Manski's \((1-\tau)\)-absolute loss \citep[Sec.~3.3]{Manski:1988}.
The corresponding population minimizer is the
Bayes decision rule under that loss.

By contrast, a soft-margin SVM minimizes a regularized empirical hinge-loss
criterion. The hinge loss function is convex and a ``reasonable surrogate''
for the classification loss function
\citep[pp.~36--38]{steinwart2008support}. It is not the same as the
classification loss function, but
\citet[Proposition~3.1]{Babii:Chen:Ghysels:Kumar:2026} give conditions under
which the unrestricted population minimizers of suitably loss-weighted hinge
and other convex surrogate losses yield the correct classification rule at
covariate values in the support of $X$. However, these population results do
not determine how a covariate value $X_*$ outside the support of $X$ should be
classified. With discrete data, there are in general many linear decision
boundaries (classification rules) consistent with the maximum score
restrictions. Our method quantifies the resulting uncertainty in how a given
out-of-support $X_*$ should be classified.

\section{Model}\label{sec:model}

This section describes the model we treat and the parameters of interest.
We begin with definitions and notation.

Let $X \in \mathbb{R}^q$ be a random or non-stochastic (fixed) vector.
If $X$ is random, assume for now that it is discretely distributed with $J$ mass points $\left\{x_j: j=1, \ldots, J\right\}$ and probability mass function $p\left(x_j\right)>0$.
Continuously distributed components are treated in Section~\ref{sec:hypercube-xv}.
If $X$ is fixed, let it have $J$ possible values $\left\{x_j: j=1, \ldots, J\right\}$.
In this case, we treat a design in which there are $n$ observations of $(Y,X)$ and $n_j > 0$ of these observations have $X=x_j$. Define $p\left(x_j\right)=n_j / n$, the fraction of observations in the design for which $X=x_j$. For any random event $A$, let $P\left(A \mid X=x_j\right)$ denote the probability of $A$ conditional on $X=x_j$ if $X$ is random and the probability of $A$ at $X=x_j$ if $X$ is fixed.

The model is
\begin{align*}
&Y =
\left \{
\begin{array}{cc}
1 &\text{ if } \beta'X - U > 0 \\
0 &\text{ if } \beta'X - U \leq 0
\end{array}
\right.,
\end{align*}
where \(\beta \in \mathbb R^q\) is an unknown vector of constant parameters, the conditional distribution of \(U\) given \(X=x_j\) is absolutely continuous with respect to Lebesgue measure, \(\mathbb{P}(U < 0 \mid X=x_j)=\tau\) for all \(j=1,\ldots,J\), and \(0<\tau<1\).  We assume that this model is correctly specified unless stated otherwise.
Note that the $(1-\tau)$ quantile of $Y$ conditional on $X=x_j$ is
\begin{align}\label{model:qr:iden}
\mathbb{Q}_{1-\tau}\left(Y \mid X=x_j\right)=I\left(\beta^{\prime} x_j > 0\right),
\end{align}
where $I (\cdot)$ is the usual indicator function.
A scale normalization is needed for point or partial identification of $\beta$.
We use $\beta_1 =1$, where $\beta_1$ is the first component of $\beta$.

For each $j=1, \ldots, J$ define
\begin{align}\label{eq:dj}
f(x_j) \equiv [\mathbb{P}(Y = 1 |  X = x_j) -\tau]
\; \text{ and } \;
d_j \equiv \text{sgn}[ f(x_j) ],
\end{align}
where for a real number $u$, $\text{sgn}(u) = 1$ if $u > 0$,  $\text{sgn}(u) = -1$ if $u < 0$, and  $\text{sgn}(u) = 0$ if $u = 0$.
Let $r^{\prime} \beta$, where $r \in \mathbb{R}^q$ is a non-zero vector of constants, be a linear combination of components of $\beta$. The parameter $\beta$ is generally not point identified when $X$ is discretely distributed with finitely many mass points or fixed with finitely many possible values. The linear combination $r^{\prime} \beta$ is generally  not point identified if at least one component of $\left(r_2, \ldots, r_q\right)^{\prime}$ is non-zero, but it is interval identified. Let $\left[c_L, c_U\right]$ be the identified interval.

\subsection{Assumptions}

We assume the following regularity conditions.

\begin{assumption}\label{a:design}
One of the following conditions holds:
\begin{itemize}
\item[(a)] (i) $X$ is fixed and $p(x_j) \equiv n_j /n > 0$ is known.
(ii) For each $j=1,\ldots,J$, $\{ Y_{\ell,j} : \ell =1,\ldots,n_j \}$ is an independent
random sample from $\mathbb{P}(Y|X = x_j)$, and these cell-specific samples are
mutually independent across \(j\).

\item[(b)] (i) $X$ is random.
(ii) $\{(Y_i , X_i) : i =1,\ldots,n\}$ is an independent random sample from the joint
distribution of $(Y,X)$.
(iii) For all $j = 1,\ldots, J$, $p(x_j) \equiv \mathbb{P}  (X = x_j) > 0$.

\end{itemize}
\end{assumption}

\begin{assumption}\label{a:margin-condition}
$\left| f\left(x_j\right) \right| >0$ for all $j=1, \ldots, J$.
\end{assumption}

\begin{assumption}\label{a:para-space}
\(\beta\in\mathbf B\), where
$
\mathbf B
=
\{(1,b_2,\ldots,b_q):
\underline{b}_k\leq b_k\leq \overline{b}_k,\ k=2,\ldots,q\}
$
for known finite constants \(\underline{b}_k<\overline{b}_k\).
\end{assumption}

Under Assumption~\ref{a:para-space}, the normalized parameter space for $\beta$ is a known compact box.

Define
\begin{align*}
g_0 (x_j)=
\left \{
\begin{array}{ll}
\mathbb{E} (Y_{\ell,j} - \tau )p(x_j) &\text{ if $X$ is fixed} \\
\mathbb{E} \left[ (Y - \tau)  I( X = x_j) \right]  &\text{ if $X$ is random}
\end{array}
\right..
\end{align*}
If $X$ is random, we include the indicator function $I( X = x_j)$ in $g_0 (x_j)$ so that it can be estimated as a sample average without a random denominator.  Under Assumption~\ref{a:margin-condition}, $\mathrm{sgn}(g_0(x_j)) = d_j$, so we estimate $g_0(x_j)$ from the data.

\subsection{Partial Identification}

In model \eqref{model-intro}, $Y=1$ if and only if $\beta'X > U$, where $\mathbb{P}(U < 0 \mid X)=\tau$.  Under Assumption~\ref{a:margin-condition}, we have
\begin{align*}
& \mathbb{P}(Y=1 \mid X=x_j)>\tau \Leftrightarrow \beta'x_j >0, \\
& \mathbb{P}(Y=1 \mid X=x_j)<\tau \Leftrightarrow \beta'x_j <0.
\end{align*}
We define the strict sign-restriction set as
\[
\mathcal B^\circ
=
\{b\in\mathbf B: d_jx_j'b>0,\ j=1,\ldots,J\}
\]
and base the main population bounds on the weak-inequality outer set
\begin{align}\label{def:B0}
\mathcal B(0)
=
\{b\in\mathbf B: d_jx_j'b\geq0,\ j=1,\ldots,J\}.
\end{align}
This set contains \(\mathcal B^\circ\) and may include boundary values with
\(d_jx_j'b=0\). Throughout the main text, \(c_L\) and \(c_U\) denote the
lower and upper bounds on \(r'\beta\) over \(\mathcal B(0)\). Equivalently,
they are the optimal values in
\begin{align}\label{lp-pop-coef}
 \underset{b \in \mathbf{B}}{\operatorname{maximize}} \left( \underset{b \in \mathbf{B}}{\operatorname{minimize}} \right): \; r'b
 \end{align}
 subject to
 \begin{align}\label{d-constraint}
 d_j \left( x_{j,1} + b_2 x_{j,2} + \cdots +  b_q x_{j,q}  \right) \geq 0
\end{align}
for all \(j=1,\ldots,J\).

\begin{remark}\label{rem:ep-outer}
Assumption~\ref{a:margin-condition} implies that the true index is separated
from zero at the support points: there exists \(\varepsilon_\beta>0\) such
that \(\min_{1\leq j\leq J}|\beta'x_j|=\varepsilon_\beta\). This margin is a
feature of the unknown true parameter and is not treated as known in the
baseline analysis. The weak inequalities in \(\mathcal B(0)\) therefore give
a conservative set of weakly sign-consistent parameter values.

For a chosen \(\varepsilon>0\), one can instead impose the margin-restricted set
\[
\mathcal B(\varepsilon)
=
\{b\in\mathbf B: d_jx_j'b\geq\varepsilon,\ j=1,\ldots,J\}.
\]
Positive values of \(\varepsilon\) require separation from the boundary
\(x_j'b=0\) and generally yield narrower bounds, \(c_L(\varepsilon)\) and \(c_U(\varepsilon)\).
If $\varepsilon_\beta$ or a lower bound, $\varepsilon_{L}$, on $\varepsilon_\beta$
were known, one could replace the right-hand side of \eqref{d-constraint} with
$\varepsilon = \varepsilon_\beta$ or
$\varepsilon = \varepsilon_{L}$, thereby obtaining the narrower
identified interval $[ c_L(\varepsilon), c_U(\varepsilon) ]$. However, neither
$\varepsilon_\beta$ nor $\varepsilon_{L}$ is known in applications, and choosing
too large a value of $\varepsilon$ can cause the feasible region of the modified
\eqref{lp-pop-coef}-\eqref{d-constraint} to be empty. Setting $\varepsilon = 0$ in \eqref{d-constraint} avoids this problem.
Supplemental Appendix~\ref{sec:sensitivity-eps} explores the sensitivity of $c_L(\varepsilon)$,
$c_U(\varepsilon)$ and the identified intervals for $\beta$ to increasing
$\varepsilon$ from~0.
\end{remark}

\begin{remark}\label{rem:eq:iden}
Assumption~\ref{a:margin-condition} is needed because $d_j b^{\prime} x_j=0$ does not imply that $d_j=0$ and $b^{\prime} x_j=0$.
It excludes boundary cases where $f(x_j) = 0$ (equivalently, $x_j^{\prime}\beta = 0$) for some $j$. Although this simplifies identification,  it is not without cost. Cases with $f(x_j) = 0$ yield equality constraints ($x_j^{\prime}\beta = 0$), which provide stronger identifying power and could shrink the identified interval for $r^{\prime}\beta$.
Relaxing Assumption~\ref{a:margin-condition}, however, introduces significant practical challenges. In empirical applications, the true conditional probability $\mathbb{P}(Y = 1 \mid X = x_j)$ is typically unknown and must be estimated. Even if the true probability equals exactly $\tau$, estimators rarely match this precisely, necessitating an approximation interval $[\tau - \ell_n, \tau + \ell_n]$ for some tuning parameter $\ell_n$. Determining an appropriate $\ell_n$ and adjusting estimation and inference methods accordingly is nontrivial.
Moreover, existing finite-sample inference results suggest that exact equality constraints contribute little to inference about $\beta$ \citep[see][Appendix G]{Rosen:Ura:2024}. Thus, despite their theoretical appeal, equality constraints may offer limited practical benefit. Given these considerations, we maintain Assumption~\ref{a:margin-condition} for clarity and practicality, but acknowledge that exploring ways to incorporate or approximate these boundary cases remains an important avenue for future research.
\end{remark}

Under Assumption~\ref{a:design}, \(d_j \equiv \text{sgn}[f(x_j)]\) and
\(g_{0,j} \equiv g_0(x_j)\) have the same sign. Therefore, the baseline
linear programs \eqref{lp-pop-coef}-\eqref{d-constraint} can be written
equivalently as
\begin{align}\label{lp-pop-coef-g}
 \underset{b \in \mathbf{B}}{\operatorname{maximize}} \left( \underset{b \in \mathbf{B}}{\operatorname{minimize}} \right): \; r'b
 \end{align}
 subject to
 \begin{align}\label{g-constraint}
 g_{0,j} \left( x_{j,1} + b_2 x_{j,2} + \cdots +  b_q x_{j,q}  \right) \geq 0
\end{align}
for all $j=1,\ldots,J$.


\section{A Framework for Optimal Classification}\label{sec:decision:theory}

In this section, we present a framework for optimal classification.
We start with the case of a known, point-identified $\beta$.
Let $\hat{y} \in \{0, 1\}$ denote a binary classifier. For covariate values \(x\) within the support of $X$, the function
\[
f(x) = \mathbb{P}(Y=1 \mid X=x) - \tau
\]
is nonparametrically identified. At the population level, the optimal classification rule assigns
\begin{align}\label{class-np}
\hat{y} =
\begin{cases}
1, & \text{if } f(x) > 0,\\
0, & \text{if } f(x) < 0.
\end{cases}
\end{align}
Recall that Assumption~\ref{a:margin-condition} excludes the boundary case \(f(x)=0\).
However, when \(x\) lies outside the support of $X$, the function \(f(x)\)
is not nonparametrically identified, and direct classification based on
\eqref{class-np} is not feasible.
The maximum score specification provides a disciplined basis for
model-based extrapolation.
Under the maintained model and Assumption~\ref{a:margin-condition},
\(\operatorname{sgn}\{f(x)\}\) coincides with
\(\operatorname{sgn}(x'\beta)\) on the support of \(X\).
In contrast to an
unrestricted reduced-form object \(f\), however, the linear index \(x'\beta\)
is defined for any covariate value, including counterfactual values outside
the observed support.
If $\beta$ is known, we can therefore use the model-implied linear sign rule
\begin{align}\label{class-np-ms}
\hat{y} =
\begin{cases}
1, & \text{if } \beta' x > 0,\\
0, & \text{if } \beta' x < 0.
\end{cases}
\end{align}
Boundary cases are handled by the exact-boundary convention stated below.
This formulation enables classification at any covariate value $x$, including those not in the support of $X$.

\begin{remark}[Classification under Point Identification of $\beta$]
Suppose that the $(1-\tau)$-absolute loss assigns loss $(1-\tau)$ to a false negative
($\hat{y}=0$, $y=1$) and loss $\tau$ to a false positive
($\hat{y}=1$, $y=0$):
\[
  L(\hat{y}, y) \;=\; (1-\tau)\,I(\hat{y}=0,\, y=1) \;+\; \tau\,I(\hat{y}=1,\, y=0).
\]
The Bayes rule under this loss classifies as one if
$\mathbb P(Y=1\mid X=x) > \tau$
and zero if $\mathbb P(Y=1\mid X=x) < \tau$; at equality, both actions are
optimal. Thus the classification rules in
\eqref{class-np} and \eqref{class-np-ms} can be interpreted as optimal
decision rules minimizing this loss. This result is formally
established as Corollary 4 in \citet{Manski:1988}, who also emphasizes the
role of extrapolation beyond the support of $X$.
\end{remark}

We now move to the case of partially identified \(\beta\). When \(\beta'x\) is
only interval-identified, write
\[
\beta' x \in [c_L(x),\,c_U(x)].
\]
This interval provides information about the sign of the model-implied index,
and hence about the model-implied classification of off-support covariate
values. To account for the partial identification of $\beta'x$, we consider
two scenarios: one that includes the option to abstain, and another
that allows for random classification.

\subsection{Classification with Abstention}

First suppose that the decision-maker may abstain. The action set is
$\{1,0,\emptyset\}$: $\hat y=1$ denotes classification as $1$,
$\hat y=0$ denotes classification as $0$, and
$\hat y=\emptyset$ denotes abstention, or non-classification. Here, $f(x)>0$
is the state in which classification as 1 is oracle-correct, whereas $f(x)<0$
is the state in which classification as 0 is oracle-correct. The economic
interpretation of these actions and states depends on the application.

As an illustration, consider a setting in which a potential employer decides
whether to call back a job candidate for an interview based on a resume. This
setting is related to the empirical application considered in
Section~\ref{sec:KLS}. In this example, $\hat y=1$ means giving an applicant a
callback, $\hat y=0$ means not giving a callback, and $\hat y=\emptyset$
means deferring the decision by requesting additional information or placing
the applicant on a waiting list. The state $f(x)>0$ means that the
applicant's profile warrants a callback, while $f(x)<0$ means that it does
not.

Consider the loss function
\begin{align}\label{def:class:loss}
L(\hat y,f(x))
=\begin{cases}
C_b^+, & \hat y=1,\ f(x)>0,\\
C_b^-, & \hat y=0,\ f(x)<0,\\
C^+, & \hat y=0,\ f(x)>0,\\
C^-, & \hat y=1,\ f(x)<0,\\
C_\emptyset, & \hat y=\emptyset,\ f(x)>0,\\
C_\emptyset, & \hat y=\emptyset,\ f(x)<0.
\end{cases}
\end{align}
The loss function has five cost parameters:
$
(C_b^+, C_b^-, C^+, C^-, C_\emptyset).
$
The baseline costs $C_b^+$ and $C_b^-$ are the costs of the two
oracle-correct binary actions. They may differ because the two correct
decisions need not have the same payoff implications. The mistake
costs $C^+$ and $C^-$ may also differ because the two possible errors need
not have the same consequences. In the callback example, the cost of missing a
qualified applicant may differ from the cost of interviewing an unsuitable
applicant.
A medical example is similar: the cost of falsely classifying a well
individual as sick and needing treatment may differ from the cost of falsely
classifying a sick individual as well and not needing treatment. A common abstention
cost is natural when abstention refers to the same procedural action in either
state, such as requesting additional materials, placing the applicant on a
waiting list, sending the application for a second review, or otherwise
postponing the binary decision.

We evaluate decisions by minimax regret. Regret is measured relative to an
oracle that knows the sign of \(f(x)\): the oracle chooses $\hat y=1$ and
incurs $C_b^+$ when $f(x)>0$, and chooses $\hat y=0$ and incurs $C_b^-$ when
$f(x)<0$. Define
\begin{align}\label{def:class:regret}
\begin{split}
\mathrm{Reg}_{\mathrm{FP}}&=C^- - C_b^-, \\
\mathrm{Reg}_{\mathrm{FN}}&=C^+ - C_b^+, \\
\mathrm{Reg}_{\emptyset}
&=C_\emptyset-\min\{C_b^+,C_b^-\}.
\end{split}
\end{align}
Here $\mathrm{Reg}_{\mathrm{FP}}$ is the excess cost of a false positive:
choosing $\hat y=1$ when $\hat y=0$ is oracle-correct.
Similarly, $\mathrm{Reg}_{\mathrm{FN}}$ is the excess cost of a false
negative, and $\mathrm{Reg}_{\emptyset}$ is the worst-case excess cost of
abstention. In the callback example, these are,
respectively, the excess costs of giving a callback when the applicant does
not warrant one, not giving a callback when the applicant does warrant one,
and deferring the callback decision.

\begin{assumption}[Cost Structure]
\label{ass:low-regret-deferral}
The cost parameters satisfy
\begin{equation}\label{eq:cost-minimal}
0\le C_b^+ < C_\emptyset<\infty,\;
0\le C_b^- < C_\emptyset<\infty,\;
C_b^+ < C^+<\infty,\;
C_b^- < C^-<\infty.
\end{equation}
In addition, the regret from abstention satisfies
\begin{equation}\label{eq:deferral-regret-condition}
\mathrm{Reg}_{\emptyset}
\le
\min\{\mathrm{Reg}_{\mathrm{FP}},\mathrm{Reg}_{\mathrm{FN}}\}.
\end{equation}
Equivalently,
\[
C_\emptyset-\min\{C_b^+,C_b^-\}
\le
\min\{C^- - C_b^-,\, C^+ - C_b^+\}.
\]
\end{assumption}

Assumption~\ref{ass:low-regret-deferral} has two parts. The cost
inequalities in \eqref{eq:cost-minimal}
say that, if the sign of $f(x)$ were known, the oracle-correct
binary action would be preferred to both abstention and the wrong binary action.
The regret condition in \eqref{eq:deferral-regret-condition} says that, when the sign is
ambiguous, abstention has weakly smaller worst-case regret than either possible
mistake. A simple sufficient condition for \eqref{eq:deferral-regret-condition}
is
\begin{equation}\label{eq:deferral-sufficient-condition}
C_\emptyset+|C_b^+-C_b^-|\le \min\{C^+,C^-\}.
\end{equation}
This condition says that abstention is no more costly than either type of
mistake after accounting for the possible difference between the two
oracle-correct baseline costs.

Because the set \(\mathcal B(0)\) defined in \eqref{def:B0} may include
parameter values \(b\) satisfying \(b'x=0\), we define
the regret of either binary action to be zero at such values. If
\([c_L(x),c_U(x)]=\{0\}\), we apply the maintained tie-breaking rule and
classify as \(0\).
The following theorem gives the main result of this subsection.

\begin{theorem}[Optimal Classification with Abstention]\label{thm:selective}
Let $[c_L(x),c_U(x)]$ be the identified lower and upper bounds on
$\beta'x$. Consider the loss function in \eqref{def:class:loss}.
Under Assumptions~\ref{a:margin-condition} and~\ref{ass:low-regret-deferral},
a minimax-regret optimal classification rule, denoted by $\hat y^*(x)$, is
\[
\hat y^*(x)
\;=\;
\begin{cases}
1, & \text{if } c_L(x)\ge0 \text{ and } c_U(x)>0,\\
0, & \text{if } c_U(x)\le0,\\
\emptyset, & \text{if } c_L(x)<0<c_U(x).
\end{cases}
\]
\end{theorem}

\begin{remark}[Endpoint Cases]\label{rem:degenerate-boundary}
The endpoint cases \(c_L(x)=0<c_U(x)\), \(c_L(x)<0=c_U(x)\), and
\(c_L(x)=c_U(x)=0\) are not sign-ambiguity cases. They are resolved by the
binary rule in Theorem~\ref{thm:selective}; only \(c_L(x)<0<c_U(x)\) calls for
abstention.
\end{remark}

\begin{remark}[Symmetric Costs]\label{rem:sym:costs}
As a special case, suppose
\[
C_b^+=C_b^-=C_b,\qquad C^+=C^-=C.
\]
Then
\[
\mathrm{Reg}_{\mathrm{FP}}
=
\mathrm{Reg}_{\mathrm{FN}}
=
C-C_b,
\qquad
\mathrm{Reg}_{\emptyset}=C_\emptyset-C_b.
\]
If $C_b<C_\emptyset<C$, then Assumption~\ref{ass:low-regret-deferral} holds,
so Theorem~\ref{thm:selective} applies.
In this symmetric-cost case, subtracting the same oracle-correct
baseline cost from each state-contingent loss makes minimax regret equivalent
to minimizing the worst-case cost of misclassification or abstention.
\end{remark}

\begin{remark}[If \eqref{eq:deferral-regret-condition} Fails]
Theorem~\ref{thm:selective} gives the condition under which the abstention
rule is minimax-regret optimal with asymmetric costs. Without
\eqref{eq:deferral-regret-condition}, the same comparison shows that, when
\(c_L(x)<0<c_U(x)\), any action attaining the smallest value among
$\mathrm{Reg}_{\mathrm{FP}}$, $\mathrm{Reg}_{\mathrm{FN}}$, and
$\mathrm{Reg}_{\emptyset}$ is minimax-regret optimal. Thus,
\eqref{eq:deferral-regret-condition} is the substantive condition under which
ambiguity justifies abstention.
\end{remark}

\begin{remark}[Sign-Dependent Abstention]
One could instead allow sign-dependent abstention costs $C_\emptyset^+$ and
$C_\emptyset^-$. Under $C_b^+\le C_\emptyset^+$ and
$C_b^-\le C_\emptyset^-$, the minimax-regret argument is unchanged after
replacing $\mathrm{Reg}_{\emptyset}$ by
\[
\max\{C_\emptyset^+-C_b^+,\,C_\emptyset^- - C_b^-\}.
\]
We use a common $C_\emptyset$ in the main specification because abstention is
the same procedural action in the applied examples.
\end{remark}

\begin{remark}[Relation to Selective Classification]
The rule of Theorem~\ref{thm:selective} resembles selective classification, or classification with a reject
option, in the machine-learning literature
\citep[e.g.,][]{JMLR:v11:el-yaniv10a}. The connection is useful because both
frameworks allow the classifier to abstain. The objective is different,
however: here the loss function is tied to the maximum score model and the
identified set for $\beta'x$, whereas the selective-classification
literature typically takes the loss function as primitive. For a recent
review of selective classification and related topics, see
\citet{hendrickx2024machine}. In economics, \citet{breza2025generalizability}
recently propose a method for policy learning with an abstention option.
\end{remark}

\begin{remark}[Abstention in Applied Settings]
Abstention arises naturally in a variety of empirical contexts. In the
incentivized resume rating application of Section~\ref{sec:KLS}, an employer
who is uncertain whether a candidate's profile warrants a callback may
request additional materials or place the candidate on a waiting list. A
physician who faces an ambiguous diagnosis may order further tests before
choosing between treatment and no treatment, and policymakers
at the Federal Reserve who are uncertain about the direction of the economy may
defer an interest rate decision while awaiting additional data. In each case,
\eqref{eq:deferral-regret-condition} is the condition under which deferring
is less costly, in worst-case regret terms, than either possible mistake.
\end{remark}

\begin{remark}[Dempster--Shafer]
Dempster--Shafer (DS) theory \citep{Dempster2008} extends classical
probability by introducing a third category, ``don't know'', alongside
``true'' and ``false''. It assigns a triplet $(p,q,r)$ of non-negative values
summing to one, representing evidence for, against, and uncertainty about an
assertion, respectively. In short, DS theory allows $r>0$ to reflect
ambiguity. Our classification rule is derived from the identified set for
$\beta'x$, whereas DS theory operates on probability assignments
over outcomes rather than on a structural economic model. While related in
spirit, our abstention rule remains grounded in conventional
probability.
\end{remark}

\subsection{Random Classification if Abstention Is Not Possible}\label{subsec:rand:class}

Now suppose that abstention is unavailable. In the callback example,
the employer must either give the applicant a callback or not give one.
When the identified interval does not strictly contain zero, the
rule again chooses a binary action, using the exact-boundary convention in
Remark~\ref{rem:degenerate-boundary}.
When the identified
interval strictly contains zero, both signs remain possible and randomization
is a natural response to the resulting binary decision problem under ambiguity
\citep{Manski2009}.

For this binary problem, use the loss function
\begin{align}\label{def:class:loss:rc}
L(\hat y,f(x))
=\begin{cases}
C_b^+, & \hat y=1,\ f(x)>0,\\
C_b^-, & \hat y=0,\ f(x)<0,\\
C^+, & \hat y=0,\ f(x)>0,\\
C^-, & \hat y=1,\ f(x)<0,
\end{cases}
\end{align}
where
\[
0\le C_b^+<C^+<\infty,
\qquad
0\le C_b^-<C^-<\infty.
\]
Thus the mandatory binary problem allows both the oracle-correct costs and
the mistake costs to differ across signs. Since $\emptyset$ is not
available, there is no abstention cost. The relevant regret indices are
\[
\mathrm{Reg}_{\mathrm{FP}}=C^- - C_b^-,
\qquad
\mathrm{Reg}_{\mathrm{FN}}=C^+ - C_b^+.
\]

Let $p$ denote the probability of classifying as $1$. In the callback
example, $p$ is the probability of giving a callback. If $f(x)>0$, the
randomized rule makes a false negative with probability $1-p$, so its regret
is $(1-p)\mathrm{Reg}_{\mathrm{FN}}$. If $f(x)<0$, the randomized rule makes
a false positive with probability $p$, so its regret is
$p\,\mathrm{Reg}_{\mathrm{FP}}$.

\begin{theorem}[Random Classification]\label{thm:selective:rc}
Let $[c_L(x),c_U(x)]$ be the identified lower and upper bounds on
$\beta'x$.
Let Assumption~\ref{a:margin-condition} hold and assume that
\(0\le C_b^+<C^+<\infty\) and \(0\le C_b^-<C^-<\infty\).
Under the loss function in
\eqref{def:class:loss:rc}, a minimax-regret optimal binary rule classifies as
$1$ if \(c_L(x)\ge0\) and \(c_U(x)>0\), classifies
as $0$ if \(c_U(x)\le0\), and, if
$c_L(x)<0<c_U(x)$, classifies as $1$ with probability $p^*$ and as $0$ with
probability $1-p^*$, where
\[
p^*
=
\frac{\mathrm{Reg}_{\mathrm{FN}}}
     {\mathrm{Reg}_{\mathrm{FP}}+\mathrm{Reg}_{\mathrm{FN}}}.
\]
\end{theorem}

\begin{remark}[Interpreting the Randomization Probability]
The formula for $p^*$ has a direct callback interpretation. If the regret
from a false negative is large relative to the regret from a false positive,
then $p^*$ is closer to one: the employer gives a callback with higher
probability because missing a qualified applicant is especially costly. If
the regret from a false positive is large, then $p^*$ is closer to zero: the
employer gives a callback with lower probability because interviewing an
unsuitable applicant is especially costly.
\end{remark}

\begin{remark}[Symmetric Costs]
In the special case of symmetric costs in Remark~\ref{rem:sym:costs},
$p^*=1/2$, so Theorem~\ref{thm:selective:rc} gives randomization
with probabilities of 0.5.
As in Remark~\ref{rem:sym:costs}, minimax regret
is then equivalent to minimizing the worst-case cost of misclassification.
\end{remark}

Theorems~\ref{thm:selective} and~\ref{thm:selective:rc} characterize the
optimal classification rules when the identified set $[c_L(x),c_U(x)]$ is
known.
Section~\ref{sec:estimation:inference} addresses the empirical
problem of replacing the identified set with confidence bounds
$[\hat c_L(x),\hat c_U(x)]$ estimated from data. The oracle classifier
defined in Section~\ref{sec:decision:theory} serves as the
population-level benchmark against which the feasible classifiers are
evaluated.

\section{Estimation and Inference}\label{sec:estimation:inference}

Section~\ref{subsec:est} describes estimation of \(c_L\) and \(c_U\).
Sections~\ref{subsec:inference:finite} and~\ref{subsec:inference:asymp}, respectively,
describe finite-sample and asymptotic inference about \([c_L,c_U]\).

\subsection{Estimation}\label{subsec:est}

In applications, $g_{0, j}$ is unknown. We estimate it by one of two methods, depending on whether $X$ is fixed or random. If $X$ is fixed, estimate $g_{0, j}$ by
\begin{align*}
\hat{g}\left(x_j\right)
&= n_j^{-1} \sum_{\ell=1}^{n_j} \left(Y_{\ell, j} -\tau\right) p(x_j).
\end{align*}
If $X$ is random, estimate $g_{0, j}$ by
\[
\hat{g}(x_j)=n^{-1} \sum_{i=1}^n\left(Y_i-\tau\right) I\left(X_i=x_j\right).
\]
Both estimators are unbiased:
\[
\mathbb{E} \left[\hat{g}\left(x_j\right)\right]=g_0\left(x_j\right) ; \quad j=1, \ldots, J.
\]

Now let $\mathcal{G}_\alpha$ be a confidence region for $g_0 \equiv (g_{0,1},\ldots, g_{0,J})$  satisfying
\begin{align}\label{G-coverage}
\mathbb{P} \left\{ g_0   \in \mathcal{G}_\alpha \right\} \geq 1 - \alpha
\end{align}
for some $0 < \alpha < 1$.
Here, $\mathcal{G}_\alpha$ may be a finite-sample or asymptotic confidence region.  In the latter case, ``asymptotic'' refers to the limit as $n \to \infty$, so that the probability guarantee $\mathbb{P}(g_0 \in \mathcal{G}_\alpha) \geq 1-\alpha$ holds approximately for large samples.

The estimators of $c_U$ and $c_L$, denoted respectively by  $\hat{c}_U$ and $\hat{c}_L$, are the optimal values
of the objective functions of the following optimization problems:
\begin{align}\label{alt-1}
\underset{g\in\mathcal G_\alpha,\; b \in \mathbf{B}}{\operatorname{maximize}}
\left(\underset{g\in\mathcal G_\alpha,\; b \in \mathbf{B}}{\operatorname{minimize}} \right): \; r'b
 \end{align}
subject to
	 \begin{align}\label{bl-constraints}
	g_j  \left( x_{j,1} + b_2 x_{j,2} + \cdots +  b_q x_{j,q}  \right) \geq 0
	\quad \forall j=1,\ldots,J.
\end{align}

The following theorem, which is similar to Theorem 2.1 of \citet{JL:JBES}, is the basis for our inference and classification.

\begin{theorem}\label{thm1}
Let \eqref{G-coverage} hold. Then,
\begin{align*}
\mathbb{P} \left[ \hat{c}_L \leq c_L \leq r'\beta \leq c_U \leq \hat{c}_U\right] \geq
\mathbb{P} \left\{ g_0  \in \mathcal{G}_\alpha \right\} \geq 1-\alpha.
\end{align*}
\end{theorem}

Theorem~\ref{thm1} shows that  if \eqref{G-coverage} holds, the estimated interval $[\hat{c}_L, \hat{c}_U]$ contains the population interval
$[c_L, c_U]$  with finite-sample or asymptotic probability at least $(1-\alpha)$,
depending on whether $\mathcal{G}_\alpha$ is a finite-sample or asymptotic confidence region.
Theorem~\ref{thm1} implies that
to obtain a confidence
region for $[c_L, c_U]$, it suffices to obtain the confidence region $\mathcal{G}_\alpha$.

\begin{remark}[Misclassification]
Theorem~\ref{thm1} gives coverage of the population interval by the estimated
interval. Misclassification bounds require a stronger sign-stability
argument. The key observation is simple. The optimization problems use \(g\)
only through the signs of its components. If all vectors in
\(\mathcal G_\alpha\) have the same coordinatewise signs as \(g_0\), then the
sample and population problems impose the same sign restrictions. In that
event, \(\hat c_L=c_L\) and \(\hat c_U=c_U\), so the estimated and population
classifiers agree, both in the abstention and random-classification cases.
Thus misclassification can occur only when the confidence region permits a
sign pattern different from the population sign pattern. Section~\ref{sec:regret}
uses the global sign-agreement event to bound misclassification probabilities.
Supplemental Appendix~\ref{subsec:refined-misclassification} gives
refinements and extends the argument to expected excess
regret.
\end{remark}

\subsection{Finite-Sample Inference}\label{subsec:inference:finite}

This section obtains finite-sample confidence regions $\mathcal{G}_\alpha$ for fixed and random $X$.

\subsubsection{Fixed Design}\label{sec:fixed:iid}

Let $X$ be fixed and Assumption~\ref{a:design}(a) hold. Then
\begin{align}\label{eq:bounded:rv}
-\tau p\left(x_j\right) \leq  (Y_{\ell, j} - \tau )p(x_j) \leq(1-\tau) p\left(x_j\right)
\end{align}
because $Y = 0$ or 1.
 An application of Hoeffding's inequality yields
\[
\mathbb{P} \left[\left|\hat{g}\left(x_j\right)-g_0\left(x_j\right)\right| \geq p\left(x_j\right) t_j \right] \leq 2 \exp \left(-2 n_j t_j^2\right)
\]
for any $t_j>0$ and each $j=1, \ldots, J$. Moreover,
\begin{align*}
\hat{g}\left(x_j\right)-p\left(x_j\right) t_j \leq g_0\left(x_j\right) \leq \hat{g}\left(x_j\right)+p\left(x_j\right) t_j
\end{align*}
for all $j=1, \ldots, J$ with probability at least $1-2 \sum_{j=1}^J \exp \left(-2 n_j t_j^2\right)$. Set
\begin{align}\label{cv:fixed}
t_j=\left(\frac{1}{2 n_j} \log \frac{2 J}{\alpha}\right)^{1/2} \equiv t_{j,\alpha}
\end{align}
and define
\begin{align}\label{G:fixed:box}
\mathcal{G}_\alpha =
\left\{
(g_1,\ldots, g_J): \hat{g}(x_j)  - p \left(x_j\right) t_{j,\alpha}
\leq g_j \leq
\hat{g}(x_j)  + p \left(x_j\right) t_{j,\alpha} \; \forall j
\right\}.
\end{align}
Then $\mathbb{P}(g_0 \in \mathcal{G}_\alpha) \geq 1-\alpha$.

\subsubsection{Random Design}\label{sec:random:iid}

Let $X$ be random and Assumption~\ref{a:design}(b) hold. Then
\[
-\tau  \leq  (Y_i - \tau ) I\left(X_i=x_j\right) \leq 1-\tau.
\]
Another application of
Hoeffding's inequality gives
\[
\mathbb{P} \left[\left|\hat{g}\left(x_j\right)-g_0\left(x_j\right)\right| \geq t\right] \leq 2 \exp \left(-2 n t^2\right)
\]
for each $j=1,\ldots,J$ and any $t > 0$. Therefore,
\[
\hat{g}\left(x_j\right)-t \leq g_0\left(x_j\right) \leq \hat{g}\left(x_j\right)+t
\]
for all $j=1, \ldots, J$  and any $t > 0$ with probability at least $1-2 J \exp \left(-2 n t^2\right)$.
Now set
\begin{align}\label{cv_random_design}
t=\left(\frac{1}{2 n} \log \frac{2 J}{\alpha}\right)^{1/2} \equiv t_{\alpha}
\end{align}
and define
\begin{align}\label{G:random:box}
\mathcal{G}_\alpha =
\left\{
(g_1,\ldots, g_J): \hat{g}(x_j)  -  t_\alpha
\leq g_j \leq
\hat{g}(x_j)  +  t_\alpha \; \forall j
\right\}.
\end{align}
Then $\mathbb{P}(g_0 \in \mathcal{G}_\alpha) \geq 1-\alpha$.

To compare confidence regions in \eqref{G:fixed:box} and \eqref{G:random:box}, let Assumption~\ref{a:design}(a) hold so that
$p(x_j) = n_j/n$. Then,
\begin{align*}
p \left(x_j\right) t_{j,\alpha}
&= [ p \left(x_j\right) ]^{1/2} t_\alpha \le t_\alpha,
\end{align*}
which implies that \([\hat c_L,\hat c_U]\) is weakly wider under the random
design than under the fixed design.
This happens because of the need to account for randomness of $X$ when we apply Hoeffding's inequality to the random design.

\subsection{Asymptotic Inference}\label{subsec:inference:asymp}

This section gives asymptotic confidence regions $\mathcal{G}_\alpha$ for fixed and random designs. To
minimize computational complexity (see Section~\ref{sec:computation}) we focus on (hyper) rectangular regions based
on the Bonferroni inequality. Ellipsoidal regions give narrower bounds $[\hat{c}_L, \hat{c}_U]$ under certain
conditions but make computation much more difficult.

\subsubsection{Fixed Design}

As in Section~\ref{sec:fixed:iid},
let $X$ be fixed and Assumption~\ref{a:design}(a) hold. Consider a sequence of
fixed designs with \(J\) fixed and \(n_j/n\) bounded away from zero for each
\(j\). It follows from the Lindeberg--Lévy central limit theorem that
for each $j=1,\ldots,J$,
\[
\left(n / n_j^{1 / 2}\right)\left(\hat{g}_j-g_{0, j}\right) \rightarrow_d N\left(0, \sigma_j^2\right),
\]
where
\[
\sigma_j^2 \equiv \mathbb{P}\left(Y=1 \mid X=x_j\right)\left[1-\mathbb{P}\left(Y=1 \mid X=x_j\right)\right].
\]
Estimate $\sigma_j^2$ by
\begin{align*}
\hat{\sigma}_j^2
&\equiv
\left\{  n_j^{-1} \sum_{\ell=1}^{n_j} Y_{\ell, j} \right\} \left\{  n_j^{-1} \sum_{\ell=1}^{n_j} \left(1-Y_{\ell, j} \right) \right\},
\end{align*}
and let $z_{1-\alpha}$ denote the $(1-\alpha)$ quantile of the standard normal distribution.
Set
\begin{align}\label{G:fixed:box:asymp}
\mathcal{G}_\alpha =
\left\{
(g_1,\ldots, g_J): \hat{g}(x_j)  -  \frac{n_j^{1/2} \hat{\sigma}_j}{n}  z_{1 - \alpha / (2J)}
\leq g_j \leq
\hat{g}(x_j)  + \frac{n_j^{1/2} \hat{\sigma}_j}{n}  z_{1 - \alpha / (2J)} \; \forall j
\right\}.
\end{align}
Then the Bonferroni inequality gives $\mathbb{P}(g_0 \in \mathcal{G}_\alpha) \geq 1-\alpha$ asymptotically.

\subsubsection{Random Design}

Let \(X\) be random and Assumption~\ref{a:design}(b) hold. Then the
Lindeberg--Lévy theorem gives
\[
n^{1 / 2} \left(\hat{g}_j-g_{0, j}\right) \rightarrow_d N\left(0, s_j^2\right),
\]
where
\[
s_j^2 \equiv \mathrm{Var} \left[ \left(Y_i-\tau\right) I\left(X_i=x_j\right) \right].
\]
Estimate $s_j^2$ by
\begin{align*}
\hat{s}_j^2
&=  \frac{1}{n} \sum_{i=1}^n \left[  \left(Y_i-\tau\right) I\left(X_i=x_j\right) - \frac{1}{n} \sum_{i=1}^n  \left(Y_i-\tau\right) I\left(X_i=x_j\right) \right]^2.
\end{align*}
Set
\begin{align}\label{G:random:box:asymp}
\mathcal{G}_\alpha =
\left\{
(g_1,\ldots, g_J): \hat{g}(x_j)  -  \frac{\hat{s}_j}{n^{1/2}}  z_{1 - \alpha / (2J)}
\leq g_j \leq
\hat{g}(x_j)  + \frac{\hat{s}_j}{n^{1/2}}  z_{1 - \alpha / (2J)} \; \forall j
\right\}.
\end{align}
Then, the Bonferroni inequality gives $\mathbb{P}(g_0 \in \mathcal{G}_\alpha) \geq 1-\alpha$ asymptotically.
For each \(j\), the random-design marginal interval is wider than the
corresponding fixed-design marginal interval if and only if
\[
n\hat s_j^2>n_j\hat\sigma_j^2.
\]
Consequently, the random-design rectangle contains the fixed-design rectangle
if \(n\hat s_j^2\geq n_j\hat\sigma_j^2\) for every \(j\). In general, neither
rectangle need contain the other.

\begin{remark}[Alternatives to the Bonferroni Correction]
The Bonferroni-based confidence regions in \eqref{G:fixed:box:asymp} and \eqref{G:random:box:asymp} are chosen for their simplicity and because they require no assumptions beyond those already imposed. The correction can be replaced by other multiple-testing procedures. For example, bootstrap-based methods or resampling approaches could be used to construct tighter joint confidence sets for $g_0 = (g_0(x_1),\ldots,g_0(x_J))$, potentially yielding narrower bounds $[\hat{c}_L, \hat{c}_U]$ in some settings.
\end{remark}

\section{Computational Algorithms}\label{sec:computation}

In this section, we describe how to solve the optimization problems given in \eqref{alt-1}-\eqref{bl-constraints}.
In all the cases we considered, $\mathcal{G}_\alpha$ has the form
\begin{align}\label{G:box:generic}
\mathcal{G}_\alpha =
\left\{
(g_1,\ldots, g_J): \hat{g}(x_j)  -  \hat{s}(x_j, \alpha)
\leq g_j \leq
\hat{g}(x_j)  +  \hat{s}(x_j, \alpha) \; \forall j
\right\}
\end{align}
for an appropriate choice of $\hat{s}(x_j, \alpha) > 0$.
For example, for asymptotic inference under the random design, we have
$\hat{s}(x_j, \alpha) = n^{-1/2} \hat{s}_j z_{1 - \alpha / (2J)}$ (see the form of $\mathcal{G}_\alpha$ defined in \eqref{G:random:box:asymp}).
The constraints in \eqref{bl-constraints} are bilinear and could potentially create computational difficulties;
however, it turns out that it is easy to solve \eqref{alt-1}-\eqref{bl-constraints} with $\mathcal{G}_\alpha$ in the form of \eqref{G:box:generic}.
Specifically, the following theorem shows that it suffices to solve the sign-restricted optimization problem below. Since Assumption~\ref{a:para-space} makes \(\mathbf B\) a box and all additional restrictions in \eqref{g-constraint-est} are linear, this problem is a linear program (LP).

\begin{theorem}\label{thm2}
Solutions to \eqref{alt-1}-\eqref{bl-constraints} with $\mathcal{G}_\alpha$ in the form of \eqref{G:box:generic} can be obtained by solving
\begin{align}\label{alt1-box-LP}
\underset{b \in \mathbf{B}}{\operatorname{maximize}} \left(\underset{b \in \mathbf{B}}{\operatorname{minimize}} \right): \; r'b
 \end{align}
 subject to
 \begin{subequations}\label{g-constraint-est}
\begin{align}
 \left( x_{j,1} + b_2 x_{j,2} + \cdots +  b_q x_{j,q}  \right) \geq 0 \;&\text { if $\hat{g}(x_j) - \hat{s}(x_j, \alpha)  >  0$},  \label{g-constraint-est-1} \\
\left( x_{j,1} + b_2 x_{j,2} + \cdots +  b_q x_{j,q}  \right) \leq 0 \;&\text{ if $\hat{g}(x_j)  + \hat{s}(x_j, \alpha) < 0$}. \label{g-constraint-est-2}
\end{align}
\end{subequations}
\end{theorem}

Thus a constraint is imposed for index \(j\) only when the confidence interval
for \(g_j\) lies strictly above or strictly below zero. If
\(\hat g(x_j)\in[-\hat s(x_j,\alpha),\hat s(x_j,\alpha)]\), no constraint is
imposed for that index.
In what follows, we refer to the LP representation in
Theorem~\ref{thm2}, for either endpoint, as the \emph{confidence-region linear
program}.

\section{Misclassification}\label{sec:regret}

We define misclassification to mean a difference between the feasible
classification based on the estimates \((\hat c_L,\hat c_U)\) and the
classification based on the unknown population quantities \((c_L,c_U)\).
Misclassification can occur because \((\hat c_L,\hat c_U)\) is subject to
random sampling errors and, therefore, may differ from \((c_L,c_U)\). This
section describes how the difference affects the classification rules
described in Section~\ref{sec:decision:theory}. The relevant
issue is whether the signs of \(\hat c_L\) and \(\hat c_U\) agree with those
of \(c_L\) and \(c_U\). We call this issue \emph{sign stability}. The entire
probability distributions of
\(\hat c_L-c_L\) and \(\hat c_U-c_U\) are not relevant to misclassification.
The remainder of this section describes the sign-stability issues in more
detail and presents a uniform upper bound on the probability of misclassification.
Refined bounds for a given covariate value are developed in
Supplemental Appendix~\ref{subsec:refined-misclassification}.

\subsection{Sign Stability}
\label{subsec:local-sign-events}

Under Assumptions~\ref{a:design}~and~\ref{a:margin-condition},
\(g_{0,j}\neq0\), so \(d_j=\operatorname{sgn}(g_{0,j})\in\{-1,1\}\).
The population program \eqref{lp-pop-coef}--\eqref{d-constraint} imposes the
sign constraint \(d_j b'x_j\geq0\) at support point \(j\).
By contrast, the formulation in \eqref{alt-1}--\eqref{bl-constraints} jointly
optimizes over \(g\in\mathcal G_\alpha\) and \(b\), so the sign restriction at
support point \(j\) is determined by the feasible signs of \(g_j\).

The coverage event
\(g_0\in\mathcal G_\alpha\) gives endpoint coverage by
Theorem~\ref{thm1}, but it does not by itself imply
\(\hat c_L(r)=c_L(r)\) and \(\hat c_U(r)=c_U(r)\).  If a confidence interval
for \(g_j\) contains zero, the confidence-region linear program \eqref{alt1-box-LP}--\eqref{g-constraint-est} drops
the corresponding sign constraint.

Define the global sign-agreement event
\[
\Salpha
=
\{g_jg_{0,j}>0\text{ for every }g\in\G_\alpha
\text{ and every }j=1,\ldots,J\}.
\]
On \(\Salpha\), the confidence-region and population programs impose the same
sign constraints, so
\begin{align}\label{eq:global:agr}
\hat c_L(r)=c_L(r),\qquad \hat c_U(r)=c_U(r),
\end{align}
and the feasible and population decision rules agree for every \(r\).

For the rectangular confidence region \eqref{G:box:generic} and the confidence-region LP
\eqref{alt1-box-LP}--\eqref{g-constraint-est}, there are two ways in which
global sign agreement can fail.
First, the population program
imposes sign constraints at indices \(j\) for which
\(\lvert\hat g(x_j)\rvert\leq\hat s(x_j,\alpha)\), but the confidence-region
LP does not. We call these \emph{dropped constraints}. Define the
dropped-constraint set
\[
\Jhat=\{j:|\hat g(x_j)|\leq \hat s(x_j,\alpha)\}.
\]
The weak inequality in the definition of \(\Jhat\) is deliberate: when zero
is feasible for \(g_j\), the bilinear constraint does not restrict the sign of
\(b'x_j\).

Second, the confidence interval for \(g_j\) can lie strictly on the opposite
side of zero from \(g_{0,j}\).  Let
\[
\Ehat
=
\left\{
j:
\big(d_j=1,\ \hat g(x_j)+\hat s(x_j,\alpha)<0\big)
\text{ or }
\big(d_j=-1,\ \hat g(x_j)-\hat s(x_j,\alpha)>0\big)
\right\}
\]
denote the set of wrong-sign constraints.  If \(j\in\Ehat\),
the confidence-region LP reverses the direction of the corresponding
population sign constraint. We call this a \emph{sign reversal}. Such a
reversal can occur only if \(g_0\notin\mathcal G_\alpha\). The strict
inequalities in the definition of
\(\Ehat\) are exactly the conditions under which the confidence interval for
\(g_j\) lies strictly on the wrong side of zero.

\begin{lemma}[Sign-Agreement Decomposition]
\label{lem:no-sign-error}
Suppose Assumptions~\ref{a:design} and~\ref{a:margin-condition} hold, and
let \(\mathcal G_\alpha\) be the rectangular confidence region in
\eqref{G:box:generic}. Then
\[
\Salpha=\{\Jhat=\emptyset\}\cap\{\Ehat=\emptyset\}.
\]
\end{lemma}

Lemma~\ref{lem:no-sign-error} identifies the two possible failures of global
sign agreement. If neither a constraint is dropped nor a wrong-sign constraint
is imposed, the population and confidence-region programs have the same
feasible set, so their endpoints and classification rules agree for every
value of \(X_*\).
Failure of \(\Salpha\) need not imply misclassification at a
given covariate value.

\subsection{A Misclassification Bound}
\label{subsec:misclassification-events}

Misclassification means a discrepancy between the population classification
rule based on \([c_L,c_U]\) and the feasible rule based on
\([\hat c_L,\hat c_U]\); it is a sampling error relative to the population
maximum-score rule, not a prediction error relative to the realized outcome
\(Y\). Let \(M_{\mathrm{OR}}^A(x)\) and \(M_{\mathrm{MS}}^A(x)\) denote,
respectively, the population and feasible abstention rules from
Section~\ref{sec:decision:theory}. Let \(M_{\mathrm{OR}}^R(x;\xi)\) and
\(M_{\mathrm{MS}}^R(x;\xi)\) denote the corresponding random-classification
rules, where the population and feasible rules use the same
Bernoulli\((p^*)\) draw \(\xi\), independent of the estimation sample and of
\(X_*\). Define
\[
\begin{aligned}
\mathcal M_A(X_*)
&=
I\{M_{\mathrm{MS}}^A(X_*)\neq M_{\mathrm{OR}}^A(X_*)\},\\
\mathcal M_R(X_*;\xi)
&=
I\{M_{\mathrm{MS}}^R(X_*;\xi)\neq M_{\mathrm{OR}}^R(X_*;\xi)\}.
\end{aligned}
\]
On \(\Salpha\), the population and feasible endpoints are identical for every
value of \(X_*\). Hence both disagreement indicators equal zero, so each
misclassification event is contained in \(\Salpha^c\).


For brevity, we focus on the fixed design, finite-sample
confidence region in \eqref{G:fixed:box}. The method of proof of Theorem~\ref{thm:global-misclassification-bound} can be adapted to the other
confidence regions described in this paper.
For each \(j\), let
\[
h_{j,\alpha,n}
=
\left(
|g_{0,j}|-\frac{n_j}{n}t_{j,\alpha}
\right)_+,
\qquad
(a)_+=\max\{a,0\}.
\]

\begin{theorem}[Misclassification Bound]
\label{thm:global-misclassification-bound}
Suppose Assumptions~\ref{a:design}(a) and~\ref{a:margin-condition}
hold, and consider the fixed-design finite-sample confidence region
in \eqref{G:fixed:box}. For a fixed classification target
\(X_*\),
\[
\begin{aligned}
\mathbb P\{\mathcal M_A(X_*)=1\}
&\leq \mathbb P(\Salpha^c),\\
\mathbb P\{\mathcal M_R(X_*;\xi)=1\}
&\leq \mathbb P(\Salpha^c),
\end{aligned}
\]
where the second inequality uses the common randomization device. Moreover,
\begin{align}\label{eq:global-sign-bound}
\mathbb P(\Salpha^c)
\leq
\sum_{j=1}^J
\exp\left\{
-\frac{2n^2}{n_j}
\left(
|g_{0,j}|+\frac{n_j}{n}t_{j,\alpha}
\right)^2
\right\}
+
\sum_{j=1}^J
\exp\left(-\frac{2n^2h_{j,\alpha,n}^2}{n_j}\right).
\end{align}
\end{theorem}

For fixed \(\alpha\) and \(J\), along a sequence of fixed
designs for which the cell shares \(n_j/n\) are bounded away from zero, the
right-hand side of \eqref{eq:global-sign-bound} decreases exponentially in
\(n\).
Because \(\Salpha\) guarantees equality of the endpoint programs for every
value of \(X_*\), the bound is uniform in \(X_*\) and therefore also applies
when \(X_*\) is random and independent of the estimation sample.

By Lemma~\ref{lem:no-sign-error},
\[
\Salpha^c
=
\{\Ehat\neq\emptyset\}\cup\{\Jhat\neq\emptyset\}.
\]
Thus,
\[
\mathbb P(\Salpha^c)
\leq
\mathbb P(\Ehat\neq\emptyset)
+
\mathbb P(\Jhat\neq\emptyset).
\]
The first sum in \eqref{eq:global-sign-bound} bounds
\(\mathbb P(\Ehat\neq\emptyset)\), while the second bounds
\(\mathbb P(\Jhat\neq\emptyset)\).
Refined bounds for a given covariate value \(X_*\) are developed in
Supplemental Appendix~\ref{subsec:refined-misclassification}, together with
their implications for expected excess regret.

\section{Continuous Covariates}\label{sec:hypercube-xv}

In this section, we extend our framework to continuous covariates by
discretizing them and following the bounds approach of
\citet{manski2002inference}, hereafter MT.

\subsection{Model}

Let $X$ be a discretely distributed random vector with $\operatorname{dim}(X)=q_X$ and $V$ be a continuously distributed random vector with $\operatorname{dim}(V)=q_V$. The model is
given by
\begin{align*}
& Y=\left\{
\begin{array}{cc}
1 &\text { if } \beta^{\prime} X+\delta^{\prime} V-U > 0 \\
0 &\text { if } \beta^{\prime} X+\delta^{\prime} V-U \leq 0
\end{array}
\right.,
\end{align*}
where the conditional distribution of $U$ given \((X,V)\) is absolutely continuous with respect to Lebesgue measure, and $\mathbb{P} \left(U < 0 \mid X, V \right)=\tau$ almost surely for some constant $0<\tau<1$.
We use the scale/sign normalization $\delta_1=1$, where $\delta_1$ is the first element of $\delta$.
We consider two cases. First, \(V\) may be interval observed: the
researcher observes \(V_0\) and \(V_1\), with \(V_0 \leq V \leq V_1\)
component-wise, but not the exact value of \(V\). Second, \(V\) may be
observed continuously and then discretized by the researcher into hypercubes,
with \(V_0\) and \(V_1\) denoting the lower and upper endpoints of the cell
containing \(V\). In both cases, the observed covariates for this section are
\(W=(X',V_0',V_1')'\).

In this setting, $\theta \equiv (\beta', \delta')'$ is not point identified.
The objective here is to estimate upper and lower bounds on \(r^{\prime}\theta\), where \(r\) is a constant vector whose components are not all zero.

We impose the following regularity conditions.

\begin{assumption}\label{a:design:xv}
\begin{enumerate}
\item $W$ is discretely distributed.
\item $\left\{ (Y_i, W_i): i=1, \ldots, n\right\}$ is an independent random sample of $\left(Y, W\right)$.
\item
$\theta \equiv (\beta',\delta')' \in \Theta \subset \mathbb{R}^{q_X + q_V}$, where
\(\Theta\) is the known box that imposes
\(\underline{\beta}_k\leq\beta_k\leq\overline{\beta}_k\) for
\(k=1,\ldots,q_X\), \(\delta_1=1\), and
\(0\leq\delta_\ell\leq\overline{\delta}_\ell\) for
\(\ell=2,\ldots,q_V\), for known finite constants
\(\underline{\beta}_k<\overline{\beta}_k\) and
\(0<\overline{\delta}_\ell<\infty\).

\end{enumerate}
\end{assumption}

We now make the following modification of MT’s interval, monotonicity, and mean
independence (IMMI) assumptions.


\begin{assumption}[IMMI]\label{assumption:IMMI}
\begin{enumerate}
\item Interval (I): $V_0 \leq V \leq V_1$ if $V_0$ and $V_1$ are non-stochastic, and
$\mathbb{P} \left(V_0 \leq V \leq V_1\right)=1$ if they are random, where the inequalities hold component-wise.
\item Monotonicity (M): $\mathbb{E} (Y \mid X, V)$ exists and is weakly increasing in each component of $V$.
\item Mean independence (MI): $\mathbb{E} \left(Y \mid X, V, V_0, V_1\right)=\mathbb{E} (Y \mid X, V)$.
\end{enumerate}
\end{assumption}

MT's Proposition 1 uses monotonicity of \(\mathbb{E}(Y\mid X,V)\),
not the restriction that  the components of $\delta$ are non-negative.
In the semiparametric index
specification above, the signs of the coefficients on the components of \(V\)
are assumed to be known, as is often natural in applications (for example, when
\(V\) is a price). This known index-sign restriction is imposed through
\(\Theta\): after changing the sign of any component whose coefficient is known
to be negative, we write the corresponding coefficients as non-negative. With
an unrestricted conditional distribution of \(U\), monotonicity of
\(\mathbb{E}(Y\mid X,V)\) alone need not determine the signs of \(\delta\).


\subsection{Identification Result}

Let $\mathcal{W} \equiv \{ w_j \equiv (x_j, v_{0 j}, v_{1 j}): j=1, \ldots, J \}$ denote the mass points of $W$.
The following lemma is the componentwise endpoint inequality
implied by the MT argument. For scalar \(V\), MT prove sharpness of the
corresponding envelope bounds for \(\mathbb E(Y\mid X=x,V=v)\); the present
argument uses the endpoint inequalities to form sign restrictions for the
maximum score problem.

\begin{lemma}\label{lem:MT}
Let the IMMI assumptions hold. Then, the following inequalities hold for all $w \equiv (x,v_0,v_1) \in \mathcal{W}$:
\begin{align}\label{eq:MT}
\mathbb{E}\left(Y \mid X=x, V=v_0\right)
\leq \mathbb{E} \left(Y \mid W=w \right)
\leq \mathbb{E}\left(Y \mid X=x, V=v_1\right).
\end{align}
\end{lemma}

Now define
\begin{align*}
f(w_j) &\equiv \mathbb{P}\left(Y=1 \mid X=x_j, V_0=v_{0 j}, V_1=v_{1 j}\right)-\tau
\end{align*}
and
\begin{align*}
d_j = \operatorname{sgn}\left[f(w_j)\right] =\left\{
\begin{array}{cc}
1  &\text { if } f(w_j)>0 \\
0  &\text { if } f(w_j)=0 \\
-1 &\text { if } f(w_j)<0
\end{array}.
\right.
\end{align*}

We now make an analog of Assumption~\ref{a:margin-condition}.

\begin{assumption}\label{ass:f-xv}
\(\left|f(w_j)\right|>0\) for all \(j=1,\ldots,J\).
\end{assumption}

This assumption is similar to Assumption~\ref{a:margin-condition}:
it excludes boundary cases in which the conditional choice probability equals
\(\tau\).
By Lemma~\ref{lem:MT} and the implications
\begin{align*}
\left[\mathbb{P}\left(Y=1\mid X=x_j,V=v_j\right)-\tau\right]>0
&\Rightarrow
\beta'x_j+\delta'v_j>0,\\
\left[\mathbb{P}\left(Y=1\mid X=x_j,V=v_j\right)-\tau\right]<0
&\Rightarrow
\beta'x_j+\delta'v_j<0,
\end{align*}
if \(d_j=1\), then
\[
\mathbb{P}(Y=1\mid X=x_j,V=v_{1j})-\tau>0
\quad\text{and hence}\quad
\beta'x_j+\delta'v_{1j}>0.
\]
Similarly, if \(d_j=-1\), then
\[
\mathbb{P}(Y=1\mid X=x_j,V=v_{0j})-\tau<0
\quad\text{and hence}\quad
\beta'x_j+\delta'v_{0j}<0.
\]
Following the baseline analysis in Section~\ref{sec:model}, we use the
weak-inequality outer set
\begin{align*}
\Theta(0)=
\bigl\{\theta\equiv(\beta',\delta')'\in\Theta:
&\ I(d_j=1)(\beta'x_j+\delta'v_{1j})  \\
&\quad -I(d_j=-1)(\beta'x_j+\delta'v_{0j})\geq0,\;
j=1,\ldots,J
\bigr\}.
\end{align*}
This set contains the strict sign-restriction set implied by the true
parameter and may include boundary values with zero endpoint index. Positive
margin restrictions can be imposed as a sensitivity analysis, but the main
continuous-covariate analysis uses \(\varepsilon=0\).

The baseline upper and lower bounds on \(r^{\prime}\theta\) over \(\Theta(0)\) are the optimal values of the following linear program:
\begin{align}\label{lp-pop-coef-xv}
 \underset{\theta \in \Theta}{\operatorname{maximize}} \left( \underset{\theta \in \Theta}{\operatorname{minimize}} \right): \; r'\theta
 \end{align}
 subject to
 \begin{align}\label{d-constraint-xv}
 I(d_{j}=1)  (\beta^{\prime} x_j+\delta^{\prime} v_{1 j})
- I(d_{j}=-1)  (\beta^{\prime} x_j+\delta^{\prime} v_{0 j})  \geq 0
\end{align}
for all $j=1,\ldots,J$.

\subsection{Inference}

For brevity, we focus on  the random design case. It is straightforward to modify the following discussion to the fixed design case.
As before, we define
\begin{align*}
g(w_j) \equiv \left[\mathbb{P}\left(Y=1 \mid X=x_j, V_0=v_{0 j}, V_1=v_{1 j}\right)-\tau\right] p_{W}(w_j),
\end{align*}
where $p_{W}(\cdot)$ is the probability mass function of $W$. As $p_{W}(w_j) > 0$, we have that
$\operatorname{sgn}[f(w_j)] = \operatorname{sgn}[g(w_j)]$ for all $j=1,\ldots,J$.
We now define the estimator
\[
\hat{g}(w_j)=n^{-1} \sum_{i=1}^n\left(Y_i-\tau\right) I\left(W_i=w_j\right)
\]
and assume that we have the following confidence set for $g_j \equiv g(w_j)$:
\begin{align}\label{G:box:generic:xv}
\mathcal{G}_\alpha =
\left\{
(g_1,\ldots, g_J): \hat{g}(w_j)  -  \hat{s}(w_j, \alpha)
\leq g_j \leq
\hat{g}(w_j)  +  \hat{s}(w_j, \alpha) \; \forall j
\right\}
\end{align}
for an appropriate choice of $\hat{s}(w_j, \alpha) > 0$.

The same argument as in Theorem~\ref{thm2} reduces inference to the following
sign-restricted LP:

\begin{align}\label{alt1-box-LP-xv}
\underset{\theta \in \Theta}{\operatorname{maximize}} \left(\underset{\theta \in \Theta}{\operatorname{minimize}} \right): \; r'\theta
 \end{align}
 subject to
 \begin{subequations}\label{g-constraint-est-xv}
\begin{align}
  (\beta^{\prime} x_j+\delta^{\prime} v_{1 j}) \geq 0 \;&\text { if $\hat{g}(w_j) - \hat{s}(w_j, \alpha)  >  0$},  \label{g-constraint-est-1-xv} \\
 (\beta^{\prime} x_j+\delta^{\prime} v_{0 j}) \leq 0 \;&\text{ if $\hat{g}(w_j)  + \hat{s}(w_j, \alpha) < 0$}. \label{g-constraint-est-2-xv}
\end{align}
\end{subequations}
When $ v_{0 j} =  v_{1 j}$, the resulting optimization problem is the same as \eqref{alt1-box-LP}.

\subsection{Classification}

Suppose that we want to classify a new member of the population at a fixed
covariate value \((X_*,V_*)\). Equivalently, this classification problem can
be represented by the degenerate hypercube
\(W_*=(X_*',V_*',V_*')'\), where \(W_*\) may not be an
element in the support \(\mathcal W\) of \(W\).
In particular, \(V_*\) can be different from the realized values of
\((V_{0i},V_{1i})\). Then, for the classification index, set
\(r_*=(X_*',V_*')'\).
We solve \eqref{alt1-box-LP-xv} with \(r=r_*\) and let
\(c_L(X_*,V_*)\) and \(c_U(X_*,V_*)\) denote the resulting lower and upper bounds.
As before, we assign \(Y=1\) if \(c_L(X_*,V_*)\ge0\) and
\(c_U(X_*,V_*)>0\), assign \(Y=0\)
if \(c_U(X_*,V_*)\le0\), and abstain from classification or
apply random classification if \(c_L(X_*,V_*)<0<c_U(X_*,V_*)\).

\section{Empirical Illustrations}\label{sec:examples}

In Section~\ref{sec:KLS}, we apply our method to classify resumes in the context of hiring decisions.
In Section~\ref{example:gps:grouped}, we demonstrate how the approach can be extended to  interval-valued covariates using the resume classification example.
Finally, in Section~\ref{sec:CHV}, we use our method to assess the likelihood of marriage among women in developing countries.

\subsection{Incentivized Resume Rating}\label{sec:KLS}

In this section, we use data from \citet{KLS} to illustrate the methodology developed in this paper.
\citet{KLS} introduced a new experimental paradigm, called incentivized resume rating (IRR).
For each resume, the authors asked two questions, both on a Likert scale from 1 to 10:
\begin{itemize}
\item[(i)] ``How interested would you be in hiring [Name]?'' (1 = ``Not interested''; 10 = ``Very interested'');
\item[(ii)] ``How likely do you think [Name] would be to accept a job with your organization?'' (1 = ``Not likely''; 10 = ``Very likely'').
\end{itemize}

We construct the binary response variable
by setting $Y_i=1$ if the rating for the first question is greater than or
equal to 5 and $Y_i=0$ otherwise. This outcome can be viewed as a proxy for
whether a resume warrants a callback. The actions in this empirical exercise
therefore correspond to the decision problem in
Section~\ref{sec:decision:theory}: $\hat y=1$ means extending a callback,
$\hat y=0$ means declining, and $\hat y=\emptyset$ means deferring the
decision.

Among possible covariates, we include:
\begin{itemize}
\item GPA $\in \{3.0, 3.2, \ldots, 4.0\}$, that is, GPA is grouped into six categories via $\text{\texttt{ceiling}}(\text{GPA} \times 5)/5$;
\item An indicator variable for Top Internship, which refers to internships at prestigious firms;
\item Intercept.
\end{itemize}
GPA and Top Internship are the two most important predictors in the regression
analysis of \citet{KLS}; see their Table 1.
In this example, $n = 2880$ and $J = 6 \times 2 = 12$.
Out of 12 covariate groups, the minimum, median, and maximum group sizes are 110, 226,  and 347, respectively.

\begin{figure}[htbp]
\begin{center}
\includegraphics[scale=0.4]{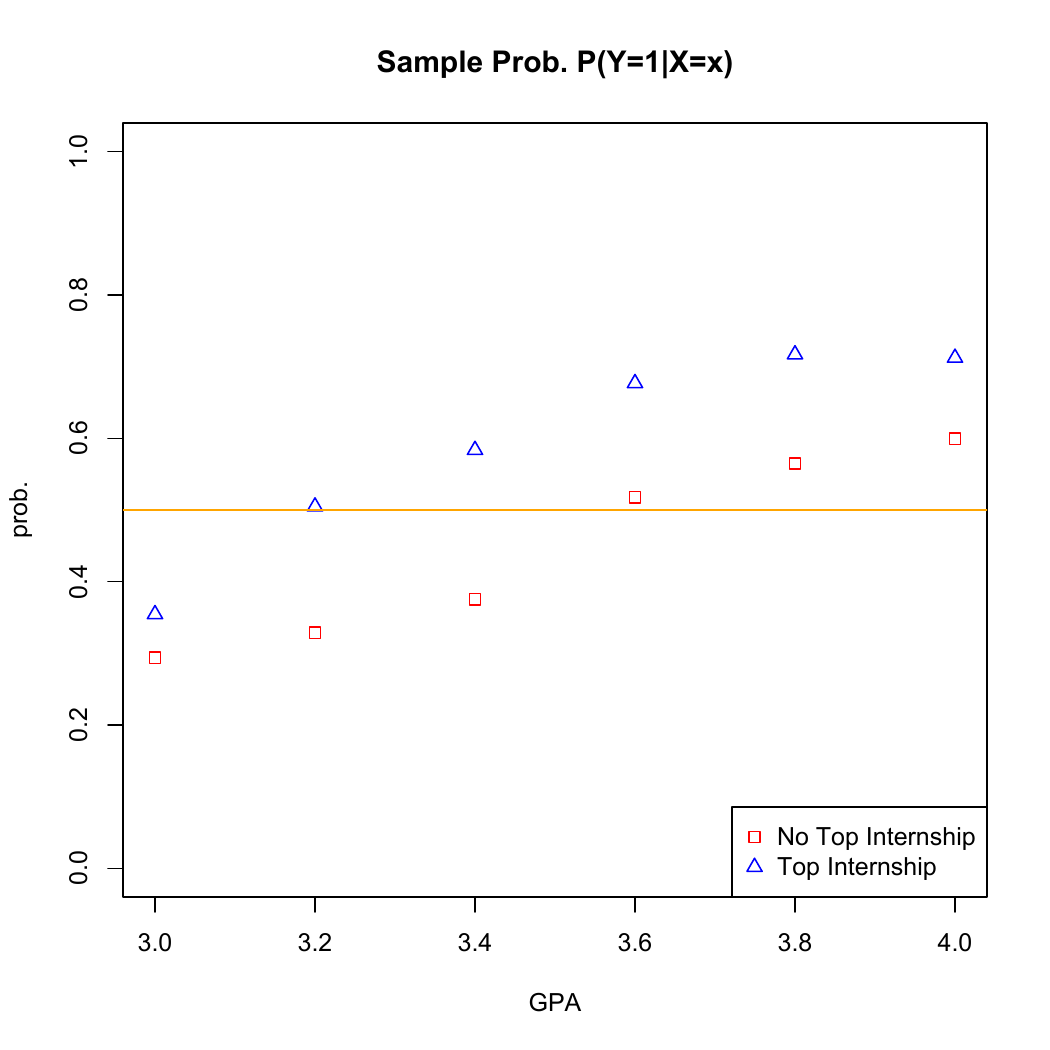}
\end{center}
\caption{Sample analog of \(\mathbb{P}(Y=1\mid X=x)\) by GPA and Top Internship}\label{fig-kls1}
\end{figure}

Figure~\ref{fig-kls1} displays the sample analog of \(\mathbb{P}(Y=1\mid X=x_j)\), the probability of receiving a
rating of 5 or higher for hiring interest, by covariate group.
Blue triangles correspond to 6 GPA groups (points of support of the covariates) with a top internship (i.e., Top Internship $= 1$)
and red squares those without a top internship (i.e., Top Internship $= 0$). It can be seen that having a top internship boosts the ratings although the vertical distance between the triangle and square is not uniform across different GPA groups.
The orange horizontal line represents the probability of 0.5. In the
estimation sample, the fraction of observations with \(Y=1\) is
\(1513/2880=0.525\), so we set \(\tau=0.5\) in this example.

In view of Figure~\ref{fig-kls1}, we consider the following simple linear specification for $\beta' x$ with $\tau = 0.5$:
\begin{align}\label{model-specification-kls}
\beta' x
&= \beta_1 \times GPA + \beta_2 \times Top \; Internship + \beta_3,
\end{align}
where $\beta_1 \equiv 1$ by scale normalization.
The main parameter of interest is $\beta_2$, which measures the effectiveness of having a top internship relative to the effect of GPA.

\input{results/table_kls_main.tex}

Column (1) of Table~\ref{tab-kls1} reports misspecification-robust asymptotic
95\% logit confidence intervals for the two normalized coefficients. Column
(2) solves \eqref{alt1-box-LP}--\eqref{g-constraint-est} after setting
\(\hat s(x_j,\alpha)=0\), thereby treating the sample sign restrictions as
known. Columns (3) and (4) instead use the asymptotic 95\% confidence regions
under fixed and random designs, respectively. For maximum score, we set
\(\tau=0.5\), normalize
\(\beta_1=1\), and restrict \(-10\leq\beta_j\leq10\), \(j=2,3\). The Top
Internship coefficient is strictly positive under logit and the sample sign
restrictions, whereas accounting for sampling uncertainty produces wider
intervals. The fixed- and random-design intervals coincide in this
application. The sample is too small for informative finite-sample bounds;
Supplemental Appendix~\ref{sec:MC} studies the sample sizes needed for such
bounds in the corresponding fixed-design simulation.

\begin{figure}[htbp]
\begin{center}
\includegraphics[scale=0.70]{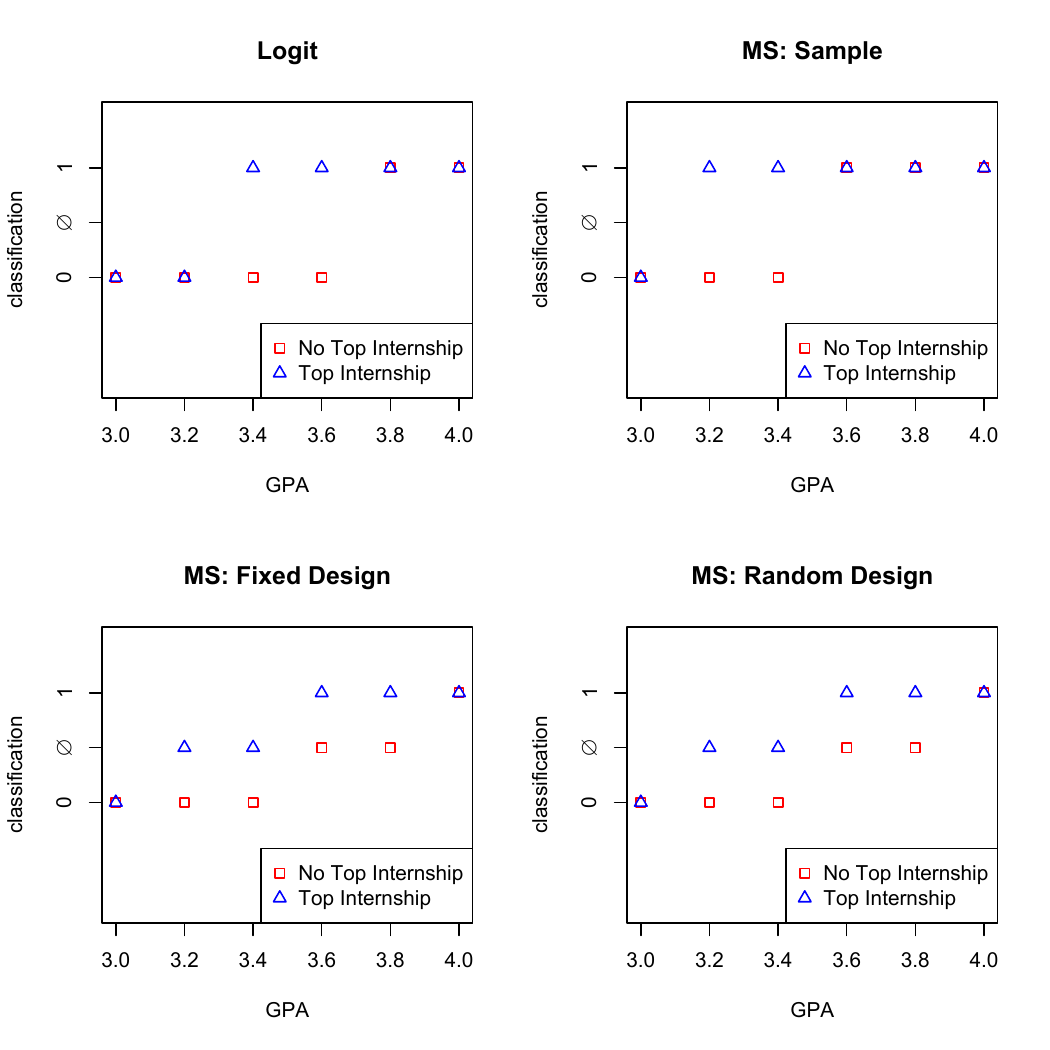}
\end{center}
\caption{Classification by Covariate Groups}\label{fig-kls2}
\end{figure}

Figure~\ref{fig-kls2} reports classification by covariate group. The top-left
panel uses the sign of \(x_j'\hat\beta_{\mathrm{Logit}}\); because
\(\tau=0.5\), it assigns \(1\) to a positive index and \(0\) otherwise. The
other panels apply
the abstention rule in Section~\ref{sec:decision:theory}, with \(\emptyset\)
denoting non-classification and exact boundaries assigned to \(0\). These
maximum score panels correspond to columns (2)--(4) of
Table~\ref{tab-kls1}.
The logit rule classifies six covariate groups as \(1\)
and six as \(0\), with Top Internship changing the classifications of the
middle GPA groups 3.4 and 3.6. Treating sample probabilities as population
probabilities, the sample maximum score rule classifies eight groups as \(1\)
and four as \(0\), with no non-classifications. Relative to no top internship,
Top Internship changes the classifications at GPA 3.2 and GPA 3.4 from \(0\)
to \(1\).
The fixed- and
random-design confidence-region panels coincide, each classifying four groups
as \(1\), four as \(0\), and four as non-classification. They classify GPA
groups of 3.6 or higher with a top internship as \(1\) and GPA groups of 3.4 or
lower without one as \(0\); all panels classify both internship statuses at
GPA 3.0 as \(0\). We omit the random-classification version because assignments
in the abstention region depend on the randomization device.

\subsection{Incentivized Resume Rating: GPA as an Interval Covariate}\label{example:gps:grouped}

In this subsection, we illustrate the extension discussed in Section~\ref{sec:hypercube-xv}.
Recall that we group GPA into six categories using
\(\text{\texttt{ceiling}}(\text{GPA}\times5)/5\). In the original data, GPA
ranges from 2.90 to 4.00 and is recorded to two decimal places. We therefore
use the following closed interval hulls for the grouped GPA values:
\[
\{[v_{0j},v_{1j}]:j=1,\ldots,6\}
=
\{[2.9,3.0],[3.0,3.2],\ldots,[3.8,4.0]\}.
\]

\input{results/table_kls_interval_main.tex}

Table~\ref{tab-kls1-xv} reports the effects of treating GPA as an interval covariate. Columns (1) and (2) reproduce the fixed- and random-design estimates from Table~\ref{tab-kls1}, where GPA is treated as discretely observed; columns (3) and (4) report the corresponding estimates when GPA is treated as interval-valued. The resulting bounds on Top Internship become wider, reflecting the additional uncertainty from interval censoring.
The interval for Top Internship widens from \([-0.2,1.0]\) to
\([-0.4,1.1]\), and the upper endpoint of the intercept interval increases
from \(-3.4\) to \(-3.2\). Thus, allowing for interval-valued GPA weakens the
sign restrictions generated by the support points, as expected.

\begin{figure}[htbp]
\begin{center}
\includegraphics[width=0.9\textwidth]{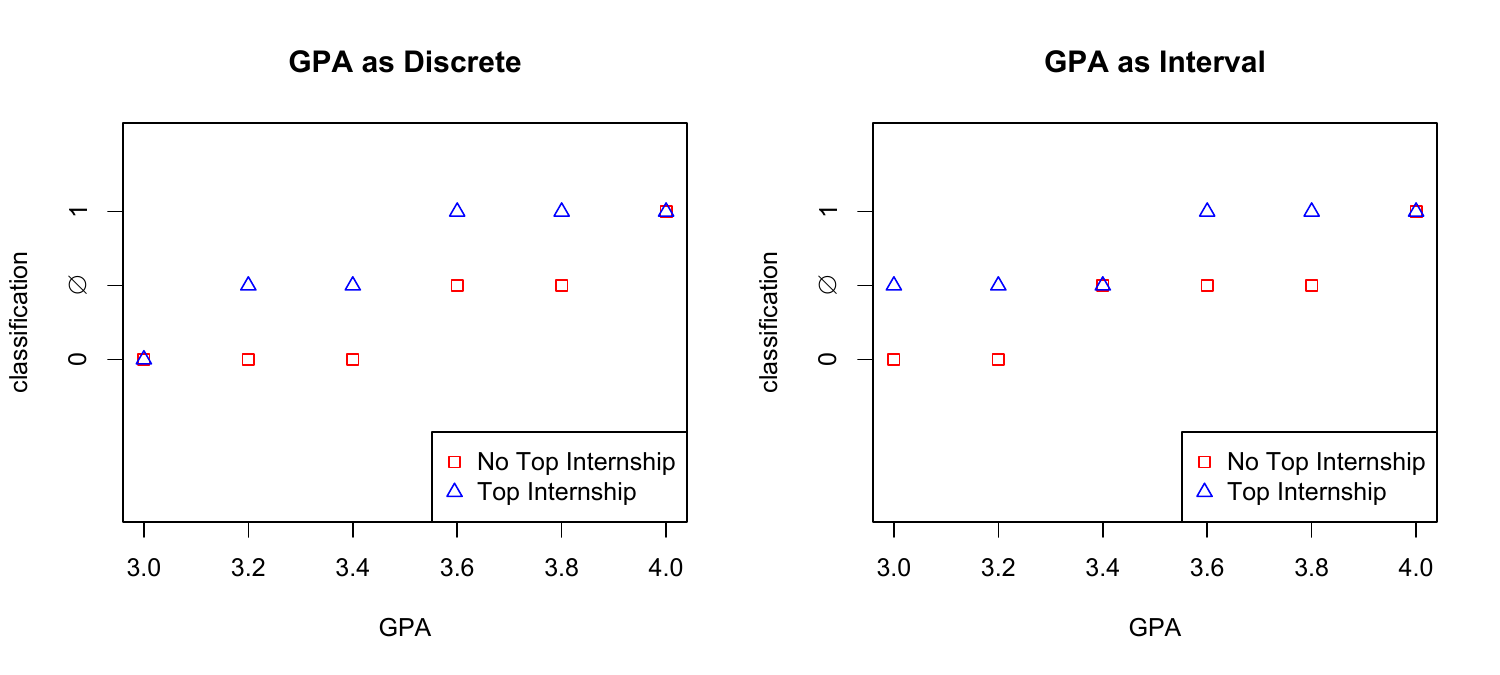}
\end{center}
\caption{Classification by Covariate Group: Discrete versus Interval-Valued GPA}\label{fig-kls2-xv}
\end{figure}

We also revisit our classification exercise.
Figure~\ref{fig-kls2-xv} compares the classification results when GPA is
treated as discretely observed and when it is treated as interval-valued. Since
the fixed- and random-design classifications coincide in this application, the
figure displays one representative classification rule for each case. Treating
GPA as interval-valued leads to more cases of non-classification
(\(\emptyset\) on the \(y\)-axis), reflecting the additional uncertainty from
interval censoring.
With discrete GPA, four resume types are classified as \(1\), four as \(0\),
and four are left unclassified. With interval-valued GPA, four resume types
are classified as \(1\), two as \(0\), and six are left unclassified.
Relative to discrete GPA, the resume type with GPA 3.4 and no top internship
and the resume type with GPA 3.0 and a top internship move from classification
as \(0\) to non-classification. These changes reflect the additional
uncertainty from interval censoring.
As before, we do not display classification outcomes under random classification.

\subsection{Age of Marriage}\label{sec:CHV}

In this subsection, we provide a second empirical example using data from \citet{Corno:ECMA:2020}.
We use their Sub-Saharan Africa sample, which contains more than two million
observations and includes survey sampling weights. Observations within the
same geographic grid cell are unlikely to be independent.
In the regression analysis of \citet{Corno:ECMA:2020}, regression coefficients are estimated with survey sampling weights and the standard errors are clustered at the grid cell level. Our framework can be extended to account for clustered dependence and survey sampling weights. Since these details are not essential for understanding the empirical results in this section, we provide them in Supplemental Appendix~\ref{sec:cluster}.
We focus on the asymptotic inference method in Supplemental
Appendix~\ref{sec:cluster:asymptotic}, because the effective sample size is
too small for the finite-sample method to be informative.

\begin{figure}[htbp]
\caption{Probability of marriage by covariates}\label{fig-chv1}
\begin{center}
\includegraphics[scale=0.70]{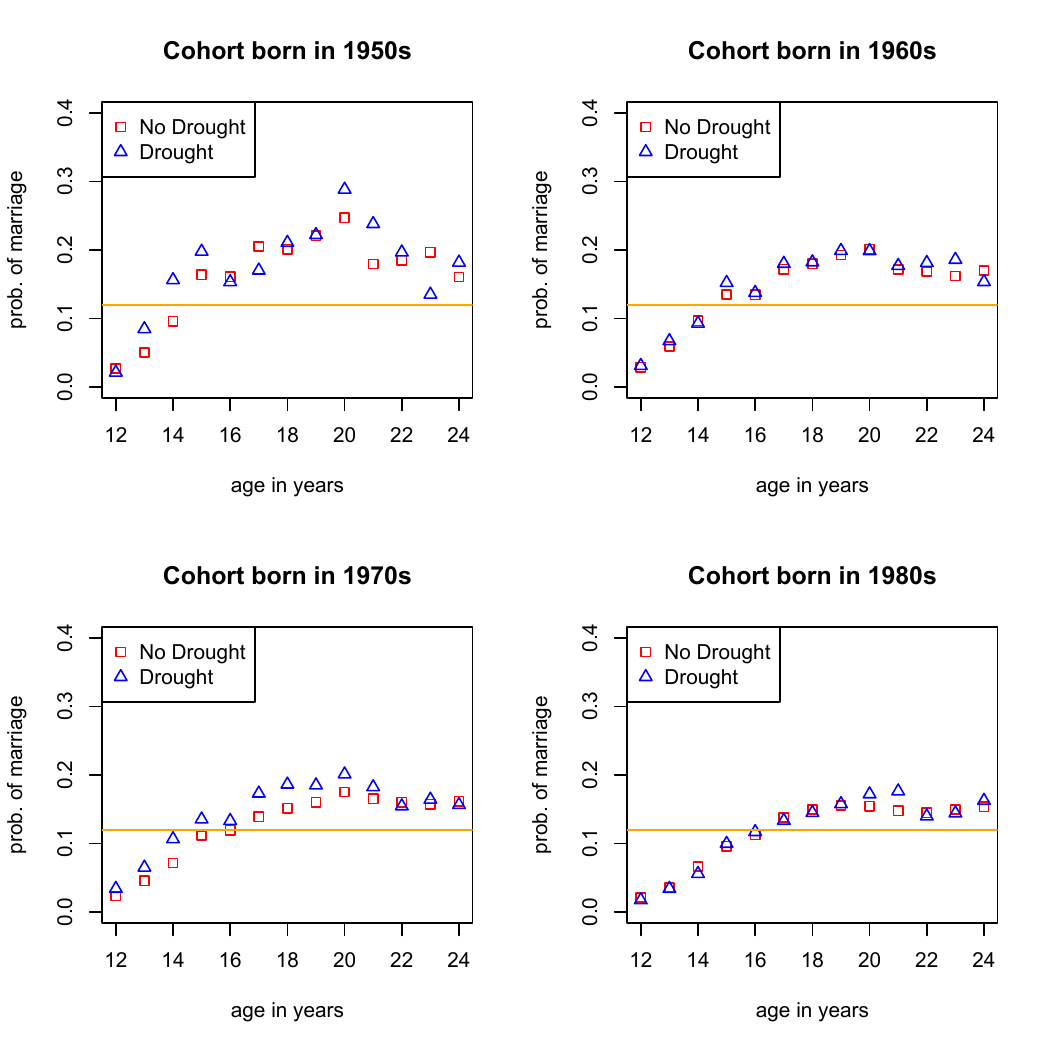}
\end{center}
\end{figure}

In this example,
the dependent variable is an indicator equal to one if a woman married at
the age corresponding to the observation and zero otherwise.
Among possible covariates, we include:
\begin{itemize}
\item age in years (minimum: 12 and maximum: 24);

\item  an indicator variable for drought, which is the main covariate in \citet{Corno:ECMA:2020};

\item three birth-decade fixed effects, with the oldest cohort omitted;

\item an intercept term.
\end{itemize}
They are all discrete covariates and the resulting $J$ is $J = 13 \times 2 \times 4 = 104$.
In the notation of Section~\ref{sec:model}, the \(J=104\)
possible values \(x_j\) are the age-drought-cohort combinations used in the
figures and tables below.
Figure~\ref{fig-chv1} displays the weighted sample analogs of
\(\mathbb P(Y=1\mid X=x_j)\), \(j=1,\ldots,J\), by age, drought, and birth
cohort.
The sample probabilities are computed using country population-adjusted survey
sampling weights. The weighted sample proportion of \(Y=1\) in the estimation
sample is 0.113. We fix \(\tau=0.12\) in this example; the orange horizontal
line in each panel shows this threshold.

It can be seen from Figure~\ref{fig-chv1} that the age effects tend to be linear up to around age 18 and become more or less flat after age 18. In view of this, we consider the following specification for $\beta' x$ with $\tau = 0.12$:
\begin{align}\label{model-specification}
\begin{split}
\beta' x
&= \beta_1 \times (age-18) I (age \leq 18) + \beta_2 \times drought
+ \beta_3 \times I ( cohort = 1960) \\
&\;\;\; + \beta_4 \times I ( cohort = 1970)
+ \beta_5 \times I ( cohort = 1980)
+ \beta_6,
\end{split}
\end{align}
where $\beta_1 \equiv 1$.

\input{results/table_chv.tex}

Table~\ref{tab-chv1} reports normalized  weighted logit confidence intervals
and maximum-score interval estimates. The logit column is reported on the same
scale as \eqref{model-specification}: the coefficient on
\((age-18)I(age\le18)\) is normalized to one. Because \(\tau=0.12\), the logit
classification rule assigns \(1\) according to whether the fitted logit index exceeds
\(\log\{\tau/(1-\tau)\}\), not zero. To put the logit coefficients on the same
zero-threshold scale as the maximum-score index, we subtract this threshold
from the raw logit intercept before normalizing. In this
application, \(\log\{\tau/(1-\tau)\}\approx-1.99\), so the adjustment is
quantitatively important. The logit intervals are based on the weighted logit estimator and
cluster-robust standard errors at the grid cell level, so both survey weights
and within-cluster dependence are handled.
Column (2) treats sample probabilities as population probabilities. Column (3)
uses the 95\% confidence region. The fixed- and random-design clustered
confidence regions select the same support-point restrictions in this
application and therefore produce the same coefficient intervals; the table
reports the common interval. Details of these clustered, survey-weighted
confidence regions are in Supplemental Appendix~\ref{sec:cluster:asymptotic}.

The weighted logit intervals indicate a positive drought coefficient and
increasingly negative cohort coefficients for later birth cohorts. These
intervals are relatively tight, reflecting the parametric logit specification.
The sample maximum-score intervals in column (2) are quite tight. Because all
regressors are integer-valued, the tightest interval is a point, as in the case
of drought, and the next tightest interval has length one, as for the other
covariates. The maximum-score intervals become wider in column (3), as
sampling uncertainty is incorporated through the confidence region
\(\mathcal G_\alpha\).

We now turn to the classification exercise. The main text reports the sample
and confidence-region maximum score rules; weighted logit classification
results are reported in Supplemental
Appendix~\ref{subsec:chv-logit-classification}. Treating sample probabilities
as population probabilities, the sample maximum score rule classifies 72
age-drought-cohort combinations as \(1\) and 32 as \(0\). For none of the 104
combinations does the admissible interval for the classification index
straddle zero, meaning that no interval has both a negative lower endpoint and
a positive upper endpoint. Consequently, the sample rule never abstains. Six
intervals equal \([0,0]\), and these cases are assigned to \(0\) under the
convention in Section~\ref{sec:decision:theory}. In these six cases, the sample
sign restrictions on the discrete covariate grid yield zero as both the
minimum and maximum admissible classification index.

\begin{figure}[htbp]
\begin{center}
\includegraphics[scale=0.70]{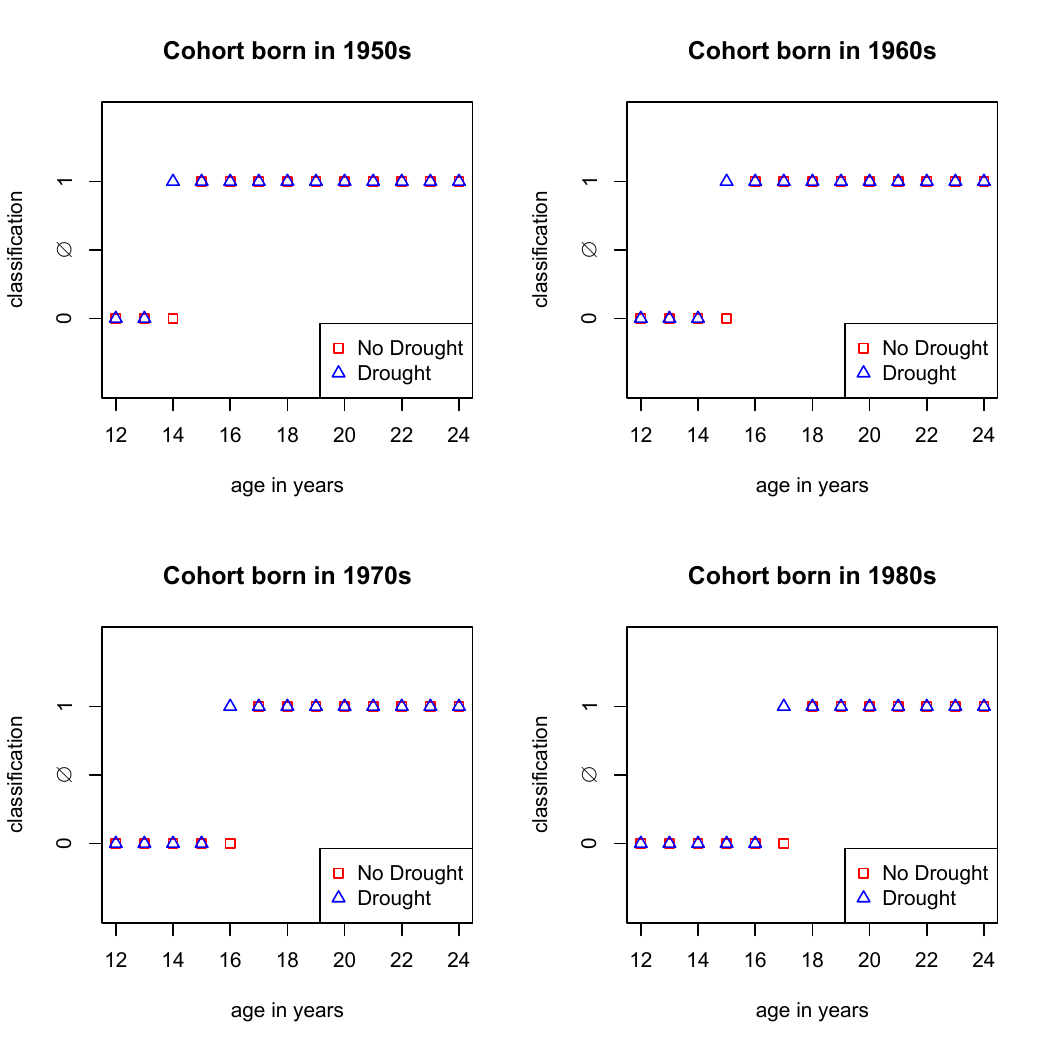}
\end{center}
\caption{Maximum Score Classification: Treating Sample Probabilities as Population Probabilities}\label{fig-chv3}
\end{figure}

The sample maximum score rule displays pronounced cohort effects:
classification as \(1\) occurs at later ages for later birth cohorts. Drought
changes the displayed classification at age 14 for the 1950 cohort, age 15 for
the 1960 cohort, age 16 for the 1970 cohort, and age 17 for the 1980 cohort.
Figure~\ref{fig-chv3} displays these classifications.

\begin{figure}[htbp]
\begin{center}
\includegraphics[scale=0.70]{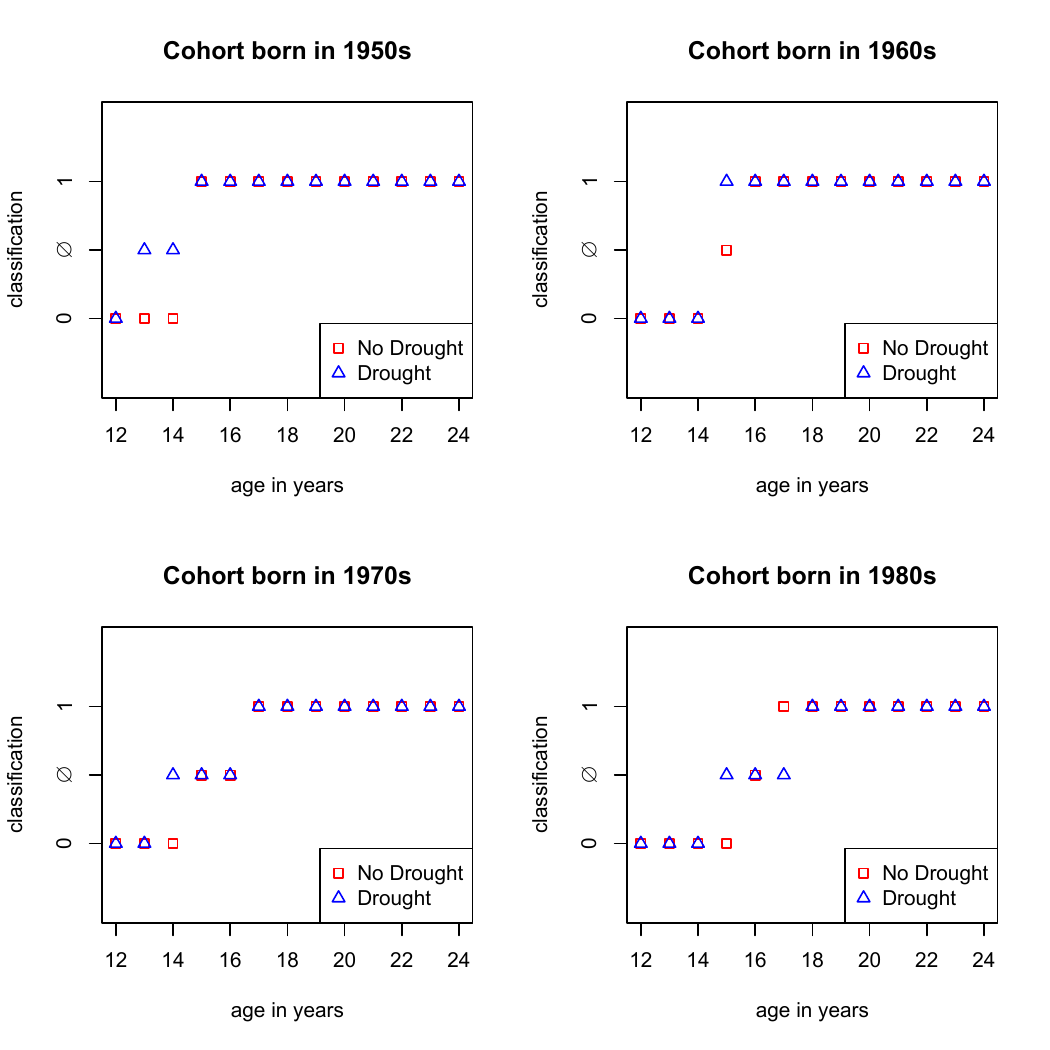}
\end{center}
\caption{Maximum Score Classification: 95\% Confidence Region}\label{fig-chv4}
\end{figure}

Figure~\ref{fig-chv4} depicts the maximum score classification rule using the
95\% confidence region. The rule classifies 70
age-drought-cohort combinations as \(1\), 22
as \(0\), and leaves 12 combinations unclassified. Drought changes the displayed
classification in six combinations. These changes occur when adding sampling uncertainty makes the
confidence-region interval for the classification index strictly contain zero
for one drought status but not the other; the rule then reports
non-classification rather than a binary label. The
random-design implementation gives the same classification in this
application, so it is not displayed separately.
As a sensitivity check, we also compare the baseline \(\tau=0.12\) results
with \(\tau=0.11\); the fixed- and random-design displayed classifications
change in 7 of 104 age-drought-cohort combinations.
Overall, the classification exercises demonstrate the usefulness of our
methodology.

\section{Monte Carlo Experiments for Misclassification}\label{sec:mc:class-errors}

To study the finite-sample behavior of the proposed classification rules, we
conduct a Monte Carlo experiment.
Specifically, we base our experimental design on the first empirical example reported in Section~\ref{sec:KLS}.
We retain the same covariate design ($X$) from Section~\ref{sec:KLS} and generate $Y$ as follows:
\begin{align*}
Y_i &= I( Y_i^\ast \geq 0), \\
Y_i^\ast &= \beta_1 \times GPA_i + \beta_2 \times \text{Top Internship}_i + \beta_3 + U_i,
\end{align*}
where \(U_i=\sigma(X_i)V_i\), \(V_i\sim N(0,1)\), and
\[
(\beta_0^{(1)},\beta_0^{(2)},\beta_0^{(3)})=(1,0.4,-3.7)'.
\]
We consider a homoskedastic design with \(\sigma(X_i)=0.5\) and a
heteroskedastic design with
\[
\sigma(X_i)
=
0.2\left[1+\left\{\frac{\mathrm{GPA}_i}{3}
+\text{Top Internship}_i\right\}^2\right].
\]
The parameter values of \((\beta_0^{(1)}, \beta_0^{(2)}, \beta_0^{(3)})\) are chosen to mimic the empirical results reported earlier and to ensure that the population values of \(g_0(x_j)\) are never zero.
The classification target is the off-support covariate value
\(X_*=(3.9,0)\), where the coordinates are GPA and Top Internship.
The quantile of interest is \(\tau = 0.5\), the reported
significance levels are \(\alpha\in\{0.01,0.05,0.10\}\), and the number of
Monte Carlo repetitions is 1,000.

\input{results/Table_mc_class_errors.tex}

Table~\ref{tab-mc-class-errors} reports misclassification frequencies using
the criterion in Section~\ref{subsec:misclassification-events} and feasible
endpoints from the fixed-design asymptotic confidence region in
\eqref{G:fixed:box:asymp}. In Panel A, both the population and feasible rules
permit abstention, and a misclassification occurs when their classifications
at \(X_*\) differ. Rows vary \(\alpha\), while columns repeat the fixed
covariate design \(c=1,2,3\) times; increasing the sample size by a factor of
\(c\) adds \(c-1\) additional copies of the original fixed
covariate design. For each Monte Carlo repetition, we compare
the confidence-region-based and population classifications and average the
resulting disagreement indicator.

Panel B instead uses the random-classification rule in
Section~\ref{subsec:rand:class} with the common randomization device and
\(p^*=1/2\), corresponding to symmetric costs. The feasible rule uses the same
device. Although the population rule differs from that in Panel A, the
definition of misclassification is unchanged.

In both panels, misclassification frequencies decrease with sample size as
the selected sign restrictions become more stable. The levels depend on the
noise design. In Panel B, conditional on an endpoint-crossing
case under common randomization, the realized error probability is
\(1-p^*\) or \(p^*\), depending on which side of zero contains the population
interval. With \(p^*=1/2\), this probability equals one half; see Supplemental
Appendix~\ref{subsec:refined-misclassification-events}. A
nonparametric sample-frequency classifier is unavailable at the off-support
target, illustrating the value of extrapolation through the maximum score
model.

\section{Conclusions}\label{sec:conclusions}

We have shown that scalable inference for the maximum score model can be
implemented using linear programming, and we have developed credible binary
classification rules that account for partial identification and sampling
uncertainty. In our empirical examples, the finite-sample methods do not
produce informative bounds, whereas the asymptotic methods do. Developing
tighter joint confidence regions than those obtained through the Bonferroni
correction is a natural direction for future research.

\appendix

\section{Proofs}\label{sec:proofs}

\begin{proof}[Proof of Theorem~\ref{thm:selective}]
For each possible sign, regret is loss relative to the oracle action for that
sign. If $f(x)>0$, the oracle action is $\hat y=1$ and the oracle loss is
$C_b^+$. If $f(x)<0$, the oracle action is $\hat y=0$ and the oracle loss is
$C_b^-$.

First suppose \(c_L(x)\geq0\) and \(c_U(x)>0\). Then every admissible value
of \(b'x\) is nonnegative and at least one admissible value is positive.
Classifying as \(1\) has zero regret at positive values and, by the
exact-boundary convention, at zero. Hence classifying as \(1\) is
minimax-regret optimal. If \(c_U(x)\leq0\), no admissible value is positive.
Classifying as \(0\) has zero regret at negative values and at zero, and is
therefore minimax-regret optimal.

Now suppose  \(c_L(x)<0<c_U(x)\). Both signs are consistent with
the identified set. If the rule classifies as $1$, its regret is zero when
$f(x)>0$ and $C^- - C_b^-=\mathrm{Reg}_{\mathrm{FP}}$ when $f(x)<0$, so its
worst-case regret is $\mathrm{Reg}_{\mathrm{FP}}$. If the rule classifies as
$0$, its regret is $C^+ - C_b^+=\mathrm{Reg}_{\mathrm{FN}}$ when $f(x)>0$
and zero when $f(x)<0$, so its worst-case regret is
$\mathrm{Reg}_{\mathrm{FN}}$. If the rule does not classify, its regret is
$C_\emptyset - C_b^+$ when $f(x)>0$ and $C_\emptyset - C_b^-$ when
$f(x)<0$, so its worst-case regret is $\mathrm{Reg}_{\emptyset}$. By
\eqref{eq:deferral-regret-condition}, abstention has weakly smaller
worst-case regret than either binary action. This proves the claim.
\end{proof}

\begin{proof}[Proof of Theorem~\ref{thm:selective:rc}]
First suppose \(c_L(x)\geq0\) and \(c_U(x)>0\). Then every admissible value
of \(b'x\) is nonnegative and at least one admissible value is positive.
Classifying as \(1\) has zero regret at positive values and, by the
exact-boundary convention, at zero. Hence classifying as \(1\) is
minimax-regret optimal. If \(c_U(x)\leq0\), no admissible value is positive.
Classifying as \(0\) has zero regret at negative values and at zero, and is
therefore minimax-regret optimal.

Now suppose  \(c_L(x)<0<c_U(x)\). Both signs are consistent with
the identified set. If the rule classifies as $1$ with probability $p$, its
regret is $(1-p)\mathrm{Reg}_{\mathrm{FN}}$ when $f(x)>0$ and
$p\,\mathrm{Reg}_{\mathrm{FP}}$ when $f(x)<0$. Its worst-case regret is
therefore
\[
\max\{(1-p)\mathrm{Reg}_{\mathrm{FN}},
      p\,\mathrm{Reg}_{\mathrm{FP}}\}.
\]
Since both regret indices are positive, this maximum is minimized by
equalizing the two terms:
\[
(1-p)\mathrm{Reg}_{\mathrm{FN}}
=
p\,\mathrm{Reg}_{\mathrm{FP}}.
\]
Solving gives
\[
p^*
=
\frac{\mathrm{Reg}_{\mathrm{FN}}}
     {\mathrm{Reg}_{\mathrm{FP}}+\mathrm{Reg}_{\mathrm{FN}}},
\]
which proves the claim.
\end{proof}

\begin{proof}[Proof of Theorem~\ref{thm1}]
Suppose that \(g_0\in\mathcal G_\alpha\). If \(b\) is feasible for the
population problem \eqref{lp-pop-coef-g}--\eqref{g-constraint}, then
\[
g_{0,j}\left(x_{j,1}+b_2x_{j,2}+\cdots+b_qx_{j,q}\right)\ge0,
\quad j=1,\ldots,J.
\]
Since \(g_0\in\mathcal G_\alpha\), the same \(b\) is feasible for the estimated
problem \eqref{alt-1}--\eqref{bl-constraints} by taking \(g=g_0\). Therefore,
the feasible region of \eqref{alt-1}--\eqref{bl-constraints} contains the
feasible region of \eqref{lp-pop-coef-g}--\eqref{g-constraint}. Consequently,
\[
\hat c_L \le c_L \le c_U \le \hat c_U
\]
on the event \(\{g_0\in\mathcal G_\alpha\}\). Hence
\[
\mathbb P\{\hat c_L \le c_L \le r'\beta \le c_U \le \hat c_U\}
\ge
\mathbb P\{g_0\in\mathcal G_\alpha\}
\ge 1-\alpha,
\]
which proves the theorem.
\end{proof}

\begin{proof}[Proof of Theorem~\ref{thm2}]
For each \(j\), the effect of the bilinear constraint in
\eqref{bl-constraints} depends on whether the coordinate confidence interval
for \(g_j\) is strictly positive, strictly negative, or contains zero. If the
interval is strictly positive, the constraint is equivalent to \(x_j'b\ge0\).
If the interval is strictly negative, it is equivalent to \(x_j'b\le0\). If the
interval contains zero, we may choose \(g_j=0\), so the constraint imposes no
restriction on \(b\).
Therefore, when $\mathcal{G}_\alpha$ is a Cartesian product of intervals as in  \eqref{G:box:generic},
solving \eqref{alt-1}-\eqref{bl-constraints}
is equivalent to solving
\begin{align}\label{alt1-box}
\underset{b \in \mathbf{B}; \; g \in \mathcal{G}_\alpha}{\operatorname{maximize}} \left(\underset{b \in \mathbf{B}; \; g \in \mathcal{G}_\alpha}{\operatorname{minimize}} \right): \; r'b
 \end{align}
 subject to \eqref{g-constraint-est} and
 \begin{align}\label{g-constraint-est-3}
 g_j \left( x_{j,1} + b_2 x_{j,2} + \cdots +  b_q x_{j,q}  \right) \geq 0
 \;&\text{ if $ -\hat{s}(x_j, \alpha) \leq \hat{g}(x_j)  \leq  \hat{s}(x_j, \alpha)$}.
\end{align}
Suppose that we want to solve the maximization problem in \eqref{alt1-box} under \eqref{g-constraint-est} and \eqref{g-constraint-est-3}.
We first start with the maximum under \eqref{g-constraint-est} only. We now verify that the same maximum is achievable under \eqref{g-constraint-est} and \eqref{g-constraint-est-3}. Let $\tilde{b}_{LP}$ denote a maximizer under the relaxed LP problem (that is, a solution under \eqref{g-constraint-est} only). If we add additional restrictions in \eqref{g-constraint-est-3}, the resulting maximum cannot be larger, and so $r'\tilde{b}_{LP}$ is the maximum value under both sets of restrictions if we can verify that $\tilde{b}_{LP}$ is feasible under \eqref{g-constraint-est-3}. For every \(j\) such that \(-\hat{s}(x_j,\alpha)\leq \hat g(x_j)\leq \hat{s}(x_j,\alpha)\), the rectangular confidence interval for \(g_j\) contains zero. Hence we can choose \(g_j=0\), in which case the bilinear constraint \(g_jx_j'\tilde b_{LP}\geq0\) is automatically satisfied, including boundary cases in which only one sign of \(g_j\) is otherwise feasible. The minimization problem is handled by the same argument. Thus, we have obtained the desired result.
\end{proof}

\begin{proof}[Proof of Lemma~\ref{lem:no-sign-error}]
For each coordinate \(j\), the interval for \(g_j\) either contains zero, lies
strictly on the population side of zero, or lies strictly on the opposite side
of zero. The condition \(\Jhat=\emptyset\) rules out the first case for every
coordinate, while \(\Ehat=\emptyset\) rules out the third. Thus every feasible
\(g_j\) has the same sign as \(g_{0,j}\) for every coordinate, which is exactly
\(\Salpha\). The converse is immediate.
\end{proof}

\begin{proof}[Proof of Theorem~\ref{thm:global-misclassification-bound}]
On \(\Salpha\), the population and confidence-region programs impose the same
sign restrictions, so their endpoints agree for every \(r\) by
\eqref{eq:global:agr}. The abstention rules therefore agree. Under the common
randomization device, the random-classification rules also agree. Thus each
misclassification event is contained in \(\Salpha^c\).

It remains to bound \(\mathbb P(\Salpha^c)\). By
Lemma~\ref{lem:no-sign-error},
\begin{align}\label{eq:global:decom}
\Salpha^c
=
\{\Ehat\neq\emptyset\}\cup\{\Jhat\neq\emptyset\}.
\end{align}

The half-width of the confidence interval for \(g_{0,j}\) is
\((n_j/n)t_{j,\alpha}\). A sign error at support point \(x_j\) can occur only
if the confidence interval for \(g_{0,j}\) lies entirely on the wrong side of
zero. If \(g_{0,j}>0\), this requires
\[
\hat g(x_j)+\frac{n_j}{n}t_{j,\alpha}<0,
\]
which implies
\[
\hat g(x_j)-g_{0,j}
<
-\left(|g_{0,j}|+\frac{n_j}{n}t_{j,\alpha}\right).
\]
If \(g_{0,j}<0\), the analogous event is
\[
\hat g(x_j)-\frac{n_j}{n}t_{j,\alpha}>0,
\]
which implies
\[
\hat g(x_j)-g_{0,j}
>
|g_{0,j}|+\frac{n_j}{n}t_{j,\alpha}.
\]
Recalling that \(d_j=\operatorname{sgn}(g_{0,j})\), both cases can be written as
\[
\left\{
d_j\left(\hat g(x_j)+d_j\frac{n_j}{n}t_{j,\alpha}\right)<0
\right\}
\subseteq
\left\{
d_j(\hat g(x_j)-g_{0,j})
<
-\left(|g_{0,j}|+\frac{n_j}{n}t_{j,\alpha}\right)
\right\}.
\]
Under the fixed design,
\[
d_j(\hat g(x_j)-g_{0,j})
=
\frac{n_j}{n}
\left[
\frac{1}{n_j}
\sum_{i:X_i=x_j}
d_j\Bigl(Y_i-\mathbb P(Y=1\mid X=x_j)\Bigr)
\right].
\]
The summands are independent, mean zero, and lie in an interval of length one.
Hence, by the one-sided Hoeffding inequality,
\[
\mathbb P
\left\{
d_j(\hat g(x_j)-g_{0,j})
<
-\left(
|g_{0,j}|+\frac{n_j}{n}t_{j,\alpha}
\right)
\right\}
\le
\exp\left\{
-\frac{2n^2}{n_j}
\left(
|g_{0,j}|+\frac{n_j}{n}t_{j,\alpha}
\right)^2
\right\}.
\]
Combining this bound with the preceding set inclusion and applying a union
bound over \(j=1,\ldots,J\) yields the bound on
\(\mathbb P(\Ehat\neq\emptyset)\).

It remains to bound the probability that at least one restriction is dropped.
For each \(j\),
\[
\{j\in\Jhat\}
=
\left\{
|\hat g(x_j)|
\leq
\frac{n_j}{n}t_{j,\alpha}
\right\}.
\]
Let \(w_j=(n_j/n)t_{j,\alpha}\). If \(g_{0,j}>w_j\), then
\[
\{|\hat g(x_j)|\leq w_j\}
\subseteq
\left\{
\hat g(x_j)-g_{0,j}\leq -(g_{0,j}-w_j)
\right\}.
\]
If \(g_{0,j}<-w_j\), then
\[
\{|\hat g(x_j)|\leq w_j\}
\subseteq
\left\{
\hat g(x_j)-g_{0,j}\geq |g_{0,j}|-w_j
\right\}.
\]
If \(|g_{0,j}|\leq w_j\), we use the trivial bound by one. Hence the
fixed-design representation of \(\hat g(x_j)-g_{0,j}\) and the one-sided
Hoeffding inequality give
\[
\mathbb P\{j\in\Jhat\}
\leq
\exp\left(-\frac{2n^2h_{j,\alpha,n}^2}{n_j}\right),
\]
where the right-hand side equals one when \(h_{j,\alpha,n}=0\). 
A union bound over the two events in \eqref{eq:global:decom}
 then over \(j=1,\ldots,J\) gives the
displayed bound on \(\mathbb P(\Salpha^c)\).
\end{proof}

\begin{proof}[Proof of Lemma~\ref{lem:MT}]
Let $W \equiv (X, V_0, V_1)$ denote the observed covariates and $w \equiv (x,v_0,v_1)$.
Let \(m_x(v)\equiv\mathbb{E}(Y\mid X=x,V=v)\).
By assumption I, conditional on \(W=w\), \(v_0\leq V\leq v_1\) component-wise.
By assumption M,
\[
m_x(v_0)\leq m_x(V)\leq m_x(v_1)
\]
almost surely conditional on \(W=w\). By the law of iterated expectations and
assumption MI,
\begin{align*}
\mathbb{E}(Y\mid W=w)
&=
\mathbb{E}\left\{\mathbb{E}(Y\mid X,V,V_0,V_1)\mid W=w\right\}\\
&=
\mathbb{E}\left\{m_x(V)\mid W=w\right\}.
\end{align*}
Taking conditional expectations of the preceding monotonicity inequalities
establishes \eqref{eq:MT}.
\end{proof}

\begingroup
\setstretch{1.20}
\bibliographystyle{econometrica}
\bibliography{MScore}
\endgroup

\newpage
\clearpage

\pagenumbering{roman}
\setcounter{page}{1}

\renewcommand{\thesection}{S-\arabic{section}}
\renewcommand{\theHsection}{supp.\arabic{section}}
\renewcommand{\theHequation}{supp.\arabic{section}.\arabic{equation}}
\setcounter{section}{0}

\begin{center}
\Large{Supplemental Appendix to ``Binary Classification with the Maximum Score Model and Linear Programming''}
\end{center}
\bigskip
\begin{center}
\begin{tabular}{ccc}
Joel L. Horowitz & \qquad & Sokbae Lee  \\
Northwestern University & \qquad  & Columbia University
\end{tabular}
\end{center}
\bigskip

\section[Sensitivity Analysis: Estimation, Inference, and Empirical Results with Positive epsilon]{Sensitivity Analysis: Estimation, Inference, and Empirical Results
         with Positive~\texorpdfstring{$\varepsilon$}{epsilon}}
\label{sec:sensitivity-eps}

This appendix extends the estimation and inference procedures to
margin-restricted bounds indexed by a tuning parameter \(\eps\geq0\).

\subsection[Sensitivity of the Population Bounds to Positive epsilon]{Sensitivity of the Population Bounds to~\texorpdfstring{$\varepsilon$}{epsilon}}
\label{subsec:sens:anal:eps}

\begingroup
The population linear programming problems \eqref{lp-pop-coef}--\eqref{d-constraint}
with \(\varepsilon\geq0\) can be written as
\begin{align}
\max_{b\in\mathbf B}\; r'b
\qquad\text{or}\qquad
\min_{b\in\mathbf B}\; r'b
\end{align}
subject to
\begin{align}\label{pop:lp:eps}
g_{0,j}\left( x_{j,1}+b_2x_{j,2}+\cdots+b_qx_{j,q}\right)
\geq |g_{0,j}|\varepsilon,
\quad j=1,\ldots,J .
\end{align}
This formulation is equivalent to using
\(d_j=\operatorname{sgn}(g_{0,j})\).  Assume that the feasible region
\eqref{pop:lp:eps} is not empty, so \(\varepsilon\) cannot be too large.

Put the problem into standard form by defining \(z_{j,k}=-x_{j,k}\) and
introducing nonnegative variables \(b_k^+\) and \(b_k^-\), \(k=2,\ldots,q\),
such that \(b_k=b_k^+-b_k^-\).  Let \(b^+\) and \(b^-\) denote the corresponding
vectors.  Define the sign-restriction components of the right-hand-side vector
by
\[
d_{\varepsilon,j}=-|g_{0,j}|\varepsilon-g_{0,j}z_{j,1}.
\]
Suppressing the box-constraint rows whose right-hand sides do not vary with
\(\varepsilon\), the standard-form inequalities corresponding to
\eqref{pop:lp:eps} are
\begin{align}
g_{0,j}\left[(b_2^+-b_2^-)z_{j,2}+\cdots
+(b_q^+-b_q^-)z_{j,q}\right]
\leq d_{\varepsilon,j},
\quad j=1,\ldots,J, \label{eq:pop-eps-standard-row}\\
b^+,b^-\geq0 \quad\text{componentwise}. \nonumber
\end{align}
The finite box restrictions in \(\mathbf B\) can be written in the same
standard-form system; they are omitted from the display only because their
right-hand sides do not depend on \(\varepsilon\).  Let \(d_\varepsilon\) denote
the full right-hand-side vector after adding these rows; its sign-restriction
components are the \(d_{\varepsilon,j}\)'s displayed above, and its remaining
components do not vary with \(\varepsilon\).

After adding slack variables to convert the displayed inequalities, together
with the box restrictions, to equalities, let \(S^U_{km}\),
\(k=1,\ldots,\mathcal K^U_m\), denote the optimal basis
matrices for the maximization version of the standard-form problem on an
interval \(\varepsilon^U_{m-1}\leq\varepsilon<\varepsilon^U_m\).  Let
\(S^L_{km}\), \(k=1,\ldots,\mathcal K^L_m\), denote the corresponding optimal
basis matrices for the minimization version on an interval
\(\varepsilon^L_{m-1}\leq\varepsilon<\varepsilon^L_m\).  The optimal basis
matrices can change as \(\varepsilon\) increases, and the largest relevant
interval ends when the feasible region becomes empty.  The integers
\(\mathcal K^U_m\geq1\) and \(\mathcal K^L_m\geq1\) allow for multiple
optimal basic solutions.  The first intervals begin at \(\varepsilon=0\), so
\(\varepsilon^U_0=\varepsilon^L_0=0\).

Let \(\rho\) be the objective coefficient vector for the standard-form
variables, with entries \(r_k\) on \(b_k^+\), entries \(-r_k\) on \(b_k^-\),
\(k=2,\ldots,q\), and zeros on slack and auxiliary variables.  Let
\(\rho^U_{B,km}\) and \(\rho^L_{B,km}\) denote the subvectors of \(\rho\)
corresponding to the basic variables in \(S^U_{km}\) and \(S^L_{km}\),
respectively.  All optimal basic solutions for a given endpoint yield the
same value of the objective function.  Therefore, for
\(\varepsilon^U_{m-1}\leq\varepsilon<\varepsilon^U_m\),
\[
c_{U,km}(\varepsilon)\equiv c_{U,m}(\varepsilon)
=
r_1+(\rho^U_{B,km})'(S^U_{km})^{-1}d_\varepsilon,
\quad k=1,\ldots,\mathcal K^U_m,
\]
and, for \(\varepsilon^L_{m-1}\leq\varepsilon<\varepsilon^L_m\),
\[
c_{L,km}(\varepsilon)\equiv c_{L,m}(\varepsilon)
=
r_1+(\rho^L_{B,km})'(S^L_{km})^{-1}d_\varepsilon,
\quad k=1,\ldots,\mathcal K^L_m .
\]

Equivalently, write \(d_\varepsilon=d_0-\varepsilon a\), where the
\(j\)'th sign-restriction component of \(a\) is \(|g_{0,j}|\) and the
components corresponding to rows whose right-hand sides do not vary with
\(\varepsilon\) are zero.  On any interval over which the relevant optimal
basis remains unchanged,
\[
c_{U,km}(\varepsilon)
=
r_1+(\rho^U_{B,km})'(S^U_{km})^{-1}d_0
-\varepsilon(\rho^U_{B,km})'(S^U_{km})^{-1}a
\]
and
\[
c_{L,km}(\varepsilon)
=
r_1+(\rho^L_{B,km})'(S^L_{km})^{-1}d_0
-\varepsilon(\rho^L_{B,km})'(S^L_{km})^{-1}a .
\]
Thus the upper and lower population endpoints are affine functions of
\(\varepsilon\) between changes in the optimal basis.  Globally, if
\(0\leq\varepsilon_1\leq\varepsilon_2\) and both feasible regions are
nonempty, then the feasible sets are nested and
\[
c_L(\varepsilon_1)\leq c_L(\varepsilon_2)
\leq c_U(\varepsilon_2)\leq c_U(\varepsilon_1).
\]
\endgroup

\subsection[Estimation and Inference with Positive epsilon]{Estimation and Inference with Positive~\texorpdfstring{$\varepsilon$}{epsilon}}
\label{subsec:est:inf:eps}

\begingroup
For \(\varepsilon\geq0\), define the sample analogs
\(\hat c_U(\varepsilon)\) and \(\hat c_L(\varepsilon)\) of the population
quantities \(c_U(\varepsilon)\) and \(c_L(\varepsilon)\) as the optimal values
of
\begin{align}
  \underset{g\in\mathcal{G}_\alpha,\; b\in\mathbf{B}}{\operatorname{maximize}}
  \Bigl(
  \underset{g\in\mathcal{G}_\alpha,\; b\in\mathbf{B}}{\operatorname{minimize}}
  \Bigr):\; r'b
  \label{alt-1-eps}
\end{align}
subject to
\begin{align}
  g_j\left( x_{j,1} + b_2 x_{j,2} + \cdots + b_q x_{j,q} \right)
  \geq\varepsilon\,|g_j|,
  \quad j=1,\ldots,J. \label{bl-constraints-eps}
\end{align}
When \(\varepsilon=0\), \eqref{alt-1-eps}--\eqref{bl-constraints-eps} reduce to
the baseline formulation.  When \(\varepsilon>0\), the constraints require a
positive margin for coordinates whose sign is restricted.  In particular, if
$g_j>0$, then
\[
  \left( x_{j,1} + b_2 x_{j,2} + \cdots + b_q x_{j,q} \right)\geq\varepsilon,
\]
and if $g_j<0$, then
\[
  \left( x_{j,1} + b_2 x_{j,2} + \cdots + b_q x_{j,q} \right)\leq-\varepsilon.
\]
Thus, positive \(\varepsilon\) strengthens the imposed sign restrictions by requiring
the score to be bounded away from zero whenever the corresponding sign is
nonzero.

On the event \(g_0\in\mathcal G_\alpha\), every feasible value of \(b\) for
the population optimization problem is feasible for the corresponding sample
optimization problem by setting \(g=g_0\).  Therefore,
\[
\hat c_L(\varepsilon)\leq c_L(\varepsilon)\leq c_U(\varepsilon)\leq \hat c_U(\varepsilon),
\]
so the same endpoint-coverage argument as in Theorem~\ref{thm1} gives
\[
\mathbb P\{\hat c_L(\varepsilon)\leq c_L(\varepsilon)\leq c_U(\varepsilon)\leq\hat c_U(\varepsilon)\}
\geq
\mathbb P\{g_0\in\mathcal G_\alpha\}
\geq 1-\alpha.
\]
\endgroup

\subsection[Computational Algorithms with Positive epsilon]{Computational Algorithms with Positive~\texorpdfstring{$\varepsilon$}{epsilon}}
\label{subsec:computation:eps}

We maintain the rectangular confidence region in
\eqref{G:box:generic}.
The following theorem extends the computational reduction of
Theorem~\ref{thm2} to positive~$\eps$.

\begin{theorem}\label{thm2-eps}
Solutions to \eqref{alt-1-eps}--\eqref{bl-constraints-eps} with
\(\mathcal{G}_\alpha\) in the form of \eqref{G:box:generic} can be
obtained by solving
\begin{align}
  \underset{b\in\mathbf{B}}{\operatorname{maximize}}
  \Bigl(
  \underset{b\in\mathbf{B}}{\operatorname{minimize}}
  \Bigr):\; r'b \label{alt1-box-LP-eps}
\end{align}
subject to
\begin{subequations}\label{g-constraint-est-eps}
\begin{align}
  \left( x_{j,1} + b_2 x_{j,2} + \cdots + b_q x_{j,q} \right)
  \geq\eps
  &\quad\text{if }
  \hat{g}(x_j)-\hat{s}(x_j,\alpha)>0, \label{g-constraint-est-1-eps}\\
  \left( x_{j,1} + b_2 x_{j,2} + \cdots + b_q x_{j,q} \right)
  \leq-\eps
  &\quad\text{if }
  \hat{g}(x_j)+\hat{s}(x_j,\alpha)<0. \label{g-constraint-est-2-eps}
\end{align}
\end{subequations}
\end{theorem}

\begin{proof}[Proof of Theorem~\ref{thm2-eps}]
The constraint~\eqref{bl-constraints-eps} is
\[
g_j\left( x_{j,1} + b_2 x_{j,2} + \cdots + b_q x_{j,q} \right)\geq\eps|g_j|
\]
for each~\(j\). Since
\(\mathcal{G}_\alpha\) is a Cartesian product of intervals, each \(g_j\)
varies independently over
\[
\bigl[\hat{g}(x_j)-\hat{s}(x_j,\alpha),\;
      \hat{g}(x_j)+\hat{s}(x_j,\alpha)\bigr].
\]
We show that, for each~\(j\), the bilinear constraint is equivalent to the
corresponding linear restriction in~\eqref{g-constraint-est-eps}.

\emph{Case 1:}
\(\hat g(x_j)-\hat s(x_j,\alpha)>0\). Every feasible \(g_j\) is strictly
positive. Dividing the constraint by \(g_j>0\) gives
\[
\left( x_{j,1} + b_2 x_{j,2} + \cdots + b_q x_{j,q} \right)\geq\eps,
\]
which is~\eqref{g-constraint-est-1-eps}.

\emph{Case 2:}
\(\hat g(x_j)+\hat s(x_j,\alpha)<0\). Every feasible \(g_j\) is strictly
negative, so \(|g_j|=-g_j\). The constraint becomes
\[
g_j\left( x_{j,1} + b_2 x_{j,2} + \cdots + b_q x_{j,q} \right)\geq-\eps g_j.
\]
Dividing by \(g_j<0\) reverses the inequality and yields
\[
\left( x_{j,1} + b_2 x_{j,2} + \cdots + b_q x_{j,q} \right)\leq-\eps,
\]
which is~\eqref{g-constraint-est-2-eps}.

\emph{Case 3:}
\(\hat g(x_j)-\hat s(x_j,\alpha)\leq0\leq
\hat g(x_j)+\hat s(x_j,\alpha)\). Zero is a feasible value of \(g_j\), and
choosing \(g_j=0\) satisfies the constraint regardless of~\(b\). No
restriction on~\(b\) is needed.

Since this equivalence holds for every~$j$, the feasible sets of
\eqref{alt-1-eps}--\eqref{bl-constraints-eps} and
\eqref{alt1-box-LP-eps}--\eqref{g-constraint-est-eps} coincide, and the
two programs have the same optimal values.
\end{proof}

\subsection[Sensitivity to epsilon in the empirical application of Section 9.1]{\texorpdfstring{Sensitivity to~$\varepsilon$ in the Empirical Application of Section~\ref{sec:KLS}}{Sensitivity to epsilon in the Empirical Application of Section 9.1}}
\label{subsec:choice:eps}

The parameter $\eps$ imposes a minimum separation condition on
the normalized index: when a constraint is active, the LP requires
\[
\left|x_{j,1} + b_2 x_{j,2} + \cdots + b_q x_{j,q}\right|\geq\eps.
\]
Thus $\eps$ should be chosen relative to the scale of the normalized
index, not to the numerical scale of the outcome. In the incentivized
resume rating application of Section~\ref{sec:KLS}, the coefficient on GPA is normalized to one
and GPA is grouped in increments of 0.2. We therefore use the
pre-specified grid
\[
  \eps\in\bigl\{0,\;0.02,\;0.04\bigr\}.
\]
The two positive values correspond to 10 and 20 percent of one GPA bin in
the normalized index. The grid is intended only as a local sensitivity
check around the baseline outer-set analysis with \(\eps=0\), not as an
estimate of the unknown population margin.

\subsection{Empirical Results}
\label{subsec:empirical:eps}

Tables~\ref{tab-kls1-multi} and~\ref{tab-kls1-xv-multi} report
coefficient bounds for the incentivized resume rating application of
Section~\ref{sec:KLS} across the three values of~$\eps$.
Table~\ref{tab-kls1-multi} is the positive-\(\eps\) analog of
Table~\ref{tab-kls1}: it recomputes the maximum score coefficient bounds for
the discrete-GPA specification in the baseline analysis in
Section~\ref{sec:KLS}.
Table~\ref{tab-kls1-xv-multi} is the positive-\(\eps\) analog of
Table~\ref{tab-kls1-xv}: it recomputes the fixed- and random-design bounds
when GPA is treated either as a discrete covariate or as an interval
covariate.
The logit estimates are unchanged across panels because logit does not
use LP-based constraint selection.  The estimated maximum score
bounds narrow monotonically as \(\eps\) increases, consistent with
Subsection~\ref{subsec:est:inf:eps}: the estimated upper bound
\(\hat c_U(\eps)\) decreases and the estimated lower bound \(\hat c_L(\eps)\)
increases as the margin requirement tightens.

\input{results/table_kls_multi_eps.tex}

\input{results/table_kls_interval_multi_eps.tex}

Table~\ref{tab:kls-eps-classification-summary} summarizes the maximum score
classifications of the twelve resume groups under discrete and interval GPA.
Although positive \(\eps\) narrows the endpoint intervals, no displayed
classification changes on the reported grid.

\input{results/table_kls_eps_classification_counts.tex}

Table~\ref{tab:kls-eps-target-endpoint} gives the corresponding endpoint
diagnostic for one boundary cell. For each endpoint, the
supporting optimal basis remains unchanged over the reported grid, so both
endpoint paths are affine in \(\eps\).

\input{results/table_kls_eps_target_endpoint.tex}


\section{Refined Misclassification Bounds and Expected Excess Regret}
\label{subsec:refined-misclassification}

Section~\ref{sec:regret} gives a global misclassification bound based on the
event that every confidence interval determines the correct population sign.
That event is sufficient for the population and confidence-region programs to
agree for every value of \(X_*\), but it can be stronger than necessary for a
given covariate value. This appendix develops the corresponding refinement. It
first describes how dropped constraints change the endpoint programs,
then bounds the endpoint-crossing events that generate misclassification, and
finally applies the same event decomposition to expected excess regret.
Throughout this appendix, let
\(\Ealpha\equiv\{\Ehat\neq\emptyset\}\) denote the wrong-sign event.

\subsection{Dropped Constraints and Endpoint Representation}\label{subsec:dropped-endpoint-representation}

For any set \(D\subseteq\{1,\ldots,J\}\) of dropped sign constraints, define
\begin{align*}
c_L^D(r)
&=
\min_{b\in\mathbf B}
\left\{
r'b:
d_j b'x_j\geq0\ \text{for all } j\in D^c
\right\}, \\
c_U^D(r)
&=
\max_{b\in\mathbf B}
\left\{
r'b:
d_j b'x_j\geq0\ \text{for all } j\in D^c
\right\}.
\end{align*}
If \(D_1\subseteq D_2\), then
\[
c_L^{D_2}(r)\leq c_L^{D_1}(r)
\leq c_U^{D_1}(r)\leq c_U^{D_2}(r).
\]

\begin{lemma}[Dropped-Constraint Endpoint Representation]
\label{lem:dropped-endpoint-representation}
Suppose Assumptions~\ref{a:design} and~\ref{a:margin-condition} hold, and let
\(\mathcal G_\alpha\) be the rectangular confidence region in
\eqref{G:box:generic}. On the event \(\Ealpha^c\), the estimated endpoints
satisfy, for every \(r\),
\[
\hat c_L(r)=c_L^{\Jhat}(r),
\qquad
\hat c_U(r)=c_U^{\Jhat}(r).
\]
\end{lemma}

\begin{proof}
Fix \(b\in\mathbf B\). If \(j\in\Jhat\), then
\(|\hat g(x_j)|\leq\hat s(x_j,\alpha)\), so \(g_j=0\) is feasible in the
confidence interval. The constraint \(g_jb'x_j\geq0\) is then automatically
satisfied, regardless of \(b'x_j\), and coordinate \(j\) imposes no
restriction on \(b\).

If \(j\notin\Jhat\), the confidence interval for \(g_j\) does not contain
zero. On \(\Ealpha^c\), it also does not lie on the wrong side of zero. Hence
every feasible value of \(g_j\) has sign \(d_j\), and there exists a feasible
\(g_j\) satisfying \(g_jb'x_j\geq0\) if and only if
\(d_jb'x_j\geq0\).

Because the confidence region is rectangular, these coordinatewise choices
can be combined into a single feasible vector \(g\in\mathcal G_\alpha\).
Therefore, the confidence-region problem imposes the population sign
restrictions for \(j\notin\Jhat\) and no sign restrictions for
\(j\in\Jhat\), giving the stated endpoint representation.
\end{proof}

\subsection{Endpoint-Crossing Misclassification Events}
\label{subsec:refined-misclassification-events}

Recall the population and feasible classification rules and the disagreement
indicators \(\mathcal M_A\) and \(\mathcal M_R\) defined in
Section~\ref{subsec:misclassification-events}. For random classification, the
population and feasible rules use the same Bernoulli\((p^*)\) draw \(\xi\).
Using independent draws would create disagreement with probability
\(2p^*(1-p^*)\) even when the two endpoint intervals agree and strictly
contain zero.

On \(\Ealpha^c\), Lemma~\ref{lem:dropped-endpoint-representation} gives the
feasible endpoints \(c_L^{\Jhat}(X_*)\) and \(c_U^{\Jhat}(X_*)\). If the
population interval strictly contains zero, dropping constraints preserves
strict ambiguity:
\[
c_L^{\Jhat}(X_*)\leq c_L(X_*)<0<c_U(X_*)\leq c_U^{\Jhat}(X_*).
\]
Thus both rules abstain or both use the common randomization device.
A discrepancy can occur only through one of the endpoint-crossing events
\[
\mathcal T_1
=
\{c_L(X_*)\geq0,\ c_U(X_*)>0,\
c_L^{\Jhat}(X_*)<0\},
\]
\[
\mathcal T_0
=
\{c_L(X_*)<0,\ c_U(X_*)\leq0,\
c_U^{\Jhat}(X_*)>0\}.
\]

For a fixed \(r\), define the minimal dropped sets that can generate
these endpoint crossings by
\begin{align}\label{eq:D_L}
\mathcal D_L(r)
=
\min_{\subseteq}
\Bigl\{
D\subseteq\{1,\ldots,J\}:
c_L(r)\geq0,\ c_U(r)>0,\ c_L^D(r)<0
\Bigr\},
\end{align}
and
\begin{align}\label{eq:D_U}
\mathcal D_U(r)
=
\min_{\subseteq}
\Bigl\{
D\subseteq\{1,\ldots,J\}:
c_L(r)<0,\ c_U(r)\leq0,\ c_U^D(r)>0
\Bigr\}.
\end{align}
Here \(\min_{\subseteq}\) selects the sets in the displayed collection that
are minimal under set inclusion. If the collection has no elements, the
corresponding collection is empty. By endpoint monotonicity,
\[
\mathcal T_1
=
\{c_L(X_*)\geq0,\ c_U(X_*)>0,\
\exists D\in\mathcal D_L(X_*):D\subseteq\Jhat\},
\]
\[
\mathcal T_0
=
\{c_L(X_*)<0,\ c_U(X_*)\leq0,\
\exists D\in\mathcal D_U(X_*):D\subseteq\Jhat\}.
\]

On \(\Ealpha^c\), the event \(\mathcal T_1\) means that the population rule
classifies as \(1\), but dropped constraints make the feasible lower endpoint
negative. Similarly, \(\mathcal T_0\) means that the population rule
classifies as \(0\), but dropped constraints make the feasible upper endpoint
positive.

\begin{corollary}[Refined Misclassification Probabilities]
\label{cor:misclassification-probabilities}
Under Assumptions~\ref{a:design} and~\ref{a:margin-condition} and the
rectangular confidence-region setup in \eqref{G:box:generic}, suppose that
\[
\mathbb P\{c_L(X_*)=c_U(X_*)=0\}=0.
\]
Then, for fixed or random \(X_*\),
\[
\mathbb P\{\mathcal M_A(X_*)=1\}
\leq
\mathbb P(\Ealpha)
+\mathbb P(\Ealpha^c\cap\mathcal T_1)
+\mathbb P(\Ealpha^c\cap\mathcal T_0),
\]
and, under the common randomization device,
\[
\mathbb P\{\mathcal M_R(X_*;\xi)=1\}
\leq
\mathbb P(\Ealpha)
+(1-p^*)\mathbb P(\Ealpha^c\cap\mathcal T_1)
+p^*\mathbb P(\Ealpha^c\cap\mathcal T_0).
\]
\end{corollary}

\begin{proof}
On \(\Ealpha^c\), Lemma~\ref{lem:dropped-endpoint-representation} implies
that a discrepancy between the feasible and population rules can occur only
through \(\mathcal T_1\) or \(\mathcal T_0\). If the population interval
strictly contains zero, both rules make the same abstention decision or use
the same Bernoulli draw. The exact-boundary case has probability zero by the
maintained condition. On \(\Ealpha\), the discrepancy indicator is bounded by
one.

For abstention, this event decomposition gives the first bound. Under common
randomization, a discrepancy on \(\mathcal T_1\) occurs only when the draw
selects zero, with probability \(1-p^*\), and a discrepancy on
\(\mathcal T_0\) occurs only when the draw selects one, with probability
\(p^*\). This gives the second bound.
\end{proof}

\subsection{Refined Fixed-Design Probability Bounds}
\label{subsec:computable-local-bounds}

For the fixed-design finite-sample confidence region in
\eqref{G:fixed:box}, recall
\[
h_{j,\alpha,n}
=
\left(
|g_{0,j}|-\frac{n_j}{n}t_{j,\alpha}
\right)_+.
\]

\begin{lemma}[Fixed-Design Wrong-Sign Bound]
\label{lem:fixed-sign-agreement-bound}
Suppose Assumptions~\ref{a:design}(a) and~\ref{a:margin-condition} hold.
Then
\begin{align}\label{eq:wrong-sign-bound}
\mathbb P(\Ealpha)\leq
\sum_{j=1}^J
\exp\left\{
-\frac{2n^2}{n_j}
\left(
|g_{0,j}|+\frac{n_j}{n}t_{j,\alpha}
\right)^2
\right\}.
\end{align}
\end{lemma}

\begin{proof}
A wrong-sign constraint at support point \(x_j\) can be imposed only if its
confidence interval lies entirely on the wrong side of zero. If \(g_{0,j}>0\),
this requires
\[
\hat g(x_j)+\frac{n_j}{n}t_{j,\alpha}<0,
\]
and if \(g_{0,j}<0\), it requires
\[
\hat g(x_j)-\frac{n_j}{n}t_{j,\alpha}>0.
\]
Both cases imply a one-sided deviation of
\(d_j\{\hat g(x_j)-g_{0,j}\}\) by more than
\(|g_{0,j}|+(n_j/n)t_{j,\alpha}\). Under the fixed design,
\[
d_j\{\hat g(x_j)-g_{0,j}\}
=
\frac{n_j}{n}
\left[
\frac{1}{n_j}
\sum_{i:X_i=x_j}
d_j\{Y_i-\mathbb P(Y=1\mid X=x_j)\}
\right].
\]
The summands are independent, mean zero, and lie in an interval of length
one. The one-sided Hoeffding inequality and a union bound over
\(j=1,\ldots,J\) give \eqref{eq:wrong-sign-bound}.
\end{proof}

For \(r\in\mathbb R^q\), define
\[
J_L(r)=\bigcup_{D\in\mathcal D_L(r)}D,
\qquad
J_U(r)=\bigcup_{D\in\mathcal D_U(r)}D.
\]

\begin{lemma}[Refined Fixed-Design Endpoint-Crossing Bounds]
\label{lem:local:prob:bounds}
Suppose Assumptions~\ref{a:design}(a) and~\ref{a:margin-condition} hold.
Consider the fixed-design finite-sample confidence region in
\eqref{G:fixed:box}, and treat \(X_*\) as fixed. Then
\begin{equation}\label{eq:refined-lower-bound}
\begin{aligned}
\mathbb P(\Ealpha^c\cap\mathcal T_1)
\leq
\min\Bigg\{&
\sum_{j\in J_L(X_*)}
\exp\left(-\frac{2n^2h_{j,\alpha,n}^2}{n_j}\right),
\\
&
\sum_{D\in\mathcal D_L(X_*)}
\exp\left(
-2n^2\sum_{j\in D}\frac{h_{j,\alpha,n}^2}{n_j}
\right)
\Bigg\},
\end{aligned}
\end{equation}
and
\begin{equation}\label{eq:refined-upper-bound}
\begin{aligned}
\mathbb P(\Ealpha^c\cap\mathcal T_0)
\leq
\min\Bigg\{&
\sum_{j\in J_U(X_*)}
\exp\left(-\frac{2n^2h_{j,\alpha,n}^2}{n_j}\right),
\\
&
\sum_{D\in\mathcal D_U(X_*)}
\exp\left(
-2n^2\sum_{j\in D}\frac{h_{j,\alpha,n}^2}{n_j}
\right)
\Bigg\}.
\end{aligned}
\end{equation}
\end{lemma}

\begin{proof}
We prove the first inequality; the second follows by replacing
\(\mathcal T_1,\mathcal D_L,J_L\) with
\(\mathcal T_0,\mathcal D_U,J_U\).

Treat \(X_*\) as fixed and write \(r=X_*\). On \(\Ealpha^c\), if
\(\mathcal T_1\) occurs, there exists
\(D\in\mathcal D_L(r)\) such that \(D\subseteq\Jhat\). For each \(j\),
the one-sided Hoeffding argument used in the proof of
Theorem~\ref{thm:global-misclassification-bound} gives
\[
\mathbb P\{j\in\Jhat\}
\leq
\exp\left(-\frac{2n^2h_{j,\alpha,n}^2}{n_j}\right).
\]
A union bound over coordinates in \(J_L(r)\) yields the first expression on
the right-hand side of \eqref{eq:refined-lower-bound}.

Alternatively, take a union bound over \(D\in\mathcal D_L(r)\). Under the
fixed design, the events \(\{j\in\Jhat\}\) are mutually independent across
support points. Therefore,
\[
\mathbb P(D\subseteq\Jhat)
=
\prod_{j\in D}\mathbb P(j\in\Jhat)
\leq
\exp\left(
-2n^2\sum_{j\in D}\frac{h_{j,\alpha,n}^2}{n_j}
\right).
\]
Summing over the minimal dropped sets and taking the minimum of the two valid
bounds proves \eqref{eq:refined-lower-bound}.
\end{proof}

For fixed \(X_*\), Corollary~\ref{cor:misclassification-probabilities}
together with Lemmas~\ref{lem:fixed-sign-agreement-bound}
and~\ref{lem:local:prob:bounds} gives the refined abstention and
random-classification bounds. These results replace the global
dropped-constraint sum with probabilities involving only constraints that
can move the relevant endpoint across zero.

\subsection{Expected Excess Regret from Confidence-Region Classification}\label{subsec:expected-regret}

The same wrong-sign and endpoint-crossing events can be weighted by their
decision-theoretic regret costs. This gives a bound on the expected increase
in regret from using the feasible rule based on the confidence region rather
than the population rule.

Using the regret notation from Section~\ref{sec:decision:theory}, define the
state-specific abstention regrets
\[
\ReEmpp=C_\emptyset-C_b^+,
\qquad
\ReEmpm=C_\emptyset-C_b^-,
\qquad
\ReEmp=\max\{\ReEmpp,\ReEmpm\}.
\]
Assumption~\ref{ass:low-regret-deferral} implies
\[
\ReEmp\leq\min\{\ReFP,\ReFN\}.
\]
Let \(\mathcal R_A(a,v)\) denote the regret of action
\(a\in\{1,0,\emptyset\}\) when the true index \(\beta'x\) takes the nonzero
value \(v\), where
\[
\begin{array}{llll}
\text{if $v>0$:} &
\mathcal R_A(1,v)=0, &
\mathcal R_A(0,v)=\ReFN, &
\mathcal R_A(\emptyset,v)=\ReEmpp, \\
\text{if $v<0$:} &
\mathcal R_A(1,v)=\ReFP, &
\mathcal R_A(0,v)=0, &
\mathcal R_A(\emptyset,v)=\ReEmpm.
\end{array}
\]
Let \(\mathcal R_R\) be the restriction of \(\mathcal R_A\) to the two binary
actions. For \(z\in\mathbb R\), let \([z]_+:=\max\{z,0\}\). Although
\(M_{\mathrm{OR}}\) is the population minimax-regret rule, it does not know
the sign of \(\beta'x\) and need not minimize regret state by state.
Consequently, the feasible rule can occasionally have lower state-specific
regret than the population rule, so their raw regret difference can be
negative. We measure sampling-induced increases in regret by the positive
part of this difference.
Define
\[
\Delta_A(x)
=
\left[
\mathcal R_A\{M_{\mathrm{MS}}^A(x),\beta'x\}
-
\mathcal R_A\{M_{\mathrm{OR}}^A(x),\beta'x\}
\right]_+
\]
and
\[
\Delta_R(x;\xi)
=
\left[
\mathcal R_R\{M_{\mathrm{MS}}^R(x;\xi),\beta'x\}
-
\mathcal R_R\{M_{\mathrm{OR}}^R(x;\xi),\beta'x\}
\right]_+.
\]

\begin{theorem}[Expected excess regret from confidence-region classification]
\label{thm:expected-excess-regret}
Under Assumptions~\ref{a:design} and~\ref{a:margin-condition}
and the rectangular confidence-region setup in \eqref{G:box:generic}, suppose that
\begin{align}\label{eq:no-exact-boundary-target}
\mathbb P\{\beta'X_*=0\}=0.
\end{align}
Then, for fixed or random \(X_*\), the following bounds hold.
\begin{enumerate}
\item \emph{Abstention.} If Assumption~\ref{ass:low-regret-deferral} holds, then
\begin{align*}
\mathbb E [\Delta_A(X_*)]
&\leq
\max\{\ReFP,\ReFN\}\,\mathbb P(\Ealpha)\\
&\quad+\ReEmpp\,\mathbb P(\Ealpha^c\cap\mathcal T_1)
+\ReEmpm\,\mathbb P(\Ealpha^c\cap\mathcal T_0).
\end{align*}
\item \emph{Random classification.} Under the cost conditions in
Theorem~\ref{thm:selective:rc} and using the common randomization device,
\begin{align*}
\mathbb E\Delta_R(X_*;\xi)
&\leq
\max\{\ReFP,\ReFN\}\,\mathbb P(\Ealpha)\\
&\quad+(1-p^*)\ReFN\,\mathbb P(\Ealpha^c\cap\mathcal T_1)
+p^*\ReFP\,\mathbb P(\Ealpha^c\cap\mathcal T_0).
\end{align*}
\end{enumerate}
\end{theorem}

\begin{proof}
It is enough to analyze \(\Ealpha^c\) and then bound the remaining sign-error
event separately. On \(\Ealpha^c\),
Lemma~\ref{lem:dropped-endpoint-representation} implies that the feasible
endpoints are \(c_L^{\Jhat}\) and \(c_U^{\Jhat}\).

If the population rule assigns \(1\), then \(c_L(X_*)\geq0\) and
\(c_U(X_*)>0\). Since the true parameter belongs to \(\mathcal B(0)\),
the maintained boundary condition implies \(\beta'X_*>0\) almost surely.
Dropping constraints can change the action only if
\(c_L^{\Jhat}(X_*)<0\), which is the endpoint-crossing event
\(\mathcal T_1\). On \(\Ealpha^c\cap\mathcal T_1\), the feasible rule moves
from the population action \(1\) to abstention, with excess regret
\(\ReEmpp\), or to randomization, with excess regret
\((1-p^*)\ReFN\) after averaging over \(\xi\).

If the population rule assigns \(0\), then \(c_U(X_*)\leq0\). Since the true
parameter belongs to \(\mathcal B(0)\), the maintained boundary condition
implies \(\beta'X_*<0\) almost surely and hence \(c_L(X_*)<0\).
Dropping constraints can change the action only if
\(c_U^{\Jhat}(X_*)>0\), which is the endpoint-crossing event
\(\mathcal T_0\). On \(\Ealpha^c\cap\mathcal T_0\), the feasible rule moves
from the population action \(0\) to abstention, with excess regret
\(\ReEmpm\), or to randomization, with excess regret \(p^*\ReFP\) after
averaging over \(\xi\).

If the population interval strictly contains zero, dropping constraints
preserves strict ambiguity:
\[
c_L^{\Jhat}(X_*)\leq c_L(X_*)<0<c_U(X_*)\leq c_U^{\Jhat}(X_*).
\]
Hence both rules choose abstention or use the common randomization device, and
there is no excess regret on \(\Ealpha^c\).

It remains to cover \(\Ealpha\). Under random classification, every realized
binary action has regret bounded by \(\max\{\ReFP,\ReFN\}\), and the
population rule has nonnegative regret, so any positive excess regret is
bounded by the same quantity. Under abstention,
Assumption~\ref{ass:low-regret-deferral} implies
\(\ReEmp\leq\min\{\ReFP,\ReFN\}\), so abstention also has regret bounded by
\(\max\{\ReFP,\ReFN\}\). Hence the same bound applies to all transitions
involving abstention. Combining these cases gives the two displayed bounds.
\end{proof}

\begin{remark}[Condition \eqref{eq:no-exact-boundary-target}]
The condition excludes target covariate values for which the true index lies
exactly on the classification boundary. For fixed \(X_*\), it reduces to
\(\beta'X_*\neq0\); for random \(X_*\), it requires the distribution of
\(\beta'X_*\) to place no mass at zero.
\end{remark}


\section{Clustered Dependence and Survey Sampling Weights}\label{sec:cluster}

In applications, observations are often dependent within clusters, and
survey sampling weights are often non-uniform.
The second empirical example in Section~\ref{sec:CHV} has both features.
This appendix constructs confidence regions for the weighted sign vector
\(g_0=(g_0(x_1),\ldots,g_0(x_J))\). Once such a region satisfies
\(\mathbb P\{g_0\in\mathcal G_\alpha\}\geq 1-\alpha\), the estimation and
inference arguments in Section~\ref{sec:estimation:inference}, the
misclassification arguments in Section~\ref{sec:regret},
and the refined misclassification and regret extensions in Supplemental
Appendix~\ref{subsec:refined-misclassification} apply without further changes.

In this appendix, \(\tau\in(0,1)\) denotes the fixed threshold
used throughout the paper; the weighted sign vector \(g_0\) records whether
weighted cell means lie above or below this threshold. All probability and
distributional statements are conditional on the survey weights.

We begin by defining the weighted target. In the fixed design case, let
\(Y_{\ell,j,c}\) and \(w_{\ell,j,c}\), respectively, denote the binary
outcome and survey sampling weight for individual \(\ell\) with covariate
\(x_j\) in cluster \(c\). Define
\begin{align*}
N_j
&\equiv
\sum_{c=1}^{m_j} \sum_{\ell=1}^{n_{j,c}} w_{\ell,j,c}, \qquad
N\equiv \sum_{j=1}^J N_j, \\
\Upsilon_{j,c}
&\equiv
\sum_{\ell=1}^{n_{j,c}} w_{\ell,j,c} (Y_{\ell,j,c}-\tau), \qquad
\overline{\Upsilon}_{j}
\equiv N_j^{-1} \sum_{c=1}^{m_j} \Upsilon_{j,c},
\end{align*}
where \(n_{j,c}\) is the number of observations with \(X=x_j\) in cluster
\(c\), and \(m_j\) is the number of clusters contributing observations with
\(X=x_j\). We then set
\begin{align*}
p(x_j) &\equiv  \frac{N_j}{N},  \\
\hat{g}(x_j)
&\equiv
\frac{\sum_{c=1}^{m_j} \Upsilon_{j,c} }{N}
=
\overline{\Upsilon}_{j}p(x_j), \\
g_0(x_j) &
\equiv \mathbb{E}(\overline{\Upsilon}_{j})p(x_j)
=N^{-1}\sum_{c=1}^{m_j}\mathbb{E}(\Upsilon_{j,c}).
\end{align*}
Equivalently, if
\[
\mu_j^F
\equiv
N_j^{-1}\sum_{c=1}^{m_j}\sum_{\ell=1}^{n_{j,c}}
w_{\ell,j,c}\mathbb{E}(Y_{\ell,j,c}),
\]
then \(g_0(x_j)=p(x_j)(\mu_j^F-\tau)\). Thus \(g_0\) records the signs of
weighted cell means relative to the threshold \(\tau\).

In the random design case, let \((Y_{i,c},X_{i,c},w_{i,c})\) denote the
binary outcome, covariate, and sampling weight for individual \(i\) in
cluster \(c\), where \(X_{i,c}\) takes values in
\(\{x_1,\ldots,x_J\}\). Let \(n_c\) denote the number of observations in cluster
\(c\), \(m\) the number of clusters, and
\(N\equiv \sum_{c=1}^{m}\sum_{i=1}^{n_c} w_{i,c}\). Define
\begin{align*}
\Upsilon_{j,c}
&\equiv
\sum_{i=1}^{n_c} w_{i,c}(Y_{i,c}-\tau)I(X_{i,c}=x_j), \\
\hat{g}(x_j)
&\equiv
N^{-1}\sum_{c=1}^{m}\Upsilon_{j,c}, \\
g_0(x_j)
&\equiv
N^{-1}\sum_{c=1}^{m}\mathbb{E}(\Upsilon_{j,c}).
\end{align*}
In both designs, \(\hat g(x_j)\) is an unbiased estimator of the weighted
target \(g_0(x_j)\), conditional on the weights.

To develop inference methods, we impose the following condition.

\begin{assumption}\label{a:cluster}
The number of covariate values \(J\) is fixed. Survey sampling
weights are fixed and nonnegative. In the fixed design case, \(N_j>0\) for
every \(j=1,\ldots,J\). In the random design case, \(N>0\). In the fixed design case, for each \(j\),
the cluster-level vectors
\(\{(Y_{\ell,j,c})_{\ell=1}^{n_{j,c}}:c=1,\ldots,m_j\}\) are independent
across \(c\). In the random design case, the cluster-level vectors
\(\{(Y_{i,c},X_{i,c})_{i=1}^{n_c}:c=1,\ldots,m\}\) are independent across
\(c\).
\end{assumption}

The union bounds below do not require independence across the coordinates
\(j=1,\ldots,J\). They use the marginal concentration or normal
approximation for each coordinate and then apply Bonferroni over
\(j=1,\ldots,J\).

\subsection{Finite Sample Inference}

\begin{proposition}\label{prop:cluster:finite}
Let Assumption~\ref{a:cluster} hold.
\begin{enumerate}
\item In the fixed design case, define
\[
\gamma_{j,c}\equiv N_j^{-1}\sum_{\ell=1}^{n_{j,c}}w_{\ell,j,c},
\qquad
t_{j,\alpha}^*
\equiv
\left(\frac{\sum_{c=1}^{m_j}\gamma_{j,c}^2}{2}
\log\frac{2J}{\alpha}\right)^{1/2}.
\]
Set
\begin{align}\label{G:fixed:box:cluster}
\mathcal{G}_\alpha =
\left\{
(g_1,\ldots,g_J): \hat{g}(x_j)-p(x_j)t_{j,\alpha}^*
\leq g_j \leq
\hat{g}(x_j)+p(x_j)t_{j,\alpha}^*
\; \forall j
\right\}.
\end{align}
Then \(\mathbb{P}(g_0\in\mathcal{G}_\alpha)\geq 1-\alpha\).

\item In the random design case, define
\[
\gamma_c\equiv N^{-1}\sum_{i=1}^{n_c}w_{i,c},
\qquad
t_\alpha^*
\equiv
\left(\frac{\sum_{c=1}^{m}\gamma_c^2}{2}
\log\frac{2J}{\alpha}\right)^{1/2}.
\]
Set
\begin{align}\label{G:random:box:cluster}
\mathcal{G}_\alpha =
\left\{
(g_1,\ldots,g_J): \hat{g}(x_j)-t_\alpha^*
\leq g_j \leq
\hat{g}(x_j)+t_\alpha^*
\; \forall j
\right\}.
\end{align}
Then \(\mathbb{P}(g_0\in\mathcal{G}_\alpha)\geq 1-\alpha\).
\end{enumerate}
\end{proposition}

\begin{proof}[Proof of Proposition~\ref{prop:cluster:finite}]
For the fixed design, the random variable \(N_j^{-1}\Upsilon_{j,c}\) lies
between \(-\tau\gamma_{j,c}\) and \((1-\tau)\gamma_{j,c}\). Hence its range
has length \(\gamma_{j,c}\). Hoeffding's inequality gives, for any
\(t_j>0\),
\[
\mathbb{P}\left[
\left|\overline{\Upsilon}_{j}-\mathbb{E}(\overline{\Upsilon}_{j})\right|
\geq t_j
\right]
\leq
2\exp\left(-2\frac{t_j^2}{\sum_{c=1}^{m_j}\gamma_{j,c}^2}\right).
\]
Since
\(\hat g(x_j)-g_0(x_j)=p(x_j)\{
\overline{\Upsilon}_{j}-\mathbb{E}(\overline{\Upsilon}_{j})\}\), the
fixed-design bound follows by setting \(t_j=t_{j,\alpha}^*\) and applying
the Bonferroni inequality over \(j\).

For the random design, \(N^{-1}\Upsilon_{j,c}\) lies between
\(-\tau\gamma_c\) and \((1-\tau)\gamma_c\), so the same argument gives
\[
\mathbb{P}\left[
\left|\hat g(x_j)-g_0(x_j)\right|\geq t
\right]
\leq
2\exp\left(-2\frac{t^2}{\sum_{c=1}^{m}\gamma_c^2}\right).
\]
Set \(t=t_\alpha^*\) and apply Bonferroni over \(j\).
\end{proof}

The quantities \([\sum_{c=1}^{m_j}\gamma_{j,c}^2]^{-1}\) and
\([\sum_{c=1}^{m}\gamma_c^2]^{-1}\) are the corresponding effective sample
sizes. With uniform weights and equal cell-specific cluster sizes, the
fixed-design effective sample size reduces to \(m_j\). With uniform weights
and equal cluster sizes in the random design, it reduces to \(m\). Under
unequal weights or unequal cluster sizes, the fixed- and random-design
half-widths should be compared coordinate by coordinate; neither formula
uniformly dominates the other.

\subsection{Asymptotic Inference}\label{sec:cluster:asymptotic}

The finite-sample regions above are valid without asymptotics but can be
wide when the effective number of clusters is small. We now give
cluster-level normal approximations. The coverage result is
stated for any variance estimator that is asymptotically conservative for
the corresponding cluster variance. In the empirical implementation, we use
the following centered cluster-score estimators. For the fixed design, let
\[
\overline{\Upsilon}_{j,\mathrm{cl}}^F
\equiv m_j^{-1}\sum_{c=1}^{m_j}\Upsilon_{j,c},
\qquad
\hat V_j^F
\equiv
\frac{1}{N_j}\frac{m_j}{m_j-1}
\sum_{c=1}^{m_j}
\left(\Upsilon_{j,c}-\overline{\Upsilon}_{j,\mathrm{cl}}^F\right)^2.
\]
For the random design, let
\[
\overline{\Upsilon}_{j,\mathrm{cl}}^R
\equiv m^{-1}\sum_{c=1}^{m}\Upsilon_{j,c},
\qquad
\hat V_j^R
\equiv
\frac{1}{N}\frac{m}{m-1}
\sum_{c=1}^{m}
\left(\Upsilon_{j,c}-\overline{\Upsilon}_{j,\mathrm{cl}}^R\right)^2.
\]
In the random-design formula, clusters with no observations at \(x_j\)
contribute \(\Upsilon_{j,c}=0\). These estimators subtract the empirical
cluster mean of the score contribution, instead of using the more
conservative uncentered second moment.

All limits below are taken along sequences in which the number
of clusters diverges: \(\min_j m_j\to\infty\) in the fixed design case and
\(m\to\infty\) in the random design case. The threshold \(\tau\) is fixed.

\begin{assumption}\label{a:cluster:asymp}
For the maintained sampling design and for each \(j=1,\ldots,J\), the
corresponding cluster triangular array satisfies the following conditions.
\begin{enumerate}
\item If the fixed design is maintained, define
\[
Z_{j,c}^F\equiv \Upsilon_{j,c}-\mathbb{E}(\Upsilon_{j,c}),
\qquad
V_j^F\equiv N_j^{-1}\sum_{c=1}^{m_j}\mathbb{E}\{(Z_{j,c}^F)^2\}.
\]
Assume \(V_j^F>0\) for all sufficiently large samples and, for some
\(\eta>0\),
\[
\frac{\sum_{c=1}^{m_j}\mathbb{E}|Z_{j,c}^F|^{2+\eta}}
{\left[\sum_{c=1}^{m_j}\mathbb{E}\{(Z_{j,c}^F)^2\}\right]^{1+\eta/2}}
\rightarrow 0.
\]
The variance estimator \(\hat V_j^F\) is nonnegative and satisfies
\[
\hat V_j^F/V_j^F\rightarrow_p \kappa_j^F
\quad\text{for some }\kappa_j^F\geq 1.
\]

\item If the random design is maintained, define
\[
Z_{j,c}^R\equiv \Upsilon_{j,c}-\mathbb{E}(\Upsilon_{j,c}),
\qquad
V_j^R\equiv N^{-1}\sum_{c=1}^{m}\mathbb{E}\{(Z_{j,c}^R)^2\}.
\]
Assume \(V_j^R>0\) for all sufficiently large samples and, for some
\(\eta>0\),
\[
\frac{\sum_{c=1}^{m}\mathbb{E}|Z_{j,c}^R|^{2+\eta}}
{\left[\sum_{c=1}^{m}\mathbb{E}\{(Z_{j,c}^R)^2\}\right]^{1+\eta/2}}
\rightarrow 0.
\]
The variance estimator \(\hat V_j^R\) is nonnegative and satisfies
\[
\hat V_j^R/V_j^R\rightarrow_p \kappa_j^R
\quad\text{for some }\kappa_j^R\geq 1.
\]
\end{enumerate}
\end{assumption}

The constants \(\kappa_j^F\) and \(\kappa_j^R\) allow the variance
estimators to be conservative. Consistent cluster-robust variance estimators
correspond to \(\kappa_j^F=1\) or \(\kappa_j^R=1\). The
centered estimators displayed above are the estimators used for the
clustered empirical example in Section~\ref{sec:CHV}.

For the empirical application in Section~\ref{sec:CHV}, we use the asymptotic confidence regions in
\eqref{G:fixed:box:asymp:cluster} and
\eqref{G:random:box:asymp:cluster} with \(J=104\), \(\alpha=0.05\), survey
sampling weights, and grid-cell clusters. A support-point sign restriction is
imposed only when the corresponding Bonferroni interval for \(g_0(x_j)\) lies
strictly on one side of zero: \(x_j\) is selected as positive if
\(\hat g(x_j)-h_{j,\alpha}>0\), and selected as negative if
\(\hat g(x_j)+h_{j,\alpha}<0\), where \(h_{j,\alpha}\) is
\(h_{j,\alpha}^F\) or \(h_{j,\alpha}^R\) depending on the maintained sampling
design. The selected sign restrictions are then used in the linear programs in
Section~\ref{sec:computation}. This is the construction used for the
confidence-region maximum-score results in Table~\ref{tab-chv1} and
Figure~\ref{fig-chv4}; the finite-sample regions in
Proposition~\ref{prop:cluster:finite} are reported for completeness but are not
used in the empirical results in Section~\ref{sec:CHV}.

\begin{proposition}\label{prop:cluster:asymp}
Let Assumption~\ref{a:cluster} hold, and impose the part of
Assumption~\ref{a:cluster:asymp} corresponding to the maintained sampling
design.
\begin{enumerate}
\item In the fixed design case, set
\[
h_{j,\alpha}^F
\equiv
\frac{N_j^{1/2}(\hat V_j^F)^{1/2}}{N}
z_{1-\alpha/(2J)}
\]
where \(z_{1-\alpha/(2J)}\) denotes the
\((1-\alpha/(2J))\)-quantile of the standard normal distribution; the factor
\(2J\) reflects two-sided marginal coverage and the Bonferroni correction over
\(j=1,\ldots,J\),
and
\begin{align}\label{G:fixed:box:asymp:cluster}
\mathcal{G}_\alpha =
\left\{
(g_1,\ldots,g_J): \hat{g}(x_j)-h_{j,\alpha}^F
\leq g_j \leq \hat{g}(x_j)+h_{j,\alpha}^F
\; \forall j
\right\}.
\end{align}
Then \(\liminf \mathbb{P}(g_0\in\mathcal{G}_\alpha)\geq 1-\alpha\).

\item In the random design case, set
\[
h_{j,\alpha}^R
\equiv
\left(\frac{\hat V_j^R}{N}\right)^{1/2}
z_{1-\alpha/(2J)}
\]
and
\begin{align}\label{G:random:box:asymp:cluster}
\mathcal{G}_\alpha =
\left\{
(g_1,\ldots,g_J): \hat{g}(x_j)-h_{j,\alpha}^R
\leq g_j \leq \hat{g}(x_j)+h_{j,\alpha}^R
\; \forall j
\right\}.
\end{align}
Then \(\liminf \mathbb{P}(g_0\in\mathcal{G}_\alpha)\geq 1-\alpha\).
\end{enumerate}
\end{proposition}

\begin{proof}[Proof of Proposition~\ref{prop:cluster:asymp}]
For the fixed design,
\[
\hat g(x_j)-g_0(x_j)
=N^{-1}\sum_{c=1}^{m_j}Z_{j,c}^F.
\]
By the fixed-design part of Assumption~\ref{a:cluster:asymp} and the
Lyapunov central limit theorem,
\[
\frac{N\{\hat g(x_j)-g_0(x_j)\}}
{(N_jV_j^F)^{1/2}}
\rightarrow_d N(0,1).
\]
Since \(\hat V_j^F/V_j^F\rightarrow_p\kappa_j^F\geq 1\), the displayed
fixed-design interval has asymptotic marginal coverage at least
\(1-\alpha/J\) for each \(j\). Bonferroni gives the joint coverage claim.

For the random design,
\[
\hat g(x_j)-g_0(x_j)
=N^{-1}\sum_{c=1}^{m}Z_{j,c}^R.
\]
By the random-design part of Assumption~\ref{a:cluster:asymp}, the
Lyapunov central limit theorem gives
\[
\frac{N^{1/2}\{\hat g(x_j)-g_0(x_j)\}}{(V_j^R)^{1/2}}
\rightarrow_d N(0,1).
\]
The corresponding conservative variance condition and Bonferroni then imply
the stated joint asymptotic coverage.
\end{proof}

The sampling design under consideration determines whether
\eqref{G:fixed:box:asymp:cluster} or \eqref{G:random:box:asymp:cluster} is the
relevant confidence region; the two need not coincide in general.

\subsection{Additional Empirical Classification Result}
\label{subsec:chv-logit-classification}

\begin{figure}[!htbp]
\vspace{1ex}
\begin{center}
\includegraphics[scale=0.70]{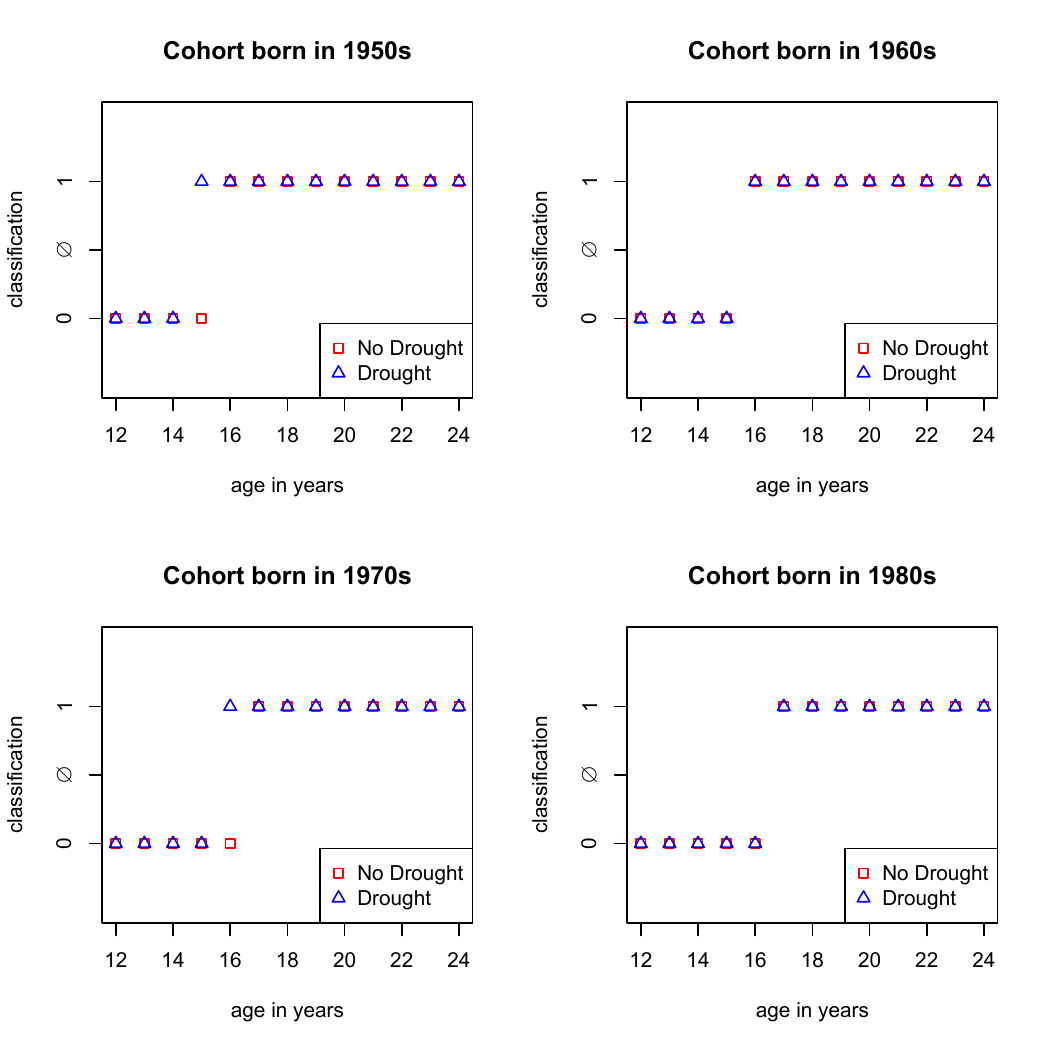}
\end{center}
\caption{Weighted Logit Classification Rule}\label{fig-chv-logit}
\end{figure}

Figure~\ref{fig-chv-logit} reports the weighted logit classification rule using
the same covariates, survey weights, and threshold \(\tau=0.12\) as the
age-of-marriage application in Section~\ref{sec:CHV}. A covariate combination
is classified as \(1\) when its fitted probability exceeds \(0.12\),
equivalently when the fitted logit index exceeds
\(\log\{\tau/(1-\tau)\}\approx-1.99\); otherwise it is classified as \(0\). Thus,
unlike the confidence-region maximum score rule, the logit rule never abstains.
It classifies 70 of the 104 age-drought-cohort combinations as \(1\) and 34 as
\(0\). Relative to the sample maximum score rule in Figure~\ref{fig-chv3}, it
differs in only four combinations. The logit rule assigns \(0\), rather than
\(1\), at age 15 for the 1950 cohort without drought, age 14 for the 1950
cohort with drought, and age 15 for the 1960 cohort with drought; it assigns
\(1\), rather than \(0\), at age 17 for the 1980 cohort without drought. The
corresponding fitted probabilities are \(0.114\), \(0.100\), \(0.111\), and
\(0.123\), respectively, so all four disagreements are localized near the
\(0.12\) threshold. Both rules display the same pronounced cohort pattern. By
comparison, the confidence-region rule in Figure~\ref{fig-chv4} leaves 12
combinations unclassified after incorporating sampling uncertainty.

\section{Comparison of Finite-Sample and Asymptotic Fixed-Design Inference}\label{sec:MC}

This section uses the same fixed covariate design and outcome-generating
process as the classification experiment in
Section~\ref{sec:mc:class-errors}, but studies the sampling behavior of the
coefficient-bound endpoints rather than classification disagreement at an
off-support covariate value. The outcome is generated by
\begin{align*}
Y_i &= I(Y_i^\ast\geq0),\\
Y_i^\ast
&= \mathrm{GPA}_i
 +0.4\,\mathrm{TopInternship}_i-3.7+U_i,
\end{align*}
where \(U_i=\sigma(X_i)V_i\) and \(V_i\sim N(0,1)\) independently across
observations. We use the homoskedastic scale \(\sigma(X_i)=0.5\) and the
heteroskedastic scale
\[
\sigma(X_i)
=
0.2\left[1+\left\{\frac{\mathrm{GPA}_i}{3}
+\mathrm{TopInternship}_i\right\}^2\right].
\]
The covariates are held fixed; only the shocks and outcomes are redrawn.

The original design has \(n_0=2{,}880\) observations and \(J=12\)
support cells. For \(c\in\{1,2,3,5,10,20,30\}\), we repeat the empirical
covariate design \(c\) times, so \(n_c=cn_0\) and \(n_{j,c}=cn_{j,0}\);
the resulting sample sizes range from \(2{,}880\) to \(86{,}400\).

The coefficient on GPA is normalized to one, and the reported coefficient is
\(\beta_2\), the coefficient on Top Internship. The population sign
restrictions at covariate values with \(\mathrm{TopInternship}=0\) imply
\(-3.8\leq\beta_3\leq-3.6\). At covariate values with
\(\mathrm{TopInternship}=1\), they imply
\(-3.4\leq\beta_2+\beta_3\leq-3.2\). Combining these restrictions, the
baseline \(\varepsilon=0\) outer set gives
\(0.2\leq\beta_2\leq0.6\).

For each error design and sample size, we use 1,000 Monte Carlo replications,
\(\tau=0.5\), and \(1-\alpha=0.95\). Each free coefficient is restricted to
\([-10,10]\). The finite-sample and asymptotic procedures are computed from
the same outcome draw within each replication. The tables report the mean and
standard deviation of the estimated lower and upper endpoints. Coverage is
the proportion of replications in which the estimated interval contains the
entire population identified interval \([0.2,0.6]\), rather than only the
generating value \(\beta_{0,2}=0.4\).

\subsection{Finite-Sample Fixed-Design Inference}

Table~\ref{tab-mc-est} uses the finite-sample fixed-design confidence region
in \eqref{G:fixed:box}. The resulting intervals are conservative at small
sample sizes, especially under heteroskedasticity. Under homoskedasticity, the
mean interval is close to the population interval by \(5n_0\) and equals it
by \(10n_0\). Under heteroskedasticity, the mean interval narrows from
\([-0.328,8.553]\) at \(n_0\) to \([0.183,0.615]\) at \(20n_0\) and
\([0.199,0.601]\) at \(30n_0\). Coverage is one in every reported design.

\input{results/table_mc_est_finite_fixed.tex}

\subsection{Asymptotic Fixed-Design Inference}

Table~\ref{tab-mc-est-asymp} repeats the exercise using the asymptotic
fixed-design confidence region in \eqref{G:fixed:box:asymp}. In this
simulation, the asymptotic intervals are tighter than the finite-sample
intervals at moderate sample sizes. Under homoskedasticity, the mean interval
is \([-0.029,0.915]\) at \(n_0\), is nearly equal to the population interval
at \(5n_0\), and equals it by \(10n_0\). Under heteroskedasticity, it narrows
from \([-0.185,6.248]\) at \(n_0\) to \([0.196,0.604]\) at \(20n_0\) and
\([0.200,0.600]\) at \(30n_0\). Coverage is again one throughout.

Here, asymptotic inference is more informative at moderate
sample sizes; the larger cell counts required by the Hoeffding guarantee
explain why finite-sample inference is omitted from the main-text application
at \(n=2{,}880\).

\input{results/table_mc_est_asymptotic_fixed.tex}

\end{document}

%% file: results/table_kls_main.tex
\begin{table}[htbp]
\caption{Interval estimates for coefficients across different methods}\label{tab-kls1}
\centering
\small
\begin{tabular}{lcccc}
  \hline
  & (1) & (2) & (3) & (4) \\
 & Logit & MS: Sample & MS: Fixed Design & MS: Random Design \\
  \hline
Top Internship & [0.32, 0.58] & [0.2, 0.6] & [$-0.2$, 1.0] & [$-0.2$, 1.0] \\
 Intercept & [$-3.73$, $-3.59$] & [$-3.6$, $-3.4$] & [$-4.0$, $-3.4$] & [$-4.0$, $-3.4$] \\
   \hline
\end{tabular}
\end{table}

%% file: results/table_kls_interval_main.tex
\begin{table}[htbp]
\caption{Coefficient Intervals with Discrete and Interval-Valued GPA}\label{tab-kls1-xv}
\centering
\small
\begin{tabular}{lcccc}
  \hline
  & (1) & (2) & (3) & (4) \\
  & \multicolumn{2}{c}{GPA as Discrete}  & \multicolumn{2}{c}{GPA as Interval Data} \\
 & Fixed Design & Random Design & Fixed Design & Random Design \\
  \hline
Top Internship & [$-0.2$, 1.0] & [$-0.2$, 1.0] & [$-0.4$, 1.1] & [$-0.4$, 1.1] \\
 Intercept & [$-4.0$, $-3.4$] & [$-4.0$, $-3.4$] & [$-4.0$, $-3.2$] & [$-4.0$, $-3.2$] \\
   \hline
\end{tabular}
\end{table}

%% file: results/table_chv.tex
\begin{table}[htbp]
\caption{Coefficient estimates and intervals}\label{tab-chv1}
\small
\begin{tabular}{lccc}
  \hline
  & (1) & (2) & (3) \\
 & Logit & MS: Sample & MS: 95\% CR \\
  \hline
Drought & [0.3, 0.6] & [1, 1] & [$-1$, 2]  \\
Birth Cohort 60 & [$-0.7$, $-0.4$] & [$-1$, 0] & [$-2$, 1]  \\
Birth Cohort 70 & [$-1.4$, $-1.0$] & [$-2$, $-1$] & [$-3$, 1]  \\
Birth Cohort 80 & [$-1.9$, $-1.4$] & [$-3$, $-2$] & [$-3$, 0]  \\
Intercept & [2.6, 3.0] & [3, 4] & [3, 4]  \\
   \hline
\end{tabular}
\end{table}

%% file: results/Table_mc_class_errors.tex
\begin{table}[hbtp]
\caption{Monte Carlo Simulation Results: Classification Errors}\label{tab-mc-class-errors}
\centering
\small
\begin{tabular}{crrrrrr}
  \hline\hline
  \multicolumn{7}{l}{Panel A. Abstention Allowed} \\
 \hline
$\alpha$ & \multicolumn{3}{c}{Homoskedastic} & \multicolumn{3}{c}{Heteroskedastic} \\
 & $n=2880$ & $2n=5760$ & $3n=8640$ & $n=2880$ & $2n=5760$ & $3n=8640$ \\
  \hline
0.01 & 0.693 & 0.231 & 0.055 & 0.722 & 0.288 & 0.094 \\
0.05 & 0.512 & 0.117 & 0.015 & 0.558 & 0.163 & 0.030 \\
0.10 & 0.426 & 0.083 & 0.007 & 0.479 & 0.127 & 0.017 \\
 \hline
  \multicolumn{7}{l}{Panel B. Random Classification} \\
 \hline
$\alpha$ & \multicolumn{3}{c}{Homoskedastic} & \multicolumn{3}{c}{Heteroskedastic} \\
 & $n=2880$ & $2n=5760$ & $3n=8640$ & $n=2880$ & $2n=5760$ & $3n=8640$ \\
  \hline
0.01 & 0.325 & 0.109 & 0.028 & 0.381 & 0.141 & 0.057 \\
0.05 & 0.240 & 0.054 & 0.003 & 0.302 & 0.073 & 0.018 \\
0.10 & 0.194 & 0.041 & 0.002 & 0.252 & 0.055 & 0.012 \\
 \hline
\end{tabular}
\vspace*{2ex}
\begin{minipage}{0.82\textwidth}
\footnotesize
Notes: Here, $n=2880$ refers to the original sample size in the Section~\ref{sec:KLS} application.
Classification errors are evaluated at the off-support target value
\(\mathrm{GPA}=3.9\), Top Internship \(=0\).
\end{minipage}
\end{table}

%% file: results/table_kls_multi_eps.tex
\begin{table}[htbp]
\caption{Coefficient Intervals across Methods and Values of $\varepsilon$}\label{tab-kls1-multi}
\centering
\small
\setlength{\tabcolsep}{4pt}
\begin{tabular}{lcccc}
\hline\hline
& (1) & (2) & (3) & (4) \\
& Logit & MS: Sample & MS: Fixed Design & MS: Random Design \\
\hline
\multicolumn{5}{l}{\textbf{$\varepsilon = 0$}} \\
\hline
Top Internship & [0.32, 0.58] & [0.20, 0.60] & [$-0.20$, 1.00] & [$-0.20$, 1.00] \\
Intercept & [$-3.73$, $-3.59$] & [$-3.60$, $-3.40$] & [$-4.00$, $-3.40$] & [$-4.00$, $-3.40$] \\
\hline
\multicolumn{5}{l}{\textbf{$\varepsilon = 0.02$}} \\
\hline
Top Internship & [0.32, 0.58] & [0.24, 0.56] & [$-0.16$, 0.96] & [$-0.16$, 0.96] \\
Intercept & [$-3.73$, $-3.59$] & [$-3.58$, $-3.42$] & [$-3.98$, $-3.42$] & [$-3.98$, $-3.42$] \\
\hline
\multicolumn{5}{l}{\textbf{$\varepsilon = 0.04$}} \\
\hline
Top Internship & [0.32, 0.58] & [0.28, 0.52] & [$-0.12$, 0.92] & [$-0.12$, 0.92] \\
Intercept & [$-3.73$, $-3.59$] & [$-3.56$, $-3.44$] & [$-3.96$, $-3.44$] & [$-3.96$, $-3.44$] \\
\hline
\end{tabular}
\end{table}

%% file: results/table_kls_interval_multi_eps.tex
\begin{table}[htbp]
\caption{Coefficient Intervals with Discrete and Interval-Valued GPA across Values of $\varepsilon$}\label{tab-kls1-xv-multi}
\centering
\small
\begin{tabular}{lcccc}
\hline\hline
& \multicolumn{2}{c}{GPA as Discrete} & \multicolumn{2}{c}{GPA as Interval Data} \\
& Fixed & Random & Fixed & Random \\
\hline
\multicolumn{5}{l}{\textbf{$\varepsilon = 0$}} \\
\hline
Top Internship & [$-0.20$, 1.00] & [$-0.20$, 1.00] & [$-0.40$, 1.10] & [$-0.40$, 1.10] \\
Intercept & [$-4.00$, $-3.40$] & [$-4.00$, $-3.40$] & [$-4.00$, $-3.20$] & [$-4.00$, $-3.20$] \\
\hline
\multicolumn{5}{l}{\textbf{$\varepsilon = 0.02$}} \\
\hline
Top Internship & [$-0.16$, 0.96] & [$-0.16$, 0.96] & [$-0.36$, 1.06] & [$-0.36$, 1.06] \\
Intercept & [$-3.98$, $-3.42$] & [$-3.98$, $-3.42$] & [$-3.98$, $-3.22$] & [$-3.98$, $-3.22$] \\
\hline
\multicolumn{5}{l}{\textbf{$\varepsilon = 0.04$}} \\
\hline
Top Internship & [$-0.12$, 0.92] & [$-0.12$, 0.92] & [$-0.32$, 1.02] & [$-0.32$, 1.02] \\
Intercept & [$-3.96$, $-3.44$] & [$-3.96$, $-3.44$] & [$-3.96$, $-3.24$] & [$-3.96$, $-3.24$] \\
\hline
\end{tabular}
\end{table}

%% file: results/table_kls_eps_classification_counts.tex
\begin{table}[htbp]
\caption{Maximum Score Classification Counts with Positive
$\varepsilon$}\label{tab:kls-eps-classification-summary}
\centering
\small
\begin{tabular}{llrrrr}
\hline\hline
$\varepsilon$ & GPA specification & $Y=1$ & $Y=0$ & $\emptyset$ &
Changed \\
\hline
$0$ & Discrete & 4 & 4 & 4 & 0 \\
$0$ & Interval & 4 & 2 & 6 & 0 \\
$0.02$ & Discrete & 4 & 4 & 4 & 0 \\
$0.02$ & Interval & 4 & 2 & 6 & 0 \\
$0.04$ & Discrete & 4 & 4 & 4 & 0 \\
$0.04$ & Interval & 4 & 2 & 6 & 0 \\
\hline
\end{tabular}
\vspace{0.5em}
\setlength{\fboxsep}{4pt}
\mbox{%
\begin{minipage}{0.90\textwidth}
\footnotesize
\emph{Note:} Rows report the twelve resume groups under discrete or interval
GPA. Fixed- and random-design rules coincide. ``Changed'' counts differences
from the corresponding $\varepsilon=0$ classification.
\end{minipage}}
\end{table}

%% file: results/table_kls_eps_target_endpoint.tex
\begin{table}[htbp]
\caption{Estimated Classification Endpoints for a Boundary Target Cell with Positive
$\varepsilon$}\label{tab:kls-eps-target-endpoint}
\centering
\small
\begin{tabular}{lrrrl}
\hline\hline
$\varepsilon$ & $\hat c_L(X_*;\varepsilon)$ & $\hat c_U(X_*;\varepsilon)$ & Width & Classification \\
\hline
$0$ & $-0.80$ & $+0.00$ & 0.80 & $Y=0$ \\
$0.02$ & $-0.78$ & $-0.02$ & 0.76 & $Y=0$ \\
$0.04$ & $-0.76$ & $-0.04$ & 0.72 & $Y=0$ \\
\hline
\end{tabular}
\vspace{0.5em}
\setlength{\fboxsep}{4pt}
\mbox{%
\begin{minipage}{0.90\textwidth}
\footnotesize
\emph{Note:} Fixed-design interval-GPA endpoints for Top Internship
$=0$ and grouped GPA $=3.2$. For each endpoint, the supporting
optimal basis remains unchanged over the reported $\varepsilon$ grid. The entry
$+0.00$ is numerically zero and is classified as $Y=0$
under the boundary convention.
\end{minipage}}
\end{table}

%% file: results/table_mc_est_finite_fixed.tex
\begin{table}[hbtp]
\caption{Monte Carlo Results: Finite-Sample Fixed-Design Inference}\label{tab-mc-est}
\centering
\small
\renewcommand{\arraystretch}{0.85}
\begin{tabular}{rccccc}
  \hline\hline
  \multicolumn{6}{l}{Top Internship coefficient: Population bounds $[0.2,0.6]$} \\
  \hline
Sample & \multicolumn{2}{c}{Mean} & \multicolumn{2}{c}{Std. Dev.} & Coverage \\
Size & Lower Bound & Upper Bound & Lower Bound & Upper Bound & \\
  \hline
  \multicolumn{6}{l}{Panel A. Homoskedastic Errors} \\
  \hline
$2{,}880$ & -0.129 & 1.710 & 0.107 & 2.546 & 1.000 \\
$5{,}760$ & 0.026 & 0.794 & 0.138 & 0.321 & 1.000 \\
$8{,}640$ & 0.134 & 0.683 & 0.102 & 0.110 & 1.000 \\
$14{,}400$ & 0.195 & 0.613 & 0.031 & 0.049 & 1.000 \\
$28{,}800$ & 0.200 & 0.600 & 0.000 & 0.000 & 1.000 \\
$57{,}600$ & 0.200 & 0.600 & 0.000 & 0.000 & 1.000 \\
$86{,}400$ & 0.200 & 0.600 & 0.000 & 0.000 & 1.000 \\
  \hline
  \multicolumn{6}{l}{Panel B. Heteroskedastic Errors} \\
  \hline
$2{,}880$ & -0.328 & 8.553 & 0.339 & 3.316 & 1.000 \\
$5{,}760$ & -0.115 & 5.550 & 0.131 & 4.570 & 1.000 \\
$8{,}640$ & -0.017 & 2.396 & 0.086 & 3.480 & 1.000 \\
$14{,}400$ & 0.027 & 0.939 & 0.070 & 1.229 & 1.000 \\
$28{,}800$ & 0.100 & 0.708 & 0.100 & 0.100 & 1.000 \\
$57{,}600$ & 0.183 & 0.615 & 0.056 & 0.053 & 1.000 \\
$86{,}400$ & 0.199 & 0.601 & 0.015 & 0.017 & 1.000 \\
  \hline\hline
\end{tabular}
\vspace*{0.5ex}
\begin{minipage}{0.90\textwidth}
\footnotesize Notes: The population identified interval is $[0.2,0.6]$.
The original fixed design from Section~\ref{sec:KLS} has $n_0=2{,}880$ observations; each reported sample size is $c n_0$ for $c\in\{1,2,3,5,10,20,30\}$.
The table reports 1,000 Monte Carlo replications. Coverage is the frequency with which the estimated interval contains the entire population identified interval.
\end{minipage}
\end{table}

%% file: results/table_mc_est_asymptotic_fixed.tex
\begin{table}[hbtp]
\caption{Monte Carlo Results: Asymptotic Fixed-Design Inference}\label{tab-mc-est-asymp}
\centering
\small
\renewcommand{\arraystretch}{0.85}
\begin{tabular}{rccccc}
  \hline\hline
  \multicolumn{6}{l}{Top Internship coefficient: Population bounds $[0.2,0.6]$} \\
  \hline
Sample & \multicolumn{2}{c}{Mean} & \multicolumn{2}{c}{Std. Dev.} & Coverage \\
Size & Lower Bound & Upper Bound & Lower Bound & Upper Bound & \\
  \hline
  \multicolumn{6}{l}{Panel A. Homoskedastic Errors} \\
  \hline
$2{,}880$ & -0.029 & 0.915 & 0.137 & 0.827 & 1.000 \\
$5{,}760$ & 0.118 & 0.688 & 0.115 & 0.115 & 1.000 \\
$8{,}640$ & 0.179 & 0.631 & 0.063 & 0.074 & 1.000 \\
$14{,}400$ & 0.199 & 0.603 & 0.011 & 0.025 & 1.000 \\
$28{,}800$ & 0.200 & 0.600 & 0.000 & 0.000 & 1.000 \\
$57{,}600$ & 0.200 & 0.600 & 0.000 & 0.000 & 1.000 \\
$86{,}400$ & 0.200 & 0.600 & 0.000 & 0.000 & 1.000 \\
  \hline
  \multicolumn{6}{l}{Panel B. Heteroskedastic Errors} \\
  \hline
$2{,}880$ & -0.185 & 6.248 & 0.157 & 4.485 & 1.000 \\
$5{,}760$ & -0.027 & 2.665 & 0.109 & 3.693 & 1.000 \\
$8{,}640$ & 0.033 & 1.158 & 0.084 & 1.854 & 1.000 \\
$14{,}400$ & 0.064 & 0.751 & 0.093 & 0.425 & 1.000 \\
$28{,}800$ & 0.145 & 0.654 & 0.089 & 0.089 & 1.000 \\
$57{,}600$ & 0.196 & 0.604 & 0.027 & 0.029 & 1.000 \\
$86{,}400$ & 0.200 & 0.600 & 0.000 & 0.000 & 1.000 \\
  \hline\hline
\end{tabular}
\vspace*{0.5ex}
\begin{minipage}{0.90\textwidth}
\footnotesize Notes: The population identified interval is $[0.2,0.6]$.
The original fixed design from Section~\ref{sec:KLS} has $n_0=2{,}880$ observations; each reported sample size is $c n_0$ for $c\in\{1,2,3,5,10,20,30\}$.
The table reports 1,000 Monte Carlo replications. Coverage is the frequency with which the estimated interval contains the entire population identified interval.
\end{minipage}
\end{table}